\pdfoutput=1  

\documentclass[aps,prd,superscriptaddress,floatfix,nofootinbib,preprintnumbers,eqsecnum,twocolumn]{revtex4-1}

\usepackage{amsmath,float,graphicx,placeins,amssymb,multirow,pgfplots,pgfplotstable}
\usepackage{tikz}
\usepackage{comment}
\usepackage{enumerate,slashed,cancel,comment,color,bbm,graphicx,rotating,placeins,mathtools,multirow,hhline}
\usepackage[normalem]{ulem}
\usepackage{array}

\usepackage{hyperref}
\usepackage{color}
 \hypersetup
 {
   colorlinks,
   linkcolor={blue!80!black},
   citecolor={blue!70!black},
   urlcolor={blue!70!black}
 }

\usepackage{lipsum}
\usepackage{diagbox}
\usetikzlibrary{shapes.geometric,arrows}
\graphicspath{}

\begin{document}
\title{Deciphering the Archaeological Record:\\
Cosmological Imprints of Non-Minimal Dark Sectors}


\def\andname{\hspace*{-0.5em}} 
\author{Keith R. Dienes}
\email[Email address: ]{dienes@email.arizona.edu}
\affiliation{Department of Physics, University of Arizona, Tucson, AZ 85721 USA}
\affiliation{Department of Physics, University of Maryland, College Park, MD 20742 USA}
\author{Fei Huang}
\email[Email address: ]{huangf4@uci.edu}
\affiliation{Department of Physics and Astronomy, University of California, Irvine, CA  92697 USA}
\affiliation{CAS Key Laboratory of Theoretical Physics, Institute of Theoretical Physics, Chinese Academy of Sciences, Beijing 100190 China}
\author{Jeff Kost}
\email[Email address: ]{jeffkost@ibs.re.kr}
\affiliation{\mbox{Center for Theoretical Physics of the Universe, Institute for Basic Science, Daejeon 34126 Korea}}
\author{Shufang Su}
\email[Email address: ]{shufang@email.arizona.edu}
\affiliation{Department of Physics, University of Arizona, Tucson, AZ 85721 USA}
\author{Brooks Thomas}
\email[Email address: ]{thomasbd@lafayette.edu}
\affiliation{Department of Physics, Lafayette College, Easton, PA 18042 USA}

\preprint{CTPU-PTC-19-35}

\begin{abstract}
Many proposals for physics beyond the Standard Model give rise to a dark sector containing many degrees of freedom.  In this work, we explore the cosmological implications of the non-trivial dynamics which may arise within such dark sectors, focusing on decay processes which take place entirely among the dark constituents.  First, we demonstrate that such decays can leave dramatic imprints on the resulting dark-matter phase-space distribution.  In particular, this distribution need not be thermal --- it can even be multi-modal, exhibiting a non-trivial pattern of peaks and troughs as a function of momentum.  We then proceed to show how these features can induce modifications to the matter power spectrum.  Finally, we assess the extent to which one can approach the archaeological ``inverse'' problem of deciphering the properties of an underlying dark sector from the matter power spectrum.  
Indeed, one of the main results of this paper is a 
remarkably simple conjectured analytic expression which permits the reconstruction of many of the important features of the dark-matter 
phase-space distribution directly from the matter power spectrum.
Our results therefore provide an interesting toolbox of methods 
for learning about, and potentially constraining, the features of non-minimal dark sectors and their dynamics in the early universe.

\end{abstract}

\maketitle
\tableofcontents


\newcommand{\PRE}[1]{{#1}} 
\newcommand{\ul}{\underline}
\newcommand{\del}{\partial}
\newcommand{\nbox}{{\,\lower0.9pt\vbox{\hrule \hbox{\vrule height 0.2 cm
\hskip 0.2 cm \vrule height 0.2 cm}\hrule}\,}}

\newcommand{\postscript}[2]{\setlength{\epsfxsize}{#2\hsize}
   \centerline{\epsfbox{#1}}}
\newcommand{\gweak}{g_{\text{weak}}}
\newcommand{\mweak}{m_{\text{weak}}}
\newcommand{\mplanck}{M_{\text{Pl}}}
\newcommand{\mstar}{M_{*}}
\newcommand{\sigmaan}{\sigma_{\text{an}}}
\newcommand{\sigmatot}{\sigma_{\text{tot}}}
\newcommand{\sigmaSI}{\sigma_{\rm SI}}
\newcommand{\sigmaSD}{\sigma_{\rm SD}}
\newcommand{\OmegaM}{\Omega_{\text{M}}}
\newcommand{\OmegaDM}{\Omega_{\text{DM}}}
\newcommand{\ipb}{\text{pb}^{-1}}
\newcommand{\ifb}{\text{fb}^{-1}}
\newcommand{\iab}{\text{ab}^{-1}}
\newcommand{\ev}{\text{eV}}
\newcommand{\kev}{\text{keV}}
\newcommand{\mev}{\text{MeV}}
\newcommand{\gev}{\text{GeV}}
\newcommand{\tev}{\text{TeV}}
\newcommand{\pb}{\text{pb}}
\newcommand{\mb}{\text{mb}}
\newcommand{\cm}{\text{cm}}
\newcommand{\m}{\text{m}}
\newcommand{\km}{\text{km}}
\newcommand{\kg}{\text{kg}}
\newcommand{\g}{\text{g}}
\newcommand{\s}{\text{s}}
\newcommand{\yr}{\text{yr}}
\newcommand{\Mpc}{\text{Mpc}}
\newcommand{\etal}{{\em et al.}}
\newcommand{\eg}{{\em e.g.}}
\newcommand{\ie}{{\em i.e.}}
\newcommand{\ibid}{{\em ibid.}}
\newcommand{\Eqref}[1]{Equation~(\ref{#1})}
\newcommand{\secref}[1]{Sec.~\ref{sec:#1}}
\newcommand{\secsref}[2]{Secs.~\ref{sec:#1} and \ref{sec:#2}}
\newcommand{\Secref}[1]{Section~\ref{sec:#1}}
\newcommand{\appref}[1]{App.~\ref{sec:#1}}
\newcommand{\figref}[1]{Fig.~\ref{fig:#1}}
\newcommand{\figsref}[2]{Figs.~\ref{fig:#1} and \ref{fig:#2}}
\newcommand{\Figref}[1]{Figure~\ref{fig:#1}}
\newcommand{\tableref}[1]{Table~\ref{table:#1}}
\newcommand{\tablesref}[2]{Tables~\ref{table:#1} and \ref{table:#2}}
\newcommand{\Dsle}[1]{\slash\hskip -0.28 cm #1}
\newcommand{\met}{{\Dsle E_T}}
\newcommand{\mpt}{\not{\! p_T}}
\newcommand{\Dslp}[1]{\slash\hskip -0.23 cm #1}
\newcommand{\Dsl}[1]{\slash\hskip -0.20 cm #1}

\newcommand{\mB}{m_{B^1}}
\newcommand{\mq}{m_{q^1}}
\newcommand{\mf}{m_{f^1}}
\newcommand{\mKK}{m_{KK}}
\newcommand{\WIMP}{\text{WIMP}}
\newcommand{\SWIMP}{\text{SWIMP}}
\newcommand{\NLSP}{\text{NLSP}}
\newcommand{\LSP}{\text{LSP}}
\newcommand{\mWIMP}{m_{\WIMP}}
\newcommand{\mSWIMP}{m_{\SWIMP}}
\newcommand{\mNLSP}{m_{\NLSP}}
\newcommand{\mchi}{m_{\chi}}
\newcommand{\mgravitino}{m_{\gravitino}}
\newcommand{\mmed}{M_{\text{med}}}
\newcommand{\gravitino}{\tilde{G}}
\newcommand{\Bino}{\tilde{B}}
\newcommand{\photino}{\tilde{\gamma}}
\newcommand{\stau}{\tilde{\tau}}
\newcommand{\slepton}{\tilde{l}}
\newcommand{\snu}{\tilde{\nu}}
\newcommand{\squark}{\tilde{q}}
\newcommand{\mgaugino}{M_{1/2}}
\newcommand{\epsEM}{\varepsilon_{\text{EM}}}
\newcommand{\mmess}{M_{\text{mess}}}
\newcommand{\lmess}{\Lambda}
\newcommand{\nmess}{N_{\text{m}}}
\newcommand{\signmu}{\text{sign}(\mu)}
\newcommand{\Omegachi}{\Omega_{\chi}}
\newcommand{\lambdafs}{\lambda_{\text{FS}}}
\newcommand{\be}{\begin{equation}}
\newcommand{\ee}{\end{equation}}
\newcommand{\bea}{\begin{eqnarray}}
\newcommand{\eea}{\end{eqnarray}}
\newcommand{\beq}{\begin{equation}}
\newcommand{\eeq}{\end{equation}}
\newcommand{\beqn}{\begin{eqnarray}}
\newcommand{\eeqn}{\end{eqnarray}}
\newcommand{\baln}{\begin{align}}
\newcommand{\ealn}{\end{align}}
\newcommand{\lsim}{\lower.7ex\hbox{{\il{\;\stackrel{\textstyle<}{\sim}\;}}}}
\newcommand{\gsim}{\lower.7ex\hbox{{\il{\;\stackrel{\textstyle>}{\sim}\;}}}}

\newcommand{\ssection}[1]{{\em #1.\ }}
\newcommand{\rem}[1]{\textbf{#1}}
\newcommand{\nn}{\nonumber}
\newcommand{\expt}[1]{\left\langle #1 \right\rangle}
\newcommand{\il}[1]{\mbox{$#1$}}
\newcommand{\abs}[1]{\left| #1 \right|}

\def\ie{{\it i.e.}\/}
\def\eg{{\it e.g.}\/}
\def\etc{{\it etc}.\/}
\def\calN{{\cal N}}
\def\calP{{\cal P}}
\def\calO{{\cal O}}

\def\mptwo{{m_{\pi^0}^2}}
\def\mp{{m_{\pi^0}}}
\def\sqtsn{\sqrt{s_n}}
\def\sqtsn{\sqrt{s_n}}
\def\sqtsn{\sqrt{s_n}}
\def\sqts0{\sqrt{s_0}}
\def\Dsqts{\Delta(\sqrt{s})}
\def\Omegatot{\Omega_{\mathrm{tot}}}
\def\tnow{t_{\mathrm{now}}}
\def\rhocrit{\rho_{\mathrm{crit}}}
\def\half{{\textstyle{1\over 2}}}
\def\quarter{{\textstyle{1\over 4}}}
\def\rhobar{{\bar{\rho}}}

\section{Introduction\label{sec:Introduction}}


Dark matter remains one of the great mysteries of modern physics.
We know that it exists, and through its gravitational interactions we know how much of 
it there has been at different epochs during the evolution of the universe.
We also know that it is approximately pressureless (with equation of state {\il{w\approx 0}}), and that it is relatively cold.
However, our knowledge of the dark sector is extremely limited.
We do not know how many species of particles comprise the dark sector,
nor do we know their masses or spins or whether they are fundamental or composite.
We likewise do not know whether the dark matter interacts non-gravitationally with the visible sector,
nor do we know much about these interactions if they do occur.
We do not even know if the dark-sector constituents interact non-gravitationally with each other.
As a result, the dark sector remains one of 
the most compelling enigmas facing physics today.

We likewise do not understand the {\it origins}\/ of the dark sector.
We have no idea what production mechanism originally populates this sector,
or whether this mechanism is even thermal.
Likewise, we do not know what kinds of non-trivial
dynamics might be involved in establishing the dark matter
that we observe today.

Tackling these problems would not be so urgent if the dark sector were not so important.
However, the total energy density of the dark sector
is approximately five times that of the visible sector.
As a result, it is primarily the dark physics which drives the evolution of the
universe through much of cosmological history.
Likewise, dark matter seeds structure
formation and thereby gives rise to the structure-filled universe 
that we observe today.

This then leads to two critical questions:
\begin{itemize}
\item  What imprints might non-trivial early-universe 
dark-sector dynamics leave in present-day observables?
\item To what extent can we {\it decipher}\/ this archaeological record,
exploiting information about the present-day universe in order
to learn about or  constrain the properties of the
dark sector?
\end{itemize}

These are clearly very broad questions, and in this paper we shall attempt to take a step
towards addressing these questions.
In particular, we shall concentrate on 
the linear matter power spectrum $P(k)$, which describes the spatial distribution of
matter.
As such, this quantity carries an immense amount of information regarding how
structure was formed in the early universe, 
as sketched below:
the matter power spectrum $P(k)$ depends on the net dark-matter phase-space
distribution $f(p)$, and this in turn 
is highly sensitive to the early-universe
dynamics we wish to constrain.

\begin{widetext}
\medskip
\begin{center}
   \fbox{ \begin{minipage}{0.9 truein}
    \begin{center}
      early-universe\\
        dynamics
    \end{center}
   \end{minipage}}
    {\il{~~\longrightarrow~~}}
   \fbox{ \begin{minipage}{1.15 truein}
    \begin{center}
     dark-matter\\
       phase-space \\ 
       distribution \\
          $f(p)$ 
     \end{center}
   \end{minipage}}
    {\il{~~\longrightarrow~~}}
   \fbox{ \begin{minipage}{1.55 truein}
    \begin{center}
      matter\\
     power spectrum\\ 
      $P(k)$
     \end{center}
  \end{minipage}}
\end{center}
\vskip -0.3 truein
\beq
\label{onepointone}
\eeq
\end{widetext}
\bigskip


Clearly, a given early-universe dynamics leads to a specific $f(p)$ and then to a
specific $P(k)$.  However, this process is not invertible;  in fact, the mappings
sketched above may not even be one-to-one. 
Nevertheless, we can ask:  To what extent can we find {\it signatures}\/ or
{\it patterns}\/ in $f(p)$ and $P(k)$ which might give us at least {\it partial}\/ 
information about the early-universe
dynamics that produced the dark matter? 

Answering this question is the primary goal of this paper.
Note that we see this exercise as having two primary motivations beyond those outlined above.
First, it is only the matter power spectrum $P(k)$ which is ultimately observable;  by contrast,
$f(p)$ and the early-universe dynamics which produces it are not observationally accessible. 
Thus, learning how to approach this ``inverse'' problem --- even in a rough, approximate way --- 
will ultimately become increasingly urgent
as further observational data is accumulated.
But perhaps even more critically, it is possible that the dark matter interacts with the visible sector
only gravitationally.
This would be unfortunate, as presumed non-gravitational interactions
between the dark and visible sectors are the underpinnings 
of all collider-based,  direct-detection, and indirect-detection dark-matter search experiments.
Thus, if the dark sector interacts with the visible sector too weakly,
it may ultimately only be through studies of quantities such as the matter power spectrum
that we will ever learn about the dark sector and its early-universe dynamics.

This paper is organized in two parts.
The first part, consisting of Sects.~\ref{sec:phase_space_evolution} and \ref{sec:perturb}, is primarily concerned with 
explorations
of the two connections sketched in Eq.~(\ref{onepointone}) above, with Sect.~\ref{sec:phase_space_evolution} devoted to explorations of how we might uncover aspects of the early-universe dynamics by studying the dark-matter phase-space distribution $f(p)$, and Sect.~\ref{sec:perturb} devoted to explorations of how we might uncover aspects of $f(p)$ given a particular matter power spectrum $P(k)$.
As discussed above, our goals are merely to observe and interpret certain patterns and signatures.
Although a complete inverse map is not possible, we shall nevertheless demonstrate
there are many ways in which we can ``invert'' certain aspects of the mappings in Eq.~(\ref{onepointone}),
and indeed in Sect.~\ref{sec:perturb} we shall conjecture a remarkably simple closed-form expression which 
will enable us to 
reconstruct many of the salient features of the underlying dark-matter phase-space distribution $f(p)$, 
given a particular matter power spectrum $P(k)$. 

By contrast, the second part of this paper, consisting of Sect.~\ref{sec:toy_model}, presents the detailed analysis of an explicit
example model which illustrates all of our main points.
In particular, we shall begin with an explicit Lagrangian describing a hypothetical non-minimal dark sector at early times,
and we shall then demonstrate that the dynamics implied by this Lagrangian indeed leaves the predicted imprints in $f(p)$ and $P(k)$.
As such, this model will permit us to perform a complete ``end-to-end'' analysis of the connections sketched in Eq.~(\ref{onepointone}). 
We shall also take the opportunity to utilize our conjectured relation from 
Sect.~\ref{sec:perturb} in order to test our ability to reconstruct many features of $f(p)$ from $P(k)$, 
and thereby demonstrate that our conjecture is indeed remarkably accurate for this purpose.
Finally, we conclude in Sect.~\ref{sec:intra_decay_conclusion} 
with a discussion of our main results and possible future research directions.

This paper also contains four Appendices.
Appendix~\ref{app:boltzmann} discusses certain aspects of the 
Boltzmann equations which underlie the physics of this paper, and 
also presents the specific Boltzmann equations 
that are used
in our analysis of the model in Sect.~\ref{sec:toy_model}.~
In Appendix~\ref{app:decay_example} we then use these Boltzmann equations
in order to provide a full numerical example of certain results quoted in Sect.~\ref{sec:phase_space_evolution}.~ 
By contrast, Appendix~\ref{app:sound_speed} 
contains a short derivation of the adiabatic sound speed associated with dark matter of a given momentum,
as background for a technical point to be discussed in Sect.~\ref{sec:perturb}.~
Finally, 
Appendix~\ref{app:time_evol} provides details concerning the time-evolution of the dark
sector in our model in Sect.~\ref{sec:toy_model} --- details which are likely to have relevance for the behavior
of non-minimal dark sectors more broadly, even beyond the specific model studied in this paper.

There are, of course, many different theoretical possibilities for early-universe cosmology, 
each giving rise to different dynamical patterns and 
different resulting behaviors for $f(p)$ and $P(k)$.
For this reason, we stress that it is not our goal in this paper to advocate for a particular
model of the early universe, or even to attempt a complete survey of all logical possibilities.
Rather, our goal 
is to develop tools which can be exploited
quite generally ---  not only to recognize and interpret the observational signatures of various dark sectors,
but also to ultimately constrain their properties.
Of course, we shall find that the particular dark-matter phase-space distributions $f(p)$ 
which interest us the most are those which exhibit 
non-standard features such as multi-modality, 
with many identifiable peaks and troughs.   As we shall demonstrate, such distributions
emerge quite naturally from {\it non-minimal}\/ dark sectors --- sectors which transcend the 
typical WIMP paradigm.
As such, many of our results will be particularly useful in such cases.
Our results, however, will be completely general, and will hold even for more minimal
theories of the early universe.


\section{Packets to packets, dust to dust: \hfill\break
    From early-universe dynamics to dark-matter phase-space distributions 
\label{sec:phase_space_evolution}}

In this section we shall discuss a variety of issues pertaining to the first connection sketched in Eq.~(\ref{onepointone}),
and the degree to which this connection might be inverted.

\subsection{The cosmological conveyor belt}  

Once dark matter has been produced in the early universe, its
properties can be described through its phase-space distribution $f(\vec x, \vec p, t)$.
The assumption of spatial homogeneity to first order
simplifies this quantity to $f(\vec p, t)$.
Given $f(\vec p,t)$, we can calculate the corresponding
number density $n(t)$,
energy density $\rho(t)$,
and pressure $P(t)$ via
\beqn
n(t) & \equiv &  g_{\rm int} \, \int {d^3 \vec p\over (2\pi)^3} \, f(\vec p, t)\nonumber\\
\rho(t) & \equiv &  g_{\rm int} \, \int {d^3 \vec p\over (2\pi)^3} \, E(p)\, f(\vec p, t)\nonumber\\
P(t) & \equiv &  g_{\rm int} \, \int {d^3 \vec p\over (2\pi)^3} \, {p^2 \over 3E(p) } \, f(\vec p, t)~,
\label{fdefs}
\eeqn
where {\il{E(p)\equiv \sqrt{p^2+m^2}}} with
{\il{p\equiv |\vec p|}}
and 
where $g_{\rm int} $ is the number of internal degrees of freedom.  
We can even calculate the corresponding equation of state {\il{w(t) \equiv P(t)/\rho(t)}}.
The dark-matter distribution $f(\vec p,t)$ is therefore a central quantity in understanding the
cosmological properties of the dark sector, telling us whether
the dark matter is cold or hot, thermal or non-thermal, and so forth.
Indeed, for non-relativistic (``cold'') dark matter we have {\il{E(p)\sim m \gg p}} and therefore {\il{w\approx 0}}, while
for relativistic (``hot'') dark matter we have {\il{E(p)\sim p\gg m}} and therefore {\il{w\approx 1/3}}.
Likewise, if the dark matter is thermal (\ie, produced while in thermal equilibrium with a heat bath), 
we expect $f(\vec p,t)$ to follow a Bose-Einstein or Fermi-Dirac distribution, while 
all other $f(\vec p,t)$ are necessarily non-thermal.
Of course, we are assuming here that the dark matter consists of only a single species.
If the dark matter consists of multiple species {\il{i=0,1,...,N}}, the dark sector
would be described through a separate phase-space distribution $f_i(\vec p,t)$ for each species.

In this section, we shall begin our study by exploring the manner in which early-universe dynamics affects
$f(\vec p, t)$.  This is the first connection sketched in Eq.~(\ref{onepointone}).
Towards this end, 
we shall start by studying 
how the distribution function $f(\vec p,t)$ evolves with time.

In the standard Friedmann-Robertson-Walker (FRW) cosmology, 
the physical distance $x(t)$ between two otherwise stationary points at time $t$ as the universe
expands is related to that at time $t'$ by 
{\il{x(t)=x(t')a(t)/a(t')}}, where $a(t)$ is the scale factor.
In order to maintain the Poisson bracket {\il{\{x,p\}=1}}, the momentum $p$ of a given free particle
therefore has to evolve inversely with respect to the coordinate $x$, implying {\il{p(t)=p(t')a(t')/a(t)}}, or
{\il{d\log p/dt= -H(t) }},
where {\il{H\equiv \dot a/a}} is the Hubble parameter.
This describes the gravitational redshift of momentum due to the expansion of the universe.
From this we see that time-evolution corresponds to additive $p$-independent shifts in the value of $\log p$,
which makes $\log p$ a particularly convenient variable for studying the time-evolution of any dark-matter
phase-space distribution.
Assuming isotropy of the phase-space distribution, quantities such as the physical number
density in Eq.~(\ref{fdefs}) can therefore be rewritten as
\beqn
n(t) &=& {g_{\rm int} \over 2\pi^2} \int_{0}^\infty dp\,  p^2 \, f(p,t)\nonumber\\ 
                &=&  {g_{\rm int} \over 2\pi^2} \int_{-\infty}^\infty  d\log p~p^3\,f(p,t)~,
\eeqn
whereupon we see that the {\it comoving}\/ number density {\il{N(t)\equiv  n a^3}} is given as
\beq
        N(t) ~=~ {g_{\rm int}\over 2\pi^2}  \int_{-\infty}^\infty  d\log p~ g(p,t) ~\equiv~ {g_{\rm int}\over 2\pi^2} \,\calN(t)~,  
\label{Ndef}
\eeq
where we have defined
\beq
    g(p,t) ~\equiv~ (ap)^3\, f(p,t)~
\label{gdef}
\eeq
and where $\calN(t)$ is the total area under the curve of $g(p,t)$ plotted versus $\log p$.

\begin{figure}[t!]
\centering
\includegraphics[keepaspectratio, width=0.49\textwidth]{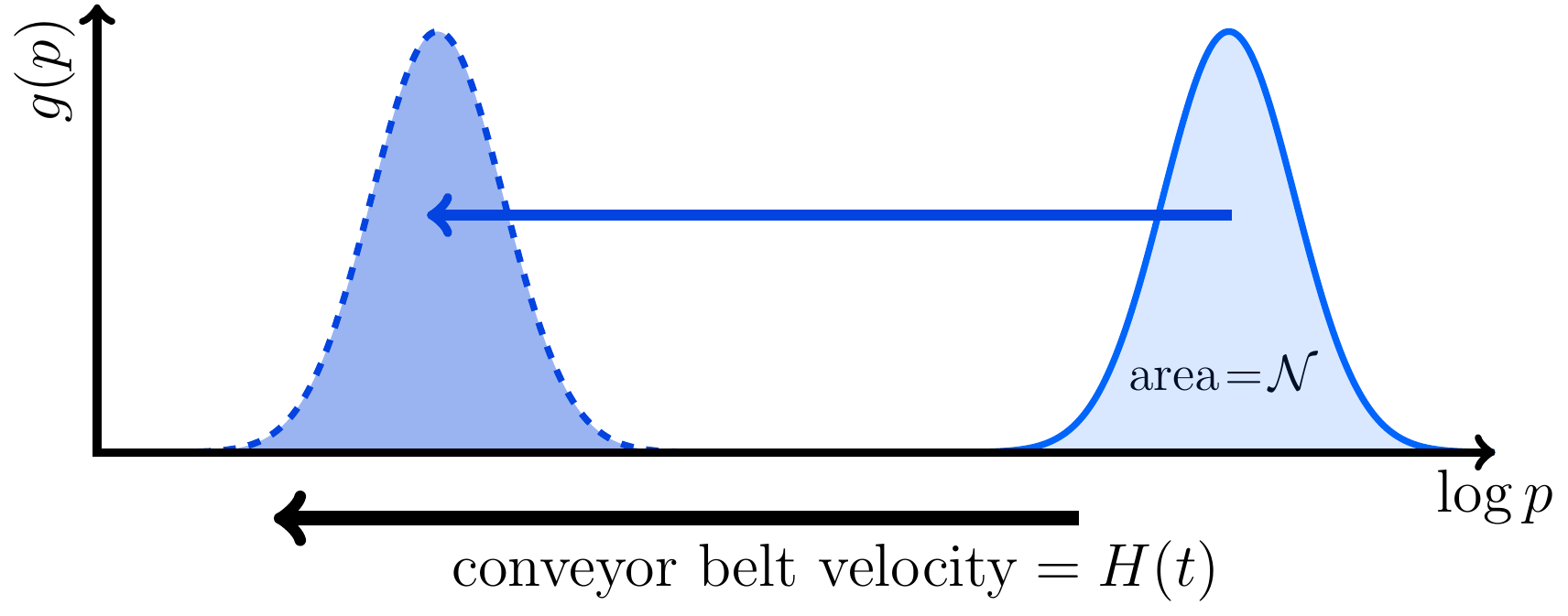} 
\caption{
The cosmological conveyor belt.  
A given dark-matter phase-space
distribution
{\il{g(p)\equiv (ap)^3 f(p)}} 
is rigidly carried towards smaller values of $\log\,p$  with 
velocity $H(t)$ as the universe expands.   
The area $\calN$ under the curve is proportional to the corresponding (fixed) comoving number
density {\il{N(t)\equiv N}}, as given in Eq.~(\ref{Ndef}).}
\label{fig:conveyor0}
\end{figure}

Particle interactions can of course change the value of the comoving number density $N(t)$. 
However, barring such interactions, this quantity should remain time-independent.
Likewise, we have already seen that $\log\, p$ merely accrues additive $p$-independent shifts  
under time-evolution, which implies that the measure $d\log\,p$ in Eq.~(\ref{Ndef}) is also unaffected.
Given the relation in Eq.~(\ref{Ndef}), this implies that the value of $g(p,t)$ should also be invariant under
time-evolution, in the sense that
\beq
             g(p(t'),t')~=~g(p(t),t)~
\eeq
for any times $t,t'$.
This implies that under time-evolution (and barring any further dark-matter creation or decay), 
the curve for $g(p)$ merely slides towards smaller values of $\log p$ without distortion, 
as if carried along on a ``conveyor belt'' moving with velocity $H(t)$.
Indeed, $\calN(t)$ --- the area under this curve --- will also be invariant.
For these reasons, we shall often concentrate on $g(p)$ rather than $f(p)$  in this paper when discussing the
dark-matter phase-space distribution. 
This situation is sketched in Fig.~\ref{fig:conveyor0}.

In this paper, we shall refer to this as the {\it cosmological conveyor belt}\/.
This conveyor belt will ultimately play an important role in our thinking.
Of course, we have already seen that 
it provides 
a useful way of conceptualizing the time-evolution of the dark-matter
phase-space distribution $g(p)$.  However, as we shall now discuss,
 it also allows us to understand how different phase-space distributions 
$g(p)$ may arise as 
the result of physical processes that occur during cosmological evolution.

As an example, let us consider the case of a minimal dark sector
consisting of a single dark-matter species subject
to a specific dark-matter production mechanism.
Accordingly, this dark-matter species is produced with a
corresponding phase-space distribution $g(p)$ which
we may regard as coming into existence on the cosmological 
conveyor belt at the time of production,
with the particular shape of $g(p)$ 
depending on the details of the production mechanism.
This distribution then simply redshifts towards smaller values of 
$\log\,p$ for all subsequent times prior to the decay of the dark matter, if any.

However, things can be very different for a non-minimal dark sector 
containing an entire {\it ensemble} of particle
species instead of a single dark-matter component.
In such cases, the phenomenology of the dark sector is not determined by the properties of any 
individual constituent alone, but is instead
determined collectively across all components.
However, for such non-minimal dark sectors,
it is possible
that the dark-matter  
production may be more complicated,
with different ``deposits'' onto the cosmological conveyor
belt occurring at different moments in cosmological history.

\begin{figure*}[t]
\centering
\includegraphics[keepaspectratio, width=0.7\textwidth]{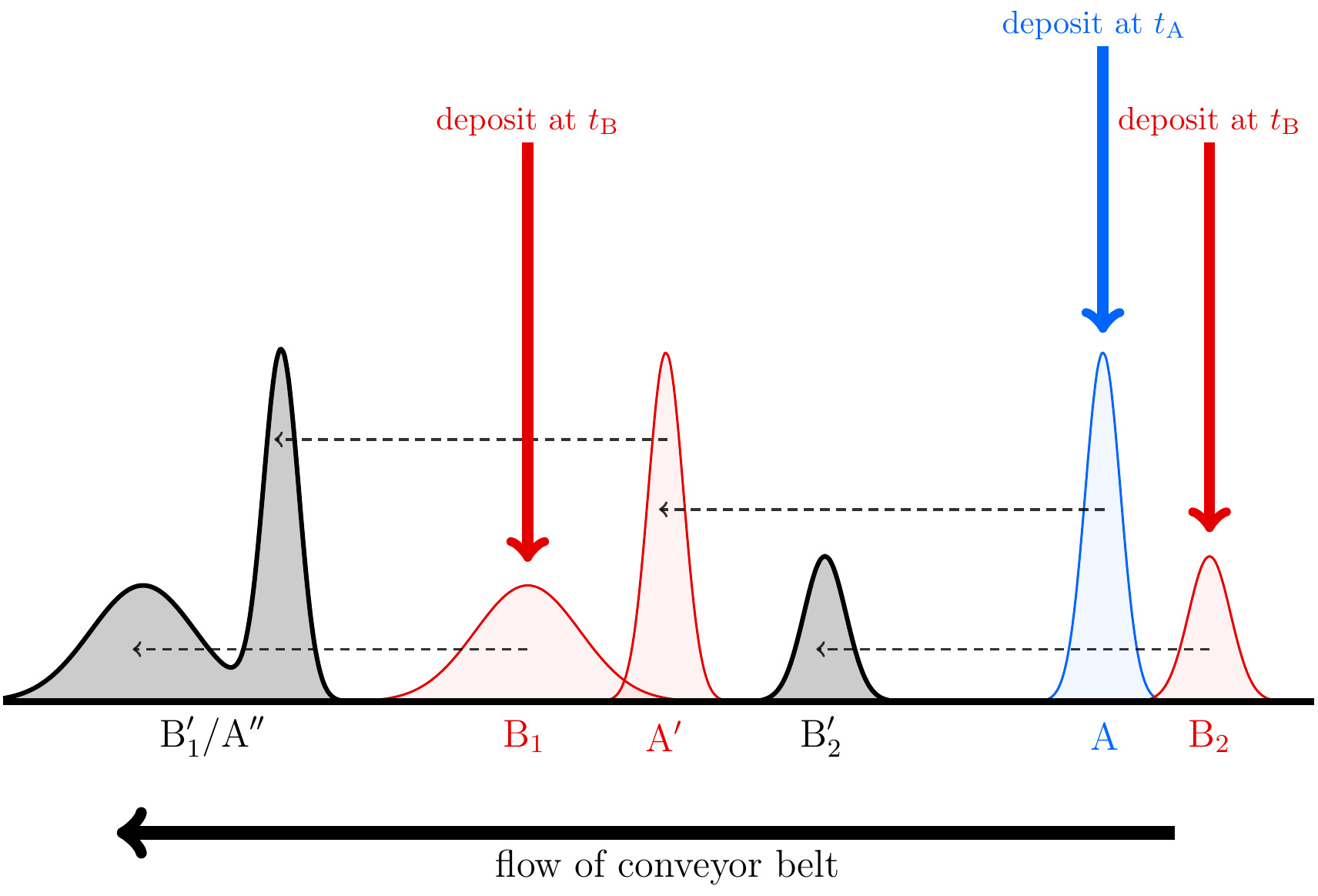} 
\caption{
Sketch of the cosmological conveyor belt at three different times:  an early time $t_A$ (blue) at which packet $A$
is deposited, a later time $t_B$ (red) at which packets $B_1$ and $B_2$ are deposited, and a final time (black).
During the time interval between $t_A$ and $t_B$,
packet $A$ has redshifted to position $A'$, where it overlaps
with the newly deposited packet $B_1$.~  After time $t_B$, 
both the combined packet $B_1/A'$ and the packet
$B_2$ continue to redshift into their final positions $B_1'/A''$ and $B_2'$.~   
Thus, the resulting dark-matter phase-space distribution consists of two disjoint contributions (black):
a warmer unimodal distribution $B_2'$ and a colder multi-modal distribution $B_1'/A''$.
However, in the absence of further information,
knowledge of this final distribution allows us to resurrect only certain aspects of this cosmological history.}
\label{fig:conveyor1}
\end{figure*}

For example, let us consider the situation
sketched in Fig.~\ref{fig:conveyor1}.~
In this scenario,
an initial dark-matter constituent $A$ 
is produced at time $t_A$ (blue), 
with a corresponding phase-space distribution
(sketched as packet $A$ in Fig.~\ref{fig:conveyor1}) deposited onto the conveyor belt at that time.
However, at a later time $t_B$ (red), two further contributions are produced, corresponding
to a colder packet $B_1$ and a warmer packet $B_2$,
both of which are also deposited onto the conveyor belt with appropriate momenta.  Of course, during the time interval  
between $t_A$ and $t_B$, our original packet $A$ has redshifted into a new location $A'$,
where it partially overlaps with $B_1$. 
These then superpose to form a non-trivial combined packet $B_1/A'$.
At all later times beyond $t_B$ (such as that sketched in black in Fig.~\ref{fig:conveyor1}),
the resulting phase-space distribution consists of two disjoint contributions:
a warmer distribution (labeled $B_2'$) which is unimodal and a colder distribution (labeled $B_1'/A''$) 
which exhibits a complex, bi-modal shape.
Thus, we see that the final resulting dark-matter distribution
can be fairly complicated and even multi-modal.  However, without further information,
it would be possible to resurrect only certain aspects of this cosmological history.

We may quantify this sort of analysis as follows.
In general, for any dark sector, the final dark-matter distribution $g(p)$ 
is realized as the accumulation of all previous
deposits occurring at all previous times during cosmological history.
Towards this end, let $\Delta(p,t)$ denote the profile
of the dark-matter deposit rate at time $t$.
Then, at any time $t$, we have
\beq
  g(p,t) ~=~ \int^t dt' \, \Delta\left( p {a(t)\over a(t')}, t'\right)~.
\label{genexp}
\eeq
Of course, if the deposits occur at discrete times $t_i$, then
\beq
  \Delta (p,t) ~=~ \sum_i \Delta_i (p) \, \delta(t-t_i)~.
\label{discrete}
\eeq
Inserting this into Eq.~(\ref{genexp}) then yields
\beq
       g(p,t) ~=~ \sum_i \Delta_i \left( p {a(t)\over a(t_i)}\right),
\label{constraints}
\eeq
demonstrating that $g(p,t)$ reflects a particular cosmological history.
Indeed, the situation sketched in Fig.~\ref{fig:conveyor1} is nothing but
the special case with
\beq
\Delta(p,t) ~=~ \Delta_A \delta(t-t_A) + \left( \Delta_{B_1} + \Delta_{B_2} \right) \delta(t-t_B)~.
\eeq

Of course, the important archaeological question 
is to determine the extent to which knowledge of $g(p,t)$ at a given time $t$ can be used to
resurrect the cosmological history.
The mathematical answer is provided by Eq.~(\ref{genexp}) or Eq.~(\ref{constraints}):   for each
value of $p$, {\it we can constrain only the corresponding sum or integral in these equations
such that this sum or integral has
the value $g(p,t)$.}\/
Note that this is an independent constraint for each value of $p$.
However, for any given time $t$,
it is clearly impossible to resurrect a unique historical deposit profile $\Delta(p,t')$ for all times {\il{t'<t}} given only
the information contained within $g(p,t)$.
To see this, let us write
Eq.~(\ref{genexp}) in the suggestive form
\beqn
  g(p,t) &=& \int_{-\infty}^\infty dt' \int_{-\infty}^\infty dp' \, \nonumber\\
           && ~~~\Delta(p',t') \, \delta\left( p'-p{a(t)\over a(t')}\right)\, \Theta(t-t')~~~~~~~~~
\label{integralform}
\eeqn
and treat $\Delta(p',t')$ as the `source' which produces the `response' $g(p,t)$.
Indeed, the factors in the integrand of this expression 
(particularly the argument of the Dirac $\delta$-function in conjunction with the Heaviside $\Theta$-function)  
make clear that $g(p)$ in some sense tallies the separate contributions along what we may call
the ``backwards FRW momentum lightcone'', in the same way as $\delta(x-ct)$ would have indicated    
a spatial lightcone in flat space.
Moreover, given the form of Eq.~(\ref{integralform}), we can identify 
\beq
        G(p,p',t,t') ~\equiv~ \delta\left( p'-p{a(t)\over a(t')}\right)\, \Theta( t-t')~~
\label{Greensfunction}
\eeq
as a Green's function connecting the source to the response.  We can  thereby convert 
Eq.~(\ref{integralform}) into the differential form
\beq
       {\partial g\over \partial t} - H p {\partial g\over \partial p} ~=~ 
    \Delta (p,t)~, 
\label{Boltzmanneq}
\eeq
where we have recognized \il{\partial/\partial t - Hp \,\partial/\partial p} as the differential
operator corresponding to the Green's function in Eq.~(\ref{Greensfunction}).
Indeed, the result in Eq.~(\ref{Boltzmanneq}) 
is nothing but the Boltzmann equation describing the evolution of the phase-space distribution $g(p,t)$
in the presence of the source term $\Delta(p,t)$, here functioning as an effective collision term.  
 {\it This demonstrates
that our conveyor-belt/deposit picture is ultimately a physical representation of the integral form of 
the Boltzmann equation.}

However, the result in Eq.~(\ref{Boltzmanneq}) also illustrates mathematically why we cannot solve for the complete
cosmological history of deposits $\Delta (p,t')$ for all times {\il{t'<t}}.
We could certainly use this relation 
to solve for $\Delta(p,t')$ 
if we had information about $g(p,t')$ at all such times.
However, we presumably only have access to $g(p,t)$ at a fixed late time $t$, long after the final deposits onto the conveyor belt have
occurred.  This then prevents us from resurrecting the desired profile history $\Delta(p,t')$. 

We have already seen in Fig.~\ref{fig:conveyor1} that 
multi-modality of the phase-space distribution suggests that
separate deposits occurred at different moments in cosmological history.
However, this immediately begs the question as to whether such a pattern
of deposits can arise naturally.
In particular, it is important to determine what
kinds of non-minimal dark sectors can give rise to
such deposit patterns.
However, as we shall now demonstrate,
such deposit patterns can easily emerge 
if our non-minimal 
dark sector contains an ensemble of states
with different masses, lifetimes, and cosmological abundances.
In such cases, it is the {\it intra-ensemble decays}\/ (\ie, decays from heavier to lighter dark-sector
components) which will naturally give rise to such deposit patterns.

\subsection{From parents to daughters:  Decays and their effects
    on dark-matter phase-space distributions\label{parent_to_daughter}}

As an initial step towards explaining how this occurs, let us first study the fundamental process in which
a parent packet decays, producing  a daughter packet.
As we shall see, this process is surprisingly subtle within an expanding universe 
because of the relativistic effects which emerge if the momenta involved are large compared to the mass of the parent. 
We shall therefore proceed to discuss this process in several steps.

In general, a parent packet describes the momentum distribution of a collection of parent particles.
Such particles ultimately decay at different times and with different momenta.
However, we can group these parent particles into subsets depending on when they decay and the momenta they
have when they decay, so that each $(p,t)$ subset consists of the particles 
which decay at a given common time $t$ with
a given common momentum $p$.   
The decays of the particles within each $(p,t)$ 
subset then collectively produce a deposit onto the daughter conveyor belt, 
and integrating these deposits with those from all other relevant $(p,t)$ subsets
(while accounting for appropriate redshifting effects)
then yields the final daughter packet $g(p)$.
It is in this way that we shall organize our discussion of the decay process through which a daughter packet
emerges through the decay of a parent packet.
Of course, the idea that we can view our daughter packet as the $(p,t)$-integral of a deposit profile $\Delta(p,t)$
is already familiar from Eq.~(\ref{genexp}).   The only difference here is that we are
reorganizing our deposits in terms of 
the momenta of the decaying parents rather than the momenta of the resulting daughters --- a
reorganization which proves particularly useful when the dark matter is produced through decays.

We shall therefore begin our discussion of the decay process by studying the properties of the
daughter deposit that emerges
from a single $(p,t)$ subset of parent particles --- {\it i.e.}\, 
from a population of parent particles  which share a common momentum  
and decay at a common time. 
Although we shall ultimately refrain from specifying a particular decay process,
for simplicity we shall initially assume that each parent particle $X$ undergoes a two-body decay of the form
{\il{X\to YY}}
with parent and daughter masses $m_X$ and $m_Y$ respectively, where {\il{m_Y < m_X/2}}.
In this case, the energy of each daughter in the rest frame of the parent is simply $m_X/2$, and likewise the
magnitude of the momentum of each daughter in the rest frame of the parent is given by $\sqrt{(m_X/2)^2 - m_Y^2}$.
Indeed, this is the non-zero momentum that is imparted to the daughters in the 
cosmological background frame (\ie, the ``lab frame'')
when the parent is at rest.
However, if the parent has momentum $p_X$
in the lab frame at the time of its decay, then the energy of each daughter in the lab frame is given by 
\beq
    E_Y ~=~  {1\over 2} \sqrt{m_X^2 + p_X^2} + {p_X \over 2}\, \sqrt{ 1- 4 m_Y^2/m_X^2}\,  \cos\theta~, 
\label{cosformula}
\eeq
where  $\theta$ is the angle between the daughter momentum in the parent rest frame and the parent momentum.  
Given that the angle $\theta$ is unfixed by the kinematics of the decay
(implying that all values of $\cos\theta$ arise with equal probability),
these daughter energies $E_Y$ will therefore vary 
uniformly within a total range of magnitude 
\beq
\Delta E_Y ~=~ E_Y^{\rm (max)} - E_Y^{\rm (min)} ~=~  p_X \sqrt{ 1- 4 m_Y^2/m_X^2}~.
\label{deltaEwidth}
\eeq
Likewise, for any daughter with energy $E_Y$ within this range, the corresponding daughter momentum
in the lab frame is simply given by {\il{p_Y= \sqrt{E_Y^2-m_Y^2}}}.
Thus, the existence of a range of possible daughter energies $\Delta E_Y$ implies 
the existence of a corresponding range of daughter momenta $\Delta p_Y$,
all of which are populated through the decays of the parent particles in the given subset.
Indeed, we observe that daughter energy range $\Delta E_Y$ --- and indeed the corresponding daughter
momentum range $\Delta p_Y$ --- both increase as functions of $p_X$.

In this context, we remark that this kinematic analysis
actually allows us to draw an even stronger conclusion
if {\il{m_Y\approx 0}}: 
our deposits will be rectangular in shape even in $p_Y$-space (and not only in $E_Y$-space),
and will actually be logarithmically {\it centered}\/ around a $p_X$-independent
``anchor'' momentum 
\beq
            p_Y^{\rm anchor} ~\approx~
            E_Y^{\rm anchor} ~\equiv~ 
       \sqrt{ E_Y^{\rm (min)} E_Y^{\rm (max)} } ~\approx~ {m_X\over 2}~.
\label{anchor}
\eeq
Every deposit, regardless of $p_X$, will then include this anchor momentum, 
even though $\langle p_Y\rangle$ and $\Delta p_Y$ will continue to grow with $p_X$.

Our main interest, of course, is not in the kinematic details of this specific decay process. 
Rather, we are interested in certain features which are exemplified above but which are generic  
across many different decay processes. 
These include processes such as 
{\il{X\to YZ}} where the daughter masses $m_Y$ and $m_Z$ do not
have a large hierarchy between them.  These also include three-body and multi-body decays such 
as {\il{X\to YZW...}} where the number of daughter particles
remains $\calO(1)$, where the daughter masses do not have large hierarchies between them,
and  where we assume ``typical'' decays within the Dalitz plot which are characterized by 
generic $\calO{(1)}$ angles between the decay products, so that
no hierarchies emerge amongst the momenta of the daughters when measured in the rest frame
of the decaying parent.

In all such cases, 
although the precise shapes of these deposits
depend on the detailed kinematics associated with the specific decay process,
certain general conclusions about the corresponding daughter deposits 
from each $(p,t)$ parent subset 
can nevertheless be drawn.
For example,
because the sum of the masses of the daughters is always less than the mass of the parent,
each daughter is produced with a non-zero momentum in the rest frame of the parent.
The magnitude of this momentum is uniquely determined for two-body decays, and is 
also determined to within an order of magnitude for the ``typical'' three-body decays discussed above.
Likewise, because the direction of the daughter momentum in the parent rest frame is
uncorrelated with the direction of the momentum of the parent in 
the lab frame,
the magnitude of the momentum of each daughter 
is broadened by the boost of the parent
into a {\it range}\/ of momentum magnitudes as measured in the lab frame.
Thus, the corresponding daughter deposit will stretch across
a range of momenta whose width generally grows as a function of the momentum of the parent at the
time of decay.
Indeed, such considerations are generic, and should hold within the classes of decay processes discussed above.

Many other properties of the possible daughter deposits can be similarly deduced from 
general considerations such as these.
As a result, given the properties of a particular daughter deposit, it is often possible to ``reconstruct''
certain generic properties of the parent particles within the corresponding $(p,t)$ subset
as well as certain generic properties of the corresponding decay process.
Our results are shown in Table~\ref{inversedeposit}, where the relevant properties of
possible daughter deposits
are listed on the left and the corresponding
parent and decay properties are listed on the right.

\begin{table*}[t!]
\begin{center}
\begin{tabular}{||c||c|c|c||c|c|c||}
\hline
\hline
 & \multicolumn{3}{c||}{Daughter deposit} & \multicolumn{1}{c|}{Parent }  &  \multicolumn{2}{c||}{Decay} \\   \cline{2-7} 
 &  \multicolumn{1}{c}{rel?} &  \multicolumn{1}{c}{~~width~~} & \multicolumn{1}{c||}{~~relative width~~} 
    & \multicolumn{1}{c|}{rel at} &  near absolute &  near relative \\ 
 ~Case~ & \multicolumn{1}{c}{~~$\langle p\rangle$ ~~}   &  \multicolumn{1}{c}{~~$\Delta p/m$~~}   
          &  \multicolumn{1}{c||}{$\Delta p/\langle p\rangle$}  & ~~decay?~~ 
              & ~~marginality?~~ & ~~marginality?~~ \\ 
\hline
\hline
~A~ & \multirow{2}{*}{~{\il{p\ll m}}~} &  \multirow{4}{*}{~~narrow~~} & narrow & \multirow{3}{*}{~non-rel~} & \multirow{2}{*}{near} & far \\ \cline{4-4}\cline{7-7} 
 B  &  &  &  $\calO(1)$ & & & $\calO(1)$ \\ \cline{2-2}\cline{4-4}\cline{6-7} 
 C  & \multirow{3}{*}{~{\il{p\sim m}}~} & & \multirow{2}{*}{narrow} & & $\calO(1)$ & far \\ \cline{5-7} 
 D  & & & & \multirow{2}{*}{  \phantom{$_\sim$}rel$_\sim$ } & near & near \\ \cline{3-4} \cline{6-7}
 E  & & $\calO(1)$ & $\calO(1)$ & & $\calO(1)$ & $\calO(1)$ \\ \cline{2-7}
 F  & \multirow{6}{*}{~{\il{p\gg m}}~} & narrow & \multirow{4}{*}{narrow} & \multirow{2}{*}{non-rel} & \multirow{2}{*}{far} & 
           ~~far ~({\il{p_{\rm parent} \ll m_{\rm daughter}}})~~ \\ \cline{3-3}\cline{7-7}
 G  &  &  \multirow{2}{*}{$\calO(1)$} & & & & far ~({\il{p_{\rm parent} \sim m_{\rm daughter}}}) \\ \cline{5-7}
 H  & & &  & \phantom{$_\gg$}rel$_\gg$ & near & near \\ \cline{3-3} \cline{5-7}
 I  &  &  \multirow{3}{*}{wide} & & non-rel & far &   far ~({\il{p_{\rm parent}\gg m_{\rm daughter}}}) \\ \cline{4-7}
 J  & & & \multirow{2}{*}{$\calO(1)$} & \phantom{$_\sim$}rel$_\sim$ & far & $\calO(1)$ \\ \cline{5-7} 
 K  & & & & {\phantom{$_\gg$}rel$_\gg$} & $\calO(1)$ or far & near \\
\hline\hline
\end{tabular}
\end{center}
\caption{ The various daughter deposits that can arise from the simultaneous decays of a population of parent particles of 
             fixed momentum.  Given the properties of the daughter deposit, we can therefore reconstruct the extent to which the parents were relativistic and the 
           extent to which the corresponding decay process was near marginality.  
       In some cases this reconstruction is unique, while in other cases several possibilities exist.
         This table nevertheless exhibits a rich pattern of correlations between daughter deposits,  parents, and associated decay properties.
           The quantities at the top of each column of this table 
           are discussed in the text, along with associated definitions and underlying assumptions.}  
\label{inversedeposit}
\end{table*}

In Table~\ref{inversedeposit}, column headings are defined as follows.
For the daughter deposit, `rel ($\langle p\rangle$)' at the top of the first column indicates whether the daughter momenta 
within the deposit --- as represented by the average momentum $\langle p\rangle$ ---
are relativistic.
Likewise, the second and third columns indicate whether the width $\Delta p$ of the daughter deposit 
is large, order-one, or small (`wide', `$\calO(1)$', or `narrow', respectively)  when compared with either
the daughter mass $m$ or the mean deposit momentum $\langle p\rangle$.  For the parent subset, `rel at decay' indicates whether the
the parent momentum is relativistic or 
non-relativistic at the time of decay, 
with `rel{\il{_{\sim}}}' and `rel$_\gg$' further specifying situations
with {\il{p_P\sim m_P}} and {\il{p_P\gg m_P}} respectively, where $m_P$ and $p_P$ are the corresponding parent mass and momentum.
By contrast, the final two columns of the table indicate the degree to which
the decay process is near marginality.  In rough terms, this refers to the degree to which the decay process endows the daughters
with additional kinetic energy beyond that which is directly inherited from the parent.
More precisely,
``absolute marginality'' and ``relative marginality'' are respectively assessed in terms of
the fractions
$p_D^{\rm rest}/m_P$ and
$p_D^{\rm rest}/p_P$,
where $p_D^{\rm rest}$ is the magnitude of the daughter momentum in the rest frame of the parent.
In the case of relative marginality, 
decays for which 
$p_D^{\rm rest}/p_P$
is much less than $1$, much greater than $1$, or $\calO(1)$ 
are respectively considered `near'-marginal, `far' from marginal, or $\calO(1)$-marginal.
By contrast, in the case of absolute marginality,
there is a maximum value of
$p_D^{\rm rest}/m_P$ which can ever kinematically arise.
This maximum value depends  on the details of the particular decay process,
and corresponds to the limit in which all of the daughter masses vanish.
(For example, this maximum value is $1/2$ for decays of the form {\il{X\to YY}}.)
We then consider the corresponding decay 
to be respectively `near' absolute marginality, `far' from absolute marginality, or exhibiting  $\calO(1)$ absolute marginality
depending on whether  
$p_D^{\rm rest}/m_P$ 
is respectively much less than $1$, very close to its maximum value, or somewhere in between.

Note that this table is designed to indicate
only those fundamental trends that emerge in limiting  hierarchical cases involving only rough
orders of magnitude.
In other words, in this table we only consider cases in which we
can cleanly identify our daughter particles as either very non-relativistic ({\il{p\ll m}}),
moderately relativistic ({\il{p\sim m}}), or ultra-relativistic ({\il{p\gg m}}), with an implied hierarchy of
momenta between these three situations.
Likewise we only consider cases in which our daughter widths $\Delta p$ are
either much less than, of approximately the same size as, 
or much bigger than $m$ or $\langle p\rangle$, with hierarchically large
separations between these different cases.
It is because we are only considering such limiting cases with large hierarchies
that we do not provide any
further information of a more specific nature within this table.
Likewise, in constructing this table,
we have made certain assumptions which are consistent
with this general purpose.
For example,
as already stated above,
these results assume that each decay produces
${\cal O}(1)$ daughters whose masses are all of the same overall scale ${\cal O}(m_D)$.
These results likewise assume
the order-of-magnitude estimate that
{\il{m_P\sim {\cal O}(m_D)}}
(respectively, {\il{m_P\gg  m_D}})
for decays which are near (respectively, far from) absolute marginality.

Despite these assumptions and restrictions, this table contains a wealth of information about the possible daughter
deposits that can emerge from the simultaneous decays of parents with a given momentum.
As an example, 
within the limits discussed above,
we learn 
that there is only one way to produce 
a non-relativistic daughter deposit 
for which the width $\Delta p$ is extremely small compared  
to both the daughter mass as well as the mean deposit momentum:
we must begin with non-relativistic parents  
experiencing a decay which is near absolute marginality 
and yet far from relative marginality (Case~A).~
Indeed, for a two-body decay of the form {\il{X\to YY}}, this would
correspond to a situation in which $\half \sqrt{1-4 m_Y^2/m_X^2}$ 
is much smaller than $1$ but much larger 
than $p_X/m_X$.
Conversely, we see that highly {\it relativistic}\/ daughter deposits with {\il{\Delta p \ll m}} and {\il{\Delta p \ll \langle p\rangle}}  
can only emerge from extremely non-relativistic parents 
experiencing decays which are {\it far}\/ from both absolute and relative marginality, 
so long as the parent momentum is significantly smaller than the daughter mass (Case~F).~
Similarly, we learn from this table that
there are only two ways of producing a
highly relativistic daughter deposit for which {\il{\Delta p\sim m}} but
{\il{\Delta p\ll \langle p\rangle}}:
the parents must either be non-relativistic when experiencing 
a decay which is far from both absolute and relative marginality (Case~G)
or highly relativistic when experiencing a decay which is near both absolute and relative
marginality (Case~H).~
Indeed, in the former case, we can further conclude that the parent momentum must
be of the same order as the daughter mass ---  otherwise our decay 
would no longer produce daughter
deposits for which {\il{\Delta p\sim m}}.
The results in this table thus provide a useful guide towards archaeological reconstruction,
at least at the level of the individual daughter deposits that emerge from $(p,t)$ parent subsets.

The results in this table are also significantly influenced by the relativistic effects 
connected with the boosting associated with highly relativistic parents.
For example, were it not for such relativistic effects,
highly relativistic parents experiencing decays which are far from absolute  
marginality but close to relative marginality 
(such as would arise for two-body decays of the form {\il{X\to YY}} when
{\il{p_X\gg m_X/2}} and {\il{m_Y\ll m_X}})
would have yielded daughter deposits
with {\il{\Delta p\gg m}} but {\il{\Delta p\ll \langle p\rangle}}.
However, it is the relativistic effects associated with this boosting
that actually broaden the width $\Delta p$ 
of the resulting daughter deposits and force {\il{\Delta p\sim \langle p\rangle}}
(as indicated within Case~K).~
Thus, even though the dark matter may be non-relativistic today as the result of the gravitational
redshifting that has occurred
since the conclusion of the decay process,
such relativistic effects may nevertheless be forever imprinted 
within the deposits that ultimately comprise the daughter packets.

Of course, as we transition across different parent subsets within any physical decay process, 
the parent and daughter masses are fixed ---  indeed, only the parent momentum at the time of decay 
varies.
Thus the absolute marginality of the decay process 
is fixed for all  parent subsets in any given decay process,
and varying the parent momentum induces transitions 
between only certain cases 
within Table~\ref{inversedeposit}.~
For example, if the parent and daughter masses correspond to a
decay process close to absolute marginality,
then we can only transition from Case~A to Case~B to Case~D to Case~H as
the parent momentum increases.
In such cases we find that the corresponding daughter deposits likewise shift from 
non-relativistic to relativistic (as expected for a decay process near
absolute marginality), but that 
{\il{\Delta p\ll m}} 
until the parent momentum becomes highly relativistic.
Interestingly, however, we see from Table~\ref{inversedeposit}
that {\il{\Delta p\ll \langle p\rangle}} 
except near a ``resonance'' 
that occurs when  $p_D^{\rm rest}$ (the daughter momentum in the rest frame of the parent)
becomes approximately equal to the parent momentum.
At that point, as illustrated in Case~B, we find  {\il{\Delta p\approx \langle p\rangle}}.

It is also possible to understand certain general aspects of the {\it shapes}\/ of these deposits.
As long as the decay process produces a unique magnitude for the momentum of the daughter in 
the rest frame of the parent, the boost due to the momentum of the parent will cause
the energy 
of the daughter to have  the general form {\il{E_D=E_1 + E_2\cos\theta}}.  Indeed, we have
already seen an example of this for the {\il{X\to YY}} decay process in Eq.~(\ref{cosformula}).
Isotropy implies that all values of $\cos \theta$ are equally likely to occur,
and thus the resulting deposit has a {\it flat}\/ profile 
when plotted in $E$-space.   In other words, when plotted in $E$-space
the resulting deposit is ``brick''-shaped.
We will discuss this further below.
However, changing variables to $p$-space 
then converts this flat profile into one that rises as a function of $p$ --- an
effect which is negligible for highly relativistic deposits (for which {\il{E\approx p}}) but
otherwise sizable.
This effect is even more dramatic when the deposit is plotted versus $\log p$.

Having discussed the individual daughter deposits 
that emerge from each $(p,t)$ subset of parent particles,
we now must combine these deposits in order to construct the
final, total daughter  packet.
Indeed, this is the only way in which we can properly discuss
the manner in which the decays of {\it all}\/ of the parent particles within a complete parent packet
produce a complete daughter packet.
However, in order to combine these deposits correctly, there are a number of 
additional effects that must be taken into account:
\begin{itemize}
\item   First, the different parent particles within the parent packet do not all decay at the same
 time.  The decay process is a probabilistic  exponential one, with a survival probability 
 scaling with time as $e^{-t/\tau}$  where {\il{\tau= 1/\Gamma}} is the proper decay lifetime and $\Gamma$ the decay width. 
Thus we can never determine precisely when a given parent particle will decay.
\item   Second, as the result of time-dilation effects, the different parent particles do not even share a common effective lifetime.
Indeed, parent particles with larger momenta within the parent packet will 
have longer effective lifetimes, thereby delaying their decays relative to the decays of those particles with smaller momenta.
\item   Finally, we must account for the continual pull of gravitational redshifting which 
persists throughout this entire process.   This affects 
the parent particles, as their momenta continually redshift until the moment of decay.   However, this also affects the daughter deposits,
as each deposit also immediately begins to redshift while waiting for subsequent deposits to appear.  
The final daughter packet is then determined only after each deposit has arrived.
\end{itemize}
All of these effects combine to render the transition from parent packet to daughter 
packet a fairly complicated affair.

In order to determine the extent to which the features described
in Table~\ref{inversedeposit} 
for the individual daughter deposits
might eventually survive for the full daughter packets ---
and also to determine what additional features might accumulate for these daughter packets
as the result of the effects itemized above ---
we shall proceed in several steps.
First, it will prove instructive to analyze the decay process while incorporating the second and third features above, 
but disregarding the first.   Then, we shall consider the opposite situation in which we incorporate the first and third features
but disregard the second.
Each of these approaches  will provide useful, complementary information concerning the final daughter packets that emerge.
Finally, we shall analyze the full decays, taking all three features into account simultaneously.

\subsubsection{Instantaneous-decay approximation with time-dilation and redshifting effects}

\begin{figure*}[t]
\centering
\includegraphics[keepaspectratio, width=.75\textwidth]{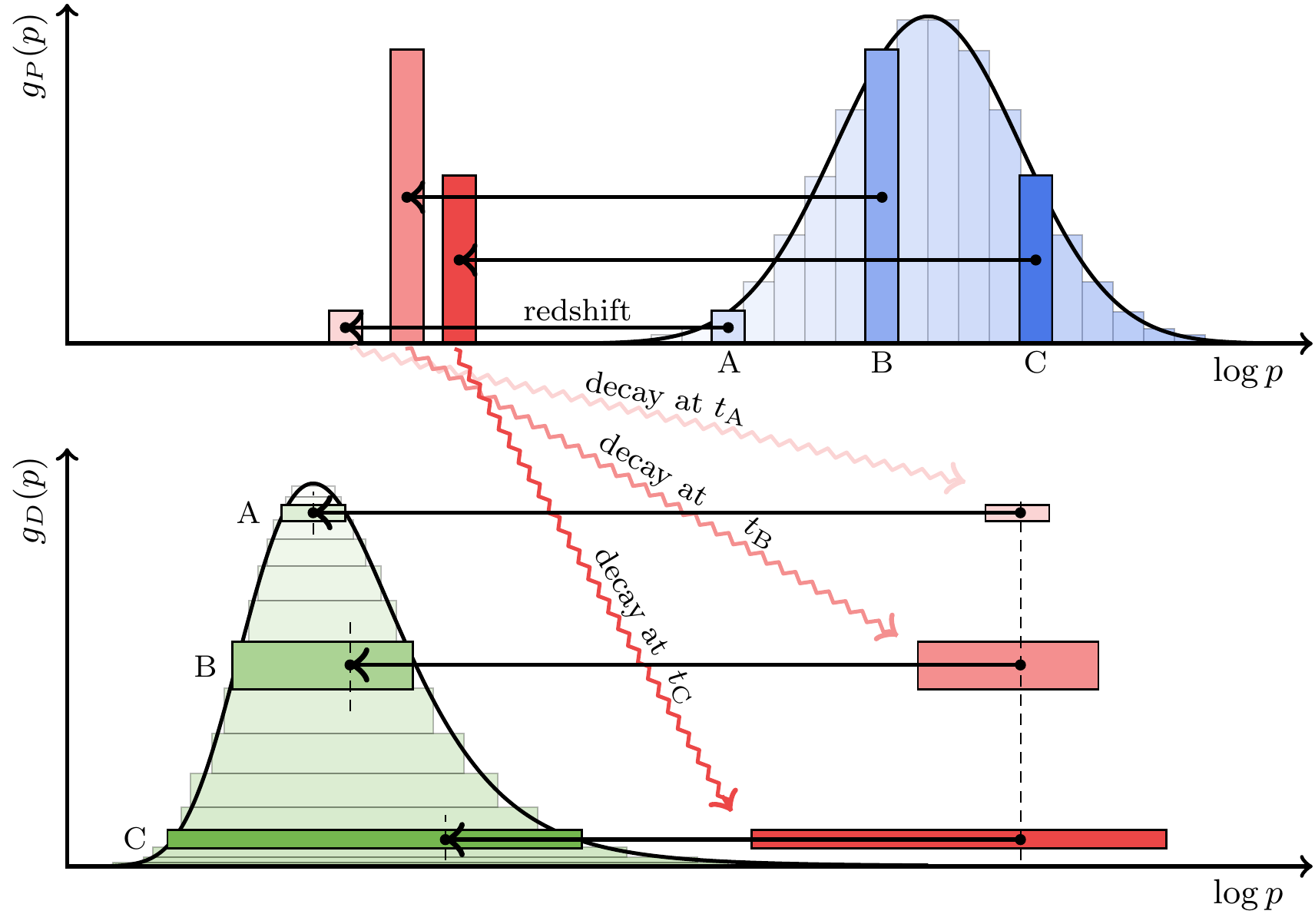} 
\caption{The process through which a parent packet $g_P(p)$ (blue) decays into a daughter packet $g_D(p)$ 
(green) in an expanding FRW cosmology,
with intermediate steps indicated (red/pink). 
Each momentum slice of the parent packet experiences a different time-dilation factor which not only
extends the lifetimes of the parent particles associated with that slice
but also increases the total accumulated cosmological
redshift experienced by that momentum slice prior to decay.
The sequential decays of these redshifted momentum slices 
then make sequential contributions to the daughter packet, with
each contribution extending over an increasingly broad range of momenta and 
experiencing its own cosmological redshift immediately upon production until further contributions arrive.   
This process ends with the decay of the last momentum slice of the parent, thereby completely depleting the parent
packet (blue) and yielding the total daughter packet (green).  Note that this illustration assumes the instantaneous-decay
approximation in which the decay of each parent particle
is treated as occurring promptly once its lifetime {\il{\tau=1/\Gamma}} is reached.}
\label{fig:conveyor3}
\end{figure*}

We begin, therefore, by considering the decay process from parent packet to daughter packet 
under the so-called instantaneous-decay approximation in which each parent particle 
is assumed to decay precisely at its lifetime {\il{\tau\equiv 1/\Gamma}}, as measured in its rest frame.
We will also assume that the parent species has mass $m$ and is produced at some time {\il{t=t_0}} with a simple, unimodal
phase-space distribution $g_P(p)$.  
The resulting decay process is sketched in Fig.~\ref{fig:conveyor3}.~

Because the parent packet $g_P(p)$ stretches across a non-zero range of momenta, each momentum slice of the parent packet will experience a different time-dilation factor.
This then increases the effective lifetime of each slice prior to decay, causing slices with larger momenta to decay later than those with smaller momenta.
However, during the time interval prior to decay, each momentum slice also experiences a cosmological redshift.
This redshift decreases the momentum of each slice and thereby partially mitigates the time-dilation effect.
Indeed, combining both effects, we see that each momentum slice with initial momentum $p_0$ at {\il{t=t_0}} decays at
\beq
        t_{\rm decay}(p_0) ~=~ t_0+ \int_{t_0}^{t_0+\tau} dt' \, \frac{\sqrt{ p_0^2 (t_0/t')^{4/[3(1+w)]} + m^2}}{m}~,~
\label{tdecay}
\eeq
where we have assumed an FRW cosmology 
with the Hubble factor scaling as \mbox{{\il{H\sim 2/[3(1+w)t]}}} 
where $w$ is the equation-of-state parameter.
Likewise, the momentum of this slice at the time of its decay is given by
\beq
        p_{\rm decay}(p_0) ~=~ p_0 \left[ \frac{t_0}{t_{\rm decay}(p_0)} \right]^{2/[3(1+w)]} ~,
\label{pdecay}
\eeq
where $t_{\rm decay}(p_0)$ is given in Eq.~(\ref{tdecay}).
Indeed, these results are completely general within the instantaneous-decay approximation.

As we move upwards within the parent packet 
towards momentum slices with increasing values of $p_0$,
the time-dilation factor increases --- but this also provides a longer time interval
during which cosmological redshifting occurs.
It is therefore important to understand the 
extent to which this extra redshifting might compensate for the greater original momentum of the momentum slice.
Indeed, at first glance it might even seem that momentum slices with greater initial momenta $p_0$ could ultimately 
be redshifted to {\it smaller}\/ 
final momenta $p_{\rm decay}$ when they actually decay, with
{\il{p_{\rm decay}(p'_0) < p_{\rm decay}(p_0)}} even when {\il{p'_0 >  p_0}}.
However, it is straightforward to verify that {\il{d p_{\rm decay}(p_0)/dp_0 >0}} for all $p_0$,
ensuring that {\il{p_{\rm decay}(p'_0) >  p_{\rm decay}(p_0)}} for all {\il{p'_0 >  p_0}}.
Thus, momentum slices with greater initial momenta $p_0$ continue to have greater momenta when they each decay. 
However, it is also straightforward to verify that 
{\il{d p_{\rm decay}(p_0)/dp_0 <1}} for all $p_0$.   
Thus, any two momentum slices whose original momenta $p_0$ differ by an amount $\Delta p_0$ will 
have decay momenta $p_{\rm decay}$ differing by an amount {\il{\Delta p_{\rm decay} < \Delta p_0}}.
Unless {\il{p_0\gg p_{\rm decay}}}, this further implies that {\il{\Delta \log p_{\rm decay} < \Delta \log p_0}}. 
This ``momentum compression'' is illustrated along the top portion of Fig.~\ref{fig:conveyor3},
where the relative horizontal spacings between the original (blue) momentum slices labeled $A$, $B$, and $C$  are
larger than the relative horizontal spacings between the 
corresponding redshifted (red/pink) momentum slices which are sketched at the (different) times of their decays.

The decay of each momentum slice of the parent then yields a contribution to the emerging 
phase-space distribution $g_D(p)$ of the daughters.
Indeed, these contributions are nothing but the deposits already discussed in Table~\ref{inversedeposit}.~
These individual deposits contributions are sketched as the red/pink ``bricks'' within the lower portion
of Fig.~\ref{fig:conveyor3}.~ 
As already noted above,
the precise shapes of these deposits onto the daughter conveyor belt
depend on the detailed kinematics associated with the decay process,
and thus our representation of these contributions in Fig.~\ref{fig:conveyor3} as rectangular bricks whose height does not vary with 
momentum should at this stage be understood merely as an approximate abstraction adopted for visual simplicity (albeit an abstraction
which becomes increasingly accurate for relativistic daughter packets). 
However, a number of features associated with these bricks are rather important and are incorporated into this sketch.
First,
we observe that 
the total area associated with each brick
is proportional to the total area of the original decaying momentum slice of the parent,
with the proportionality constant signifying the number of daughters (in this case, two) 
produced through the decay of each parent.
Thus, the vertical normalization of the lower portion of this figure need not be assumed to be the
same as that of the upper portion. 
Second, because our original parent packet $g_P(p)$ has a profile which first rises and then falls as a function of momentum, the total areas of the bricks that are deposited  first grow and then shrink as the decay process proceeds.
Third, in making this sketch we have assumed that the mass of the parent greatly exceeds the combined
masses of the daughters.  It is for this reason that we have sketched our bricks as having momenta 
which 
exceed $p_{\rm decay}$.  Indeed, as indicated in Table~\ref{inversedeposit}, 
these brick momenta will {\it greatly}\/ exceed $p_{\rm decay}$ if the parent momentum slice is 
sufficiently non-relativistic at the time of its decay, and if the decay process itself is relatively far from absolute marginality.
In such cases, the bulk of the brick momentum comes from the 
energy liberated during the decay process itself.
Finally, as discussed above,
the width of each deposited brick
grows as a function of $p_{\rm decay}$  (and therefore grows as a function of $p_0$).
Likewise, if the daughter is sufficiently relativistic as well,
each deposited brick will be ``anchored''  around a common anchor momentum $p_{\rm anchor}$.
These features are also illustrated for the red/pink bricks shown  in Fig.~\ref{fig:conveyor3},
with the black vertical dashed line indicating $p_{\rm anchor}$.

The final stage of the evolution from parent to daughter packet once again involves cosmological redshifting.
Because the narrowest bricks are deposited first, they begin redshifting towards smaller momenta before the subsequent, wider bricks have even been deposited.   Thus, as shown in Fig.~\ref{fig:conveyor3},
 the narrowest bricks experience the largest redshifts --- an effect which causes the upper portion of the emerging daughter packet $g_D(p)$ to experience  
an effective ``pull'' to the left relative to the lower portion when plotted versus momentum.
This effect will be discussed further below.

Ultimately, the final shape of the daughter packet $g_D(t)$ is then given by the sum of these different bricks at the time when the final (bottom) brick is deposited.  
Indeed, we shall take the deposit of the final daughter brick
(or equivalently the decay of the highest parent momentum slice) 
as marking 
the completion of the decay process.
The daughter packet at this moment is sketched in green in Fig.~\ref{fig:conveyor3}.~
It is important to realize that 
because the bottom brick within the green daughter packet 
has not experienced any redshift at the completion time,
it will still be anchored around the dashed black vertical anchor line.
In other words, in such 
cases, the base of the 
final green daughter packet technically extends all the way to (and well beyond) the anchor 
line, even if this behavior is not directly visible in 
Fig.~\ref{fig:conveyor3} because the bricks below brick~$C$ have increasingly small heights.
Of course, after the decay completion time,
the shape of the daughter green packet 
is fully determined;  the packet then redshifts rigidly along the momentum conveyor belt 
towards smaller momenta.

We have already noted that
relativistic time-dilation effects cause the narrowest daughter bricks to experience larger redshifts
than those experienced by the wider daughter bricks.
This relativistic effect therefore 
provides a leftward contribution to the overall ``tilt''
(or skewness) of the daughter packet.
In general, the skewness $S$ of a given $g(p)$ packet 
can be quantified as the third moment of the $g(p)$ distribution in 
$\log p$-space:
\beq
        S ~\equiv~ {1\over \sigma^3} \langle  \,(\log p-\langle \log p\rangle)^3 \,\rangle~,
\label{Sdef}
\eeq
where 
\beq
             \sigma^2 ~\equiv~ \langle  \, (\log p-\langle \log p\rangle)^2 \,\rangle
\eeq
and where all averaging is done with respect to the $g(p)$ distribution.
Thus, any effect which provides a  leftward contribution to the overall 
tilt of the $g(p)$ packet is one which tends to {\it increase}\/ the skewness $S$ of
the packet.
Indeed, this effect is particularly pronounced 
when the parent particles are highly relativistic when produced.  
By contrast, when the parent is non-relativistic at production,
there are no significant time-dilation effects within the decay process. 
Thus the upper portion of the daughter packet no longer experiences extra redshifting
relative to the lower portion,
and the corresponding extra leftward contribution to the overall tilt of the daughter packet is eliminated.

\begin{figure}[h]
\centering
\includegraphics[keepaspectratio, width=0.49\textwidth]{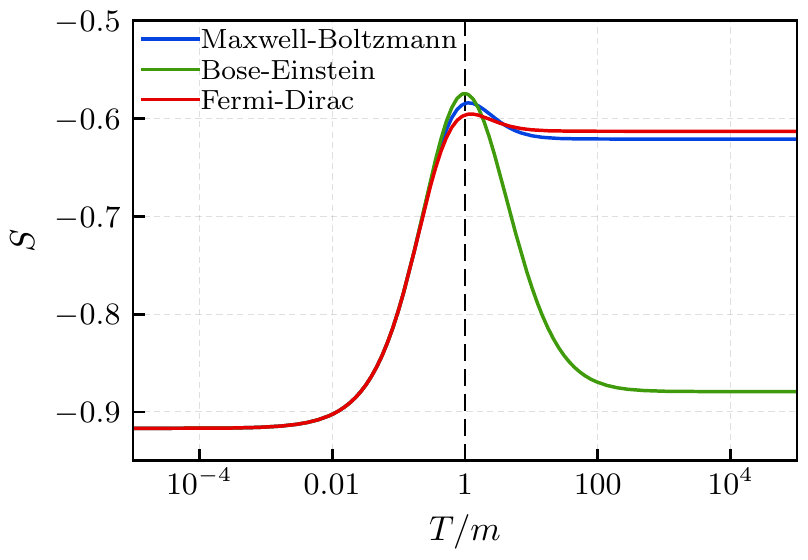} 
\caption{Many dark-matter phase-space distributions tend to have pronounced rightward tilts (negative skewness $S$) 
when plotted versus $\log p$.
Here we plot the skewness $S$ of a thermal distribution as a function of the temperature $T$ for a species of mass $m$ 
obeying Maxwell-Boltzmann, Bose-Einstein, or Fermi-Dirac statistics.} 
\label{fig:skew_thermal}
\end{figure}

This is important because there is already a strong tendency towards {\it rightward}\/ tilting
(negative skewness) for $g(p)$ packets simply because such packets are plotted versus $\log p$.
Of course, we have already seen that each of our individual deposits 
has the shape of a flat ``brick'' when plotted versus the daughter energy $E$, and thus
has no intrinsic skewness.   However, 
the change of variables from $E$ to $p$ already introduces a rightward tilt (one which 
disappears in the relativistic limit),
and the further change from $p$ to $\log p$ introduces an additional rightward tilt.
Thus, even a distribution which is completely symmetric in $E$-space 
around a mean value $\langle E\rangle$
will nevertheless exhibit a pronounced rightward tilt 
(or negative skewness) 
when plotted versus $\log p$. 
Rightward tilts thus tend to be the ``norm''
for most daughter $g(p)$ packets.
In fact, as shown in Fig.~\ref{fig:skew_thermal} for comparison purposes, 
even ``standard'' distributions
such as a thermal distribution 
exhibit a pronounced negative skewness.
Because of this strong tendency towards rightward tilting,
a parent packet will often have to be exceedingly relativistic before the 
relativistic leftward-tilt contribution due to time dilation is able to overcome the 
pre-existing contributions towards rightward tilting 
and produce a daughter packet exhibiting an overall leftward tilt, with {\il{S>0}}.

Looking over the entire process shown in Fig.~\ref{fig:conveyor3}, we see that each 
{\it vertical}\/ momentum slice in the original parent packet $g_P(p)$ ultimately gives rise to a {\it horizontal}\/ 
slice in the daughter packet $g_D(p)$.  Thus, 
while the resulting shape of the daughter packet is ultimately sensitive   
to many kinematic details associated with the decay process, 
certain general conclusions can be drawn.  Each of these can therefore play an ``archaeological'' 
role in helping us to determine the properties of the parent, given only the 
shape of the daughter.  
\begin{itemize}
\item First, for example, we see that the {\it widths}\/ of the daughter packet are correlated
with the redshifted momenta of the individual slices of the parent at the times when they decay. 
Indeed, the maximum width along the {\it base}\/ of the daughter packet 
corresponds to the {\it greatest} momentum within the parent packet,
while the narrowest width (which governs the shape of the {\it peak}\/ of the daughter packet, \ie, its ``cuspiness'')
corresponds to the {\it smallest}\/ momentum within the parent packet.
In technical terms, these two features together characterize the ``kurtosis'' of the daughter packet.
\item Second, in a similar fashion, the ascending and descending {\it slopes}\/ of the daughter packet 
are correlated with the vertical heights of the corresponding deposited bricks.  However, the height of
each brick
is correlated with its {\it area}, and this area is in turn correlated with the 
area of the originally decaying parent momentum slice and thus with the corresponding value of $g_P(p)$.
\item Finally, as discussed above, 
the relative tilt of 
the daughter packet can provide information about the relativistic nature of the
parent at the time when it was initially produced:
\beqn
  && {\rm relativistic ~parent} \nonumber\\
  &&  ~~~~\Longleftrightarrow~~
  \begin{cases}
     {\rm leftward~contribution} \\
     {\rm to~daughter~ tilt.}
  \end{cases}
\label{tiltrules}
\eeqn
Indeed, such relativistic effects 
tend to increase the skewness of the daughter packet,
thereby either providing the packet with an outright leftward tilt or decreasing the extent to which the packet 
would otherwise tilt to the right. 
\end{itemize}
Using such pieces of information, many gross features of the daughter packet can be 
exploited in order to learn about the parent ---  even independently of the detailed kinematics of the particular decay process.
Indeed, one could even proceed beyond skewness 
and kurtosis  to consider higher moments of the daughter phase-space distribution.
Of course, the results outlined above have been obtained under the instantaneous-decay approximation.
However, 
we nevertheless expect these features to continue to hold in a rough, average sense
even for a full-fledged exponential decay.

Within the instantaneous-decay approximation,
it is easy to estimate the conditions under which the relativistic leftward contribution to the tilt of 
the daughter packet will be significant in size.
As discussed above, this contribution arises when the parent is relativistic,
and results from the fact that the different momentum slices of the 
parent packet decay at different times.
This in turn implies that the upper portion of the daughter packet experiences a redshift
relative to the lower portion, thereby pulling the daughter packet to the left
and increasing its skewness. 
The fact that the parent is relativistic implies that {\il{\langle p \rangle_P/m_P\gg 1}},
where $\langle p\rangle_P$ denotes a central momentum value of the parent packet 
and where $m_P$ is the parent mass. 
Likewise, we can roughly approximate 
the time interval between the earliest and latest decays of the parent momentum slices,
or equivalently between the deposit times of the narrowest and widest daughter bricks,
as {\il{\Delta t  \approx (\Delta p)_P/(m_P \Gamma)}},
where $(\Delta p)_P$ is the width of the 
parent packet in momentum space.
However, during this interval,
the top of the daughter packet redshifts 
in $(\log \,p)$-space by an approximate distance $H\Delta t$, where $H$ represents an
average value of the Hubble parameter during this interval.  Relative to the rest of the
daughter packet, this redshift will induce a significant
contribution towards 
leftward tilt 
if $H\Delta t \gsim \Delta(\log p)_D $ 
where {\il{\Delta(\log p)_D \approx \log(1 + (\Delta p)_D/\langle p_D\rangle)}} 
is the width of the daughter packet in $(\log p)$-space
and where $(\Delta p)_D$
and $\langle p\rangle_D$ are respectively the
width and mean value of the daughter packet in momentum space.
Combining these expressions, we thus find that the resulting leftward contribution to the daughter tilt will be significant
if
\beq
      {(\Delta p)_P\over m_P} ~\gsim~ 
      {\Gamma \over H}\,
       \log \left( 1+ {(\Delta p)_D\over \langle p\rangle_D} \right)~.
\label{leftcondition}
\eeq

For the sake of later comparison,
it is also instructive to understand the limit of the decay process 
sketched in Fig.~\ref{fig:conveyor3}
when the parent packet is {\it non}\/-relativistic at the time it is produced.
In such cases,
the entire parent packet redshifts a uniform amount before decaying, with the momentum slices maintaining their
relative distances along the $\log\, p$ axis.
Likewise, under the instantaneous-decay approximation,
the momentum slices all decay at the same time, simultaneously depositing daughter bricks of exceedingly narrow
widths directly on top of each other.
(More precisely, their horizontal displacements, if any, are relatively small compared with their total widths.)
Thus, within the instantaneous-decay approximation, we expect that non-relativistic parent packets  
will ultimately give rise to extremely sharp ($\delta$-function-like) daughter packets. 
Moreover, we expect that this will remain true regardless of the shape of the parent packet.
Of course, at this stage it remains 
to determine how these two observations fare when we go beyond the instantaneous-decay approximation.
This will be discussed below.

\subsubsection{Exponential decay in the non-relativistic limit:   
A universal dark-matter distribution}

Thus far, we have analyzed the decay process from parent packet to daughter packet
while taking into account the effects of time dilation as well as cosmological redshifting.
However, it is also instructive to consider the complementary situation
in which we disregard time-dilation effects and instead properly treat
the decay of each parent particle 
as a probabilistic process following an exponential profile $e^{-t/\tau}$ where {\il{\tau\equiv 1/\Gamma}} 
is the proper lifetime of the species. 
Indeed, 
such a treatment describing the decay from parent packet to daughter packet
will actually be exact 
in the limit that the parent packet is non-relativistic.

Under these assumptions, it proves possible to proceed completely analytically.
We assume the existence of a parent packet of total number density $N_P(t_0)$ at some
initial time $t_0$. 
Ignoring the time-dilation effects associated with the different particles within the parent packet is tantamount to 
assuming that the parent particles are extremely non-relativistic.
Thus all of the parent particles 
have the same effective lifetime, 
so that the actual {\it shape}\/ of  
the parent packet is irrelevant.
Given that the parent particles all share the same effective lifetime $\tau$,
at any future time $t$ beyond $t_0$ we have
\beq
            N_P(t) ~=~ N_P(t_0) \,\exp\lbrack -\Gamma (t-t_0) \rbrack ~ \Theta(t-t_0)~,
\label{steppone}
\eeq
or equivalently
\beq
           {dN_p(t)\over dt} ~=~ -\Gamma\, N_P(t)~.
\label{stepptwo}
\eeq 
We have explicitly inserted the Heaviside $\Theta$-function into Eq.~(\ref{steppone}) 
simply as a way of  enforcing our condition that {\il{t\geq t_0}}.
Although we shall never seek to understand what might have happened at earlier times, this $\Theta$-function will
soon play an important role.  

For concreteness, let us assume that our parent decays into two identical daughters.  
(Other cases pose no special difficulties and would proceed analogously.) 
The daughter number density therefore grows with time as
\beq
          {dN_D(t)\over dt} ~=~ {+} 2\Gamma \, N_P(t)~.  
\label{steppthree}
\eeq
Indeed, this growth in the daughter number density is realized as the result 
of a continuous stream of deposits onto the daughter conveyor belt. 
In the non-relativistic limit, these deposits are infinitely narrow
and have fixed momentum $p_\ast$, where $p_\ast$ is merely a property of the decay kinematics,
depending on the parent and daughter masses alone.   Indeed, $p_\ast$ is nothing but the momentum of the daughters in the rest
frame of the parent (which is of course equivalent to the cosmological background lab frame for extremely 
non-relativistic parents), and thus serves as the fixed deposit location in momentum-space throughout the decay process.
However, although the deposits at any time $t$ have momentum $p_\ast$, each deposit redshifts while
subsequent deposits continue to arrive.
We shall let $t_f$ denote the ``final'' time at which our decay process has essentially concluded.
We then find that any deposit which occurs at time $t$ will redshift to a final momentum
given by
\beq 
                p(t) ~=~ p_\ast \left[ \frac{a(t)}{a(t_f)} \right] ~=~ p_\ast \left( \frac{t}{t_f}\right)^{\kappa/3}~,
\label{steppfour}
\eeq
where {\il{a(t)\sim t^{\kappa/3}}} is the FRW scale factor, with {\il{\kappa = 3/2}} 
for a radiation-dominated universe and {\il{\kappa=2}} for a matter-dominated universe.
In writing Eq.~(\ref{steppfour}) we are of course assuming that 
the entire decay process occurs within a time interval during which the value of $\kappa$ is constant.
Note that this final momentum $p(t)$ of the deposit depends on the time $t$ at which this deposit occurs,
and is thus a function of $t$.
Using Eq.~(\ref{steppfour}), we can therefore change variables from $t$ to $p(t)$, 
reflecting the fact that the
cosmological redshifting has induced 
the spread in decay {\it times}\/
to become a spread in final daughter {\it momenta}\/.
Indeed, from Eq.~(\ref{steppfour}) we find that {\il{d \log p(t)/dt= H(t)}}, whereupon Eq.~(\ref{steppthree})
becomes
\beqn
            {d N_D(t) \over d\log p} &=& {2\Gamma\over H(t)} N_P(t)\nonumber\\
           &=& {2\Gamma\over H(t)} 
            \,  N_P(t_0) \,\exp\lbrack -\Gamma (t-t_0) \rbrack ~ \Theta(t-t_0)~.~\nonumber\\
\eeqn    
However, from Eq.~(\ref{Ndef}) we see that 
\beq
            {d N_D(t) \over d\log p} ~=~ {g_{\rm int}\over 2\pi^2} \, g(p)~
\eeq
where $g(p)$ is the phase-space distribution of the daughter particles.
Using the relations {\il{H(t)= H(t_f) (p_\ast/p)^{3/\kappa}}} and
{\il{t/t_f= (p/p_\ast)^{3/\kappa}}} we therefore obtain our final result for the daughter 
phase-space distribution at time $t_f$:
\beqn
   g(p) &=& {4\pi^2 N_P(t_0)\over g_{\rm int}} {\Gamma\over H(t_f)} \, \left( {p\over p_\ast}\right)^{3/\kappa} \nonumber\\
           &&  \times 
           \exp\left\lbrace -\Gamma\left\lbrack t_f\left( {p\over p_\ast}\right)^{3/\kappa} -t_0\right\rbrack\right\rbrace
   \Theta\left( p - p_{\rm min}  \right)~,~\nonumber\\
\label{gnonrel}
\eeqn
where {\il{p_{\rm min} \equiv p_\ast \left( t_0/t_f\right)^{\kappa/3}}}.
Within Eq.~(\ref{gnonrel}), the Heaviside $\Theta$-function now simply indicates that there exists a 
minimum momentum $p_{\rm min}$ reached by the daughter packet.  This 
is the momentum to which the very first daughter deposit is redshifted when the
decay process concludes.
Indeed, there are no daughter particles with {\il{p<p_{\rm min}}} because no parent particles decay prior to $t_0$.
           
\begin{figure}[t!]
\centering
\includegraphics[keepaspectratio, width=0.49\textwidth]{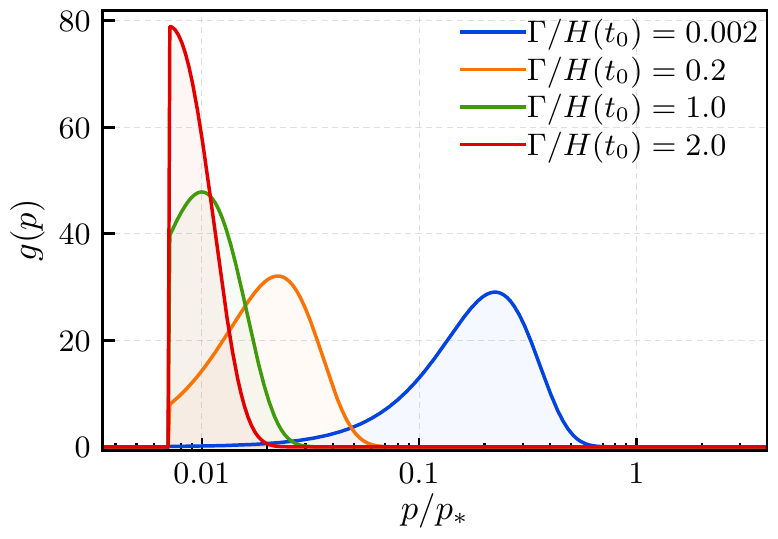} 
\caption{The universal form $g(p)$ for all decay daughters in the limit 
that the parent momentum is extremely non-relativistic.
This universal form is given in Eq.~(\ref{gnonrel}) 
and plotted at a fixed final time $t_f$ for different values of $\Gamma$, with all other quantities held fixed.
For this plot we have chosen the reference values 
{\il{\kappa=3/2}} and {\il{N_P(t_0)=g_{\rm int}=1}}.
Note, in particular, that we have held $t_0$ fixed and taken {\il{t_f= 20 \Gamma_{\rm min}^{-1}}}, where $\Gamma_{\rm min}$ is the
smallest of the decay widths for which the corresponding $g(p)$ distributions are shown in this figure.
Thus $t_f$ represents the time at which the decay process concludes for the case with the smallest $\Gamma$.~
For all other cases in this figure with larger decay widths,
we retain this value of $t_f$:
thus the corresponding
daughter packets are already fully produced prior to $t_f$ and then  redshift rigidly until {\il{t=t_f}}.  
Moreover, in the formal {\il{\Gamma/H(t_0)\gg 1}} limit,  the decay timescale is so rapid
compared with the redshifting timescale that all decay deposits occur essentially simultaneously at {\il{p=p_\ast}}
and then redshift together to {\il{p=p_{\rm min}}}.
Thus, in this limit, we find {\il{g(p)\to \delta(p-p_{\rm min})}}.}
\label{fig:gnonrel}
\end{figure}

As the above derivation demonstrates, our result in Eq.~(\ref{gnonrel}) is a {\it universal}\/ functional form --- 
a universal phase-space distribution to which
 {\it any}\/ phase-space distribution resulting from a decay must tend in the limit that the parent is extremely non-relativistic.
Indeed, the only assumption that entered into this derivation is the assumption that the 
magnitudes of the daughter momenta are all equal to a unique value $p_\ast$.  This assumption is true 
for two-body decays and is also largely valid for higher-body decays that are ``typical'' 
in the sense described earlier.
Consequently,
the only features within this functional form that are sensitive to the particular decay process 
are the value of $p_\ast$ which enters into Eq.~(\ref{gnonrel}) --- thereby merely serving 
as a reference momentum ---  and an overall numerical factor describing the daughter multiplicity.
Moreover, in the non-relativistic limit, this result is wholly independent of the shape of the parent packet.
Although this result has appeared previously in a variety of specific contexts~(see, {\it e.g.}\/,
Ref.~\cite{Kaplinghat:2005sy}),
our derivation of this result is completely general and proceeds rather simply and 
directly from our conveyor-belt picture.
This universal form for $g(p)$ in Eq.~(\ref{gnonrel}) is plotted in Fig.~\ref{fig:gnonrel} for different values of $\Gamma$.

One notable feature of this functional form for $g(p)$ is that it is not continuous at {\il{p=p_{\rm min}}}.
Indeed, {\il{g(p)=0}} for all {\il{p<p_{\rm min}}}, while 
\beq
 \lim_{(p\to p_{\rm min})^+} g(p) ~=~ {4\pi^2 N_P(t_0) \over g_{\rm int}} {\Gamma\over H(t_0)}~.
\label{discon}
\eeq
At first glance, this might seem disturbing, as we are used to phase-space packets which rise smoothly from zero at 
small momenta and fall smoothly to zero at high momenta.
However, we must recall that $g(p)$ is a {\it distribution function} --- specifically, 
a particle-number $\log p$-space {\it density}.
Just as with a Dirac $\delta$-function, there is nothing inconsistent about discontinuous densities 
because the physical quantities of interest (such as the actual numbers of physical particles, or their 
total energies, pressures, {\it etc.}\/) are {\it integrals}\/ over these densities, as in  Eq.~(\ref{fdefs}).  
Indeed, the discontinuity in $g(p)$ at {\il{p=p_{\rm min}}} ultimately stems from the discontinuity in 
the slope $dN_P/dt$ at the initial time $t_0$ --- a feature which is endemic to {\it all}\/ decay processes
which are modeled through exponentials as in Eq.~(\ref{steppone}).
Despite this discontinuity, the actual {\it number}\/ of daughter particles $N_D(t)$ rises continuously from zero at {\il{t=t_0}},
as required.

It is interesting to consider our expression in Eq.~(\ref{gnonrel}) in the 
formal {\il{\Gamma/H(t_0)\to\infty}} limit.
In this limit,
the decay timescale is so rapid
compared with the redshifting timescale that all decay deposits essentially occur at {\il{t=t_0}}
and then redshift together from
$p_\ast$ at {\il{t=t_0}} to $p_{\rm min}$ at {\il{t=t_f}}.   
It then follows that
{\il{g(p)\to \delta(p-p_\ast)}} 
as {\il{\Gamma/H(t_0)\to\infty}}.
Thus in such cases the Dirac $\delta$-function even emerges as a limiting case for $g(p)$.
However, we hasten to point out that numerous other effects enter into the game when {\il{\Gamma/H(t_0)\gg 1}}.
For example, as $\Gamma/H(t_0)$ grows large, we can no longer ignore the {\it inverse-decay}\/ process in which
daughter particles recombine to form the parent.
The presence of both processes tends to thermalize the parent and daughter packets to each other.
Likewise, 
as $\Gamma/H(t_0)$ grows large,
the 
resulting phase-space distribution $f(p)$ no longer satisfies the condition
{\il{f(p)\ll 1}} for all momenta.
Quantum statistical effects will therefore also become important.

\begin{figure}[h]
\centering
\includegraphics[keepaspectratio, width=0.49\textwidth]{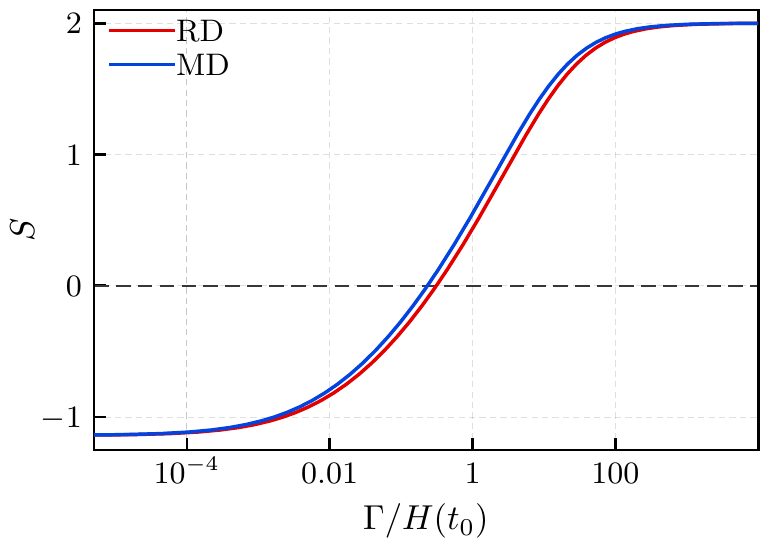} 
\caption{The skewness $S$ of the universal functional form for $g(p)$ in Eq.~(\ref{gnonrel}), plotted
as a function of $\Gamma/H(t_0)$ for {\il{\kappa=3/2}} and {\il{\kappa=2}} [corresponding to 
radiation-domination (RD) and matter-domination (MD) respectively].
We see that 
this skewness is negative (signifying rightward tilt) for {\il{\Gamma/H(t_0)\ll 1}}, but then rises to positive values (leftward tilt) 
as {\il{\Gamma/H(t_0)\to \infty}}, ultimately asymptoting at {\il{S=2}}.}
\label{fig:Snonrel}
\end{figure}

Another notable feature associated with this universal asymptotic form for $g(p)$ in Eq.~(\ref{gnonrel}) 
is its skewness $S$.
This skewness $S$ is plotted as a function of $\Gamma/H(t_0)$ in Fig.~\ref{fig:Snonrel}.
For {\il{\Gamma/H(t_0)\ll 1}}, we see that $g(p)$ is highly rightward-tilted, with {\il{S<0}}.
However, as $\Gamma/H(t_0)$ increases, we see that $S$ also begins to increase, with $g(p)$ actually becoming
leftward-tilted for $\Gamma/H(t_0)\gsim 1$.   This increasing tendency towards leftward tilting
is largely due to the Heaviside cutoff in Eq.~(\ref{gnonrel}), as this cutoff plays an increasingly
dominant role in eliminating the low-momentum tail of the $g(p)$ distribution.
Ultimately, for {\il{\Gamma/H(t_0)\gg 1}},
we see that {\il{S\to 2}}.
Given that this is the limit in which {\il{g(p)\to \delta(p-p_{\rm min})}}, 
it may seem strange at first glance that $S$ does not return to zero in this limit
(even if this limit is unphysical because of the effects described above).
However, strictly speaking, the Dirac $\delta$-function lacks a width and therefore does not have a well-defined skewness;
like the longitude of the North Pole, this quantity depends on the precise route through which the limit is approached.

Even though the discontinuity in Eq.~(\ref{discon}) is a true physical effect 
--- one which becomes increasingly prominent for large $\Gamma/H(t_0)$ ---
the existence of this discontinuity does signal a certain 
internal stress within our original assumption of a truly exponential decay beginning promptly at {\il{t=t_0}}.
As discussed above,
this discontinuity ultimately stems from our decision to 
model our decay process
through an exponential as in Eq.~(\ref{steppone}), thereby
intrinsically introducing a discontinuity
in the slope $dN_P/dt$ at the initial time $t_0$.
However, the validity of an exponential-decay process
implicitly rests upon the assumption that the {\it parent}\/ came into existence at 
sharply at {\il{t=t_0}}.
This, of course, cannot be strictly true, since the parent itself must be produced
through some process involving its own internal timescale.
There is thus a tension between our exponential-decay assumption 
and the timescale associated with the production of the parent --- a tension which becomes increasingly severe
as $\Gamma/H(t_0)$ becomes increasingly large and as the size of the discontinuity in Eq.~(\ref{discon}) grows.
Indeed, for sufficiently large $\Gamma/H(t_0)$, the decays of the parent will begin even before the parent itself
is fully produced.   Such effects can therefore be expected to soften this discontinuity. 

When performing our analyses in the rest of this paper, it will often be convenient to assume the 
existence of an initial time $t_0$ at which we may presume the validity of initial conditions involving
the prior existence of certain parent particles.   In such cases, we shall therefore either incorporate the existence
of such discontinuities into our analysis or restrict ourselves to regions of parameter space for which
\beq
        \Gamma/H(t_0) ~\ll~1~,
\label{condition}
\eeq
where $\Gamma$ is the decay width of the parent.
In such regions of parameter space, the resulting discontinuities in the resulting daughter packet will be relatively small and 
the daughter packets will have essentially
smooth tails at both low and high momenta.
However, none of the main results of this paper will depend on this assumption.
Thus our results will not depend on the existence or removal of these discontinuities.

\subsubsection{Full numerical analysis, and the interplay between decay kinematics and cosmological expansion}

\begin{table*}[t!]
\begin{center}
\begin{tabular}{||c||c|c|c||c|c|c||}
\hline
\hline
 & \multicolumn{3}{c||}{Daughter distribution } & \multicolumn{1}{c|}{Parent }  &  \multicolumn{2}{c||}{Decay} \\   \cline{2-7} 
 &  \multicolumn{1}{c}{rel?} &  \multicolumn{1}{c}{~~width~~} & \multicolumn{1}{c||}{~~relative width~~} 
    & \multicolumn{1}{c|}{rel at} &  near absolute &  near relative \\ 
 ~Case~ & \multicolumn{1}{c}{~~$\langle p\rangle$ ~~}   &  \multicolumn{1}{c}{~~$\Delta p/m$~~}   
          &  \multicolumn{1}{c||}{$\Delta p/\langle p\rangle$}  & ~~decay?~~ 
              & ~~marginality?~~ & ~~marginality?~~ \\ 
\hline
\hline
~A~ & \multirow{2}{*}{~{\il{p\ll m}}~} &  \multirow{2}{*}{~~narrow~~} & \multirow{3}*{$\calO(1)$} & \multirow{3}{*}{~non-rel~} & \multirow{2}{*}{near} & far \\ \cline{7-7} 
 B  &  &  &  & & & $\calO(1)$ \\ \cline{2-3}\cline{6-7} 
 C  &  \multirow{3}{*}{~{\il{p\sim m}}~} & $\calO(1)$ &  & & $\calO(1)$ & far \\ \cline{3-4} \cline{5-7} 
 D  & & narrow & narrow & \multirow{2}{*}{  \phantom{$_\sim$}rel$_\sim$ } & near & near \\ \cline{3-4} \cline{6-7}
 E  & & $\calO(1)$ & \multirow{3}{*}{$\calO(1)$} & & \multirow{1}{*}{$\calO(1)$} & $\calO(1)$ \\ \cline{2-3}\cline{5-7}

 F  & \multirow{6}{*}{~{\il{p\gg m}}~} & \multirow{2}{*}{wide} &  & \multirow{2}{*}{non-rel} & \multirow{2}{*}{far} & 
           ~~far ~({\il{p_{\rm parent} \ll m_{\rm daughter}}})~~ \\ \cline{7-7}
 G  &  &   & & & & far ~({\il{p_{\rm parent} \sim m_{\rm daughter}}}) \\ \cline{3-4} \cline{5-7}
 H  & & $\calO(1)$ &  narrow & \phantom{$_\gg$}rel$_\gg$ & near & near \\ \cline{3-4} \cline{5-7}
 I  &  &  \multirow{3}{*}{wide} & \multirow{3}{*}{$\calO(1)$} & non-rel & far &   far ~({\il{p_{\rm parent}\gg m_{\rm daughter}}}) \\ \cline{5-7}
 J  & & & & \phantom{$_\sim$}rel$_\sim$ & far & $\calO(1)$ \\ \cline{5-7} 
 K  & & & & {\phantom{$_\gg$}rel$_\gg$} & $\calO(1)$ or far & near \\
\hline\hline
\end{tabular}
\end{center}
\caption{ The full daughter phase-space distributions $g(p)$ that can arise from the 
    decays of a relatively narrow parent distribution, assuming a rigorous exponential decay 
    with all relativistic time-dilation and redshifting effects included.
    For concreteness in this analysis we have assumed a two-body decay with identical daughters. 
   Given the properties of the daughter distribution, 
     we can therefore reconstruct the extent to which the parent population was relativistic and the 
           extent to which the corresponding decay process was near marginality.  
       In some cases this reconstruction is unique, while in other cases several possibilities exist.
           The quantities at the top of each column of this table are the same as for Table~\ref{inversedeposit}.} 
\label{inversepacket}
\end{table*}

Finally, we marshall our forces 
and examine the process of a parent packet decaying to a daughter packet
including all relativistic, redshifting, and exponential-decay effects. 
It is most efficient to  analyze this case numerically, time-evolving our system according to the full  Boltzmann equations,
and in this way we have extracted 
general results for our final daughter {\it packets}\/ 
along the lines of the general results we previously obtained for the individual daughter {\it deposits}\/ in Table~\ref{inversedeposit}.~
For concreteness, in performing this analysis we have restricted our attention 
to the case of two-body decays with identical daughters, and we have
likewise assumed a relatively narrow parent packet, with {\il{\Delta p_P\ll m_P}}.   Other than these restrictions,
we have examined the full range of possibilities in this class, 
focusing on the fundamental trends that emerge in limiting hierarchical cases involving only rough orders of magnitude, as in 
Table~\ref{inversedeposit}.~
Our results are shown in Table~\ref{inversepacket}, where the column headings have the same definitions as in
Table~\ref{inversedeposit}.~
Note that an explicit numerical example which may help to further elucidate these general results
appears in Appendix~\ref{app:decay_example}.

This table describing the daughter packets 
exhibits many of the same sorts of correlations that we already saw in Table~\ref{inversedeposit} for the individual deposits.
For example, 
under the assumptions inherent in this table,
we see that there is only one way to achieve an ultra-relativistic daughter packet
with {\il{\Delta p\sim m}} and {\il{\Delta p\ll \langle p\rangle}}:
our parent must also have been 
ultra-relativistic and experienced decays
which are near both absolute and relative marginality (Case~H).
Likewise, we see that there are only two ways
of producing a relativistic
 daughter packet for which 
{\il{\langle p\rangle\sim \Delta p \sim m}}:  we can have
either a non-relativistic parent packet experiencing a decay which
is far from relative marginality [but exhibiting 
$\calO(1)$ absolute marginality] (Case~C), 
or a relativistic parent packet experiencing
$\calO(1)$ absolute and relative marginality (Case~E).~
Finally, and perhaps most interestingly, we see that
under the assumptions inherent in this table,
it is not possible to realize an ultra-relativistic
daughter packet with {\il{\Delta p\ll m}}.

It is instructive to compare the results of Table~\ref{inversepacket} for the final daughter packets
with those of Table~\ref{inversedeposit} for the individual deposits.
In Cases~B, D, E, H, J, and K, we see that the widths $\Delta p$  of the final daughter packets are not significantly
different from those of the individual daughter deposits. 
By contrast, in Cases~A, C, F, G, and I, we see that our final daughter packets are significantly broader 
than the individual deposits from which they were constructed. 
Indeed, comparing Tables~\ref{inversedeposit} and \ref{inversepacket}, 
we see that the most significant broadening of all occurs 
for Case~F,
in which `narrow' individual daughter deposits with {\il{\Delta p\ll m}} actually combine to 
produce a `wide' daughter packet with
{\il{\Delta p\gg m}}. 

At first glance, 
comparing Tables~\ref{inversedeposit} and \ref{inversepacket},~ 
we see that broadening feature appears to be perfectly correlated with
 the degree to which the daughter momentum in the rest frame of the parent exceeds the parent momentum at the time of decay, or equivalently
the degree to which the decay process is far from relative marginality:
Cases~A, C, F, G, and I correspond to cases which are all far from relative marginality,
while Cases~B, D, E, H, J, and K correspond to cases which either are near relative marginality or 
experience $\calO(1)$ relative marginality.
Indeed, as we might expect, Case~F is actually the farthest from relative marginality.
It is also natural to speculate 
that for Cases~B, D, E, H, J, and K, broadening fails to occur because 
the daughter deposits all essentially accrue {\it on top of each other}\/ throughout the decay process.
By contrast, for 
Cases~A, C, F, G, and I, it is natural to speculate that each daughter deposit tends to be 
horizontally displaced relative to the previous one, so that they land {\it side-by-side}\/.

\begin{table*}[t!]
\begin{center}
\begin{tabular}{ || c || c|c|c||} 
\hline
\hline
  Decay relative marginality   & \multicolumn{3}{c||}{Parent relativistic at decay?} \\
\cline{2-4}
$p_D^{\rm rest}/p_P$  &  
   {\il{p_P\ll m_P}} & {\il{p_P\sim m_P}} & {\il{p_P\gg m_P}} \\
\hline
\hline 
{\il{\displaystyle{ {p_D^{\rm rest}\over p_P} ~\ll~ {m_D\over m_P}}}} 
           &   \parbox[c]{1.55truein}{\medskip{\bf Weak Broadening}\/:  Deposits 
          are initially very narrow, 
           with $p_{\rm min}$ decreasing
            and $p_{\rm max}$ increasing slightly throughout decay process.
             Resulting daughter packet is narrow with sharp\\ cusp,  
              reflecting small \\
         phase space for decay.  \medskip}      
     &   \parbox[c]{1.5truein}{
          \medskip {\bf Weak Broadening}\/:  Deposits initially land on top of each other,
           then transition to behavior in which $p_{\rm min}$ decreases\\ 
            while $p_{\rm max}$ increases.   
          \\ {\bf (Case D)}\medskip }
     & \parbox[c]{2.1truein}{
            {\bf \smallskip 
              No Broadening}\/:\\
            Deposits land on top of each other throughout decay process.\\   
            Resulting daughter packet has\\ sharp left and right edges.  
          \\ {\bf (Cases H,K1)} \smallskip}     \\
\hline
{\il{\displaystyle{ {p_D^{\rm rest}\over p_P} ~\sim~ {m_D\over m_P}}}} 
      &  \parbox[c]{1.5truein}{
          {\bf Weak Broadening}\/:  Deposits initially land on top of
         each other,
           then transition to landing side-by-side.
          \\ {\bf (Case B)}}
     &   \parbox[c]{1.5truein}{
          {\bf Weak Broadening}\/:  Deposits initially land on top of each other,
           then transition to landing side-by-side.
          \\ {\bf (Case E)}}
     & \parbox[c]{2.1truein}{
            {\bf \medskip
              Weak Broadening}\/:\\
            Deposits initially land on top of \\
          each other, 
            then transition to behavior in which $p_{\rm max}$ remains constant
             while $p_{\rm min}$ drops to zero and then begins to grow.\\
             Decay ends before $p_{\rm min}$ reaches\\ 
            $p_{\rm max}$.
             Resulting daughter packet\\ has sharp right edge.\\
              {\bf (Case K2)} \medskip }\\
\hline
{\il{\displaystyle{ {p_D^{\rm rest}\over p_P} ~\gg~ {m_D\over m_P}}}} 
      &  \parbox[c]{1.5truein}{
          {\bf Strong Broadening}\/:  Deposits land side-by-side throughout the decay process.
          \\ {\bf (Cases A,C,F,G,I)}}
     &   \parbox[c]{1.5truein}{
           \medskip
          {\bf Weak Broadening}\/:  Deposits initially land 
           with $p_{\rm min}$ increasing and $p_{\rm max}$ remaining constant,\\
              but then transition to landing side-by-side. \\ {\bf (Case J)} \medskip } 
     & \parbox[c]{2.1truein}{
              {\bf No Broadening}\/:\\
            Deposits land on top of each other throughout decay process.\\ 
             Resulting daughter packet has\\
               sharp left and right edges.\\
              \smallskip }\\
\hline\hline
\end{tabular}
\end{center}
\caption{ Stacking dark-matter deposits:  the interplay between decay kinematics and cosmological expansion.  
               This table describes how 
            successive dark-matter deposits are stacked in order to build a complete daughter phase-space distribution,
                as seen in the comoving frame and 
            assuming the two-body decay of a parent 
               of mass $m_P$ 
with a very narrow phase-space distribution 
into two identical daughters of mass $m_D$.
             Here $p_P$ is the momentum of the parent at the decay time {\il{\tau= 1/\Gamma}}.
            As the decay process unfolds, the manner in which successive decay deposits are stacked is ultimately a function of not only 
              the degree to which the parent is relativistic at the time of its decay but also the relative 
                marginality $p_D^{\rm rest}/p_P$ of the decay process itself.  
         Note that we can equivalently express the relationships between $p_D^{\rm rest}/p_P$ and $m_D/m_P$ along the left-most
               column as relationships
             between $p_D^{\rm rest}/m_D$ and $p_P/m_P$ --- \ie, as relationships between the degree to which the
             daughter is relativistic within the rest frame of the parent and the degree to which the parent is relativistic within
              the background frame.  
          The information in this table explains how the data in Table~\ref{inversedeposit} ultimately 
                  leads to the data in Table~\ref{inversepacket}, and outlines
                   the cases in which cosmological expansion has a particularly significant impact
             on the resulting dark-matter phase-space distribution.}
\label{stacking}
\end{table*}

While these expectations are mostly correct,
the full story is actually more subtle.
It turns out that the manner in which successive decay deposits are stacked throughout the decay process
is ultimately a function of not only 
the relative marginality $p_D^{\rm rest}/p_P$ of the decay process,
but also the degree to which the parent is relativistic at the time of its decay. 
Moreover, the resulting deposit stacking 
patterns are actually more complex 
than mere ``on top of each other'' or ``side-by-side''.

In Table~\ref{stacking}, we describe the actual stacking patterns that result for a two-body decay
of a relatively narrow parent packet of mass $m_P$ into two identical daughters of mass $m_D$.
For ease of visualization, we describe these stacking patterns {\it within the comoving frame}\/ 
(\ie, as if we are riding along our cosmological momentum-space conveyor belt).
As the decay process unfolds,
the momentum of the parent continually redshifts;  we shall nevertheless identify a ``typical'' momentum $p_P$ 
as the central momentum of the (narrow) parent packet at {\il{t=1/\Gamma}} where $\Gamma$ is the decay width.
In general, the momentum of each deposit stretches between minimum and maximum values $p_{\rm min}$ and $p_{\rm max}$,
and our stacking patterns are determined by examining how these two quantities independently evolve with time 
in the comoving frame.
For the purposes of Table~\ref{stacking}, we strictly define deposits ``landing on top of each other''
as the situation in which both $p_{\rm min}$ and $p_{\rm max}$ remain constant in the comoving frame (implying that 
the deposits not only stack directly on top of each other but also share a common width).
By contrast, we define deposits ``landing side-by-side'' as the situation in which 
both $p_{\rm min}$ and $p_{\rm max}$ increase with time in the comoving frame, while again holding the deposit width 
$p_{\rm max}-p_{\rm min}$ fixed.
For ease of comparison we have 
also indicated which of the cases within Table~\ref{inversepacket} corresponds to which stacking pattern,
with K1 and K2 indicating to the two possibilities for Case~K within Table~\ref{inversepacket} with
absolute marginalities which are either $\calO(1)$ or far, respectively. 

As evident from Table~\ref{stacking}, 
several different stacking patterns are possible.   
These stacking patterns then allow us to assess the degree to which significant broadening might occur
when the individual daughter deposits are combined to form the final daughter packet.
As anticipated, 
Cases~A, C, F, G, and I all experience ``side-by-side'' stacking.
As such, these cases experience the strongest degree of broadening --- a broadening
which is sufficiently large that it can be seen in
comparing the results of Tables~\ref{inversedeposit} and \ref{inversepacket}.~
Likewise, for Cases~H and K1 the deposits are stacked ``on top of each other''.
Thus in these cases there is literally no broadening in transitioning from the width of the individual deposits
to the width of the resulting packet.
Finally, however, 
in Cases~B, D, E, J, and K2, 
we see that broadening does occur.
However, it is evident from the stacking patterns described in Table~\ref{stacking}
that this broadening is much weaker than that experienced for true ``side-by-side'' stacking.
This broadening is therefore too small 
to appear in the comparison between Tables~\ref{inversedeposit} and \ref{inversepacket}.~  

Ultimately, the results in Table~\ref{stacking} describe a subtle interplay
between the {\it kinematics} of the decay process
and the {\it cosmological expansion}\/ of the universe in which the decay is embedded.
In some cases, we see that the cosmological expansion has little overall effect on the shape
of the resulting daughter packet --- indeed, in such cases a packet of similar shape would have emerged 
even if the universe were not expanding.
Moreover, as indicated in Table~\ref{stacking}, these packets tend to have relatively sharp edges
(ultimately stemming from our assumption of a very narrow parent packet).
In other cases, by contrast, we see that the cosmological expansion plays a significant role in shaping the
resulting daughter packet.  In such cases the resulting packets tend not to have such sharp edges,
even when the parent packet is extremely narrow.

\begin{figure}[h!]
\centering
\includegraphics[keepaspectratio, width=0.49\textwidth]{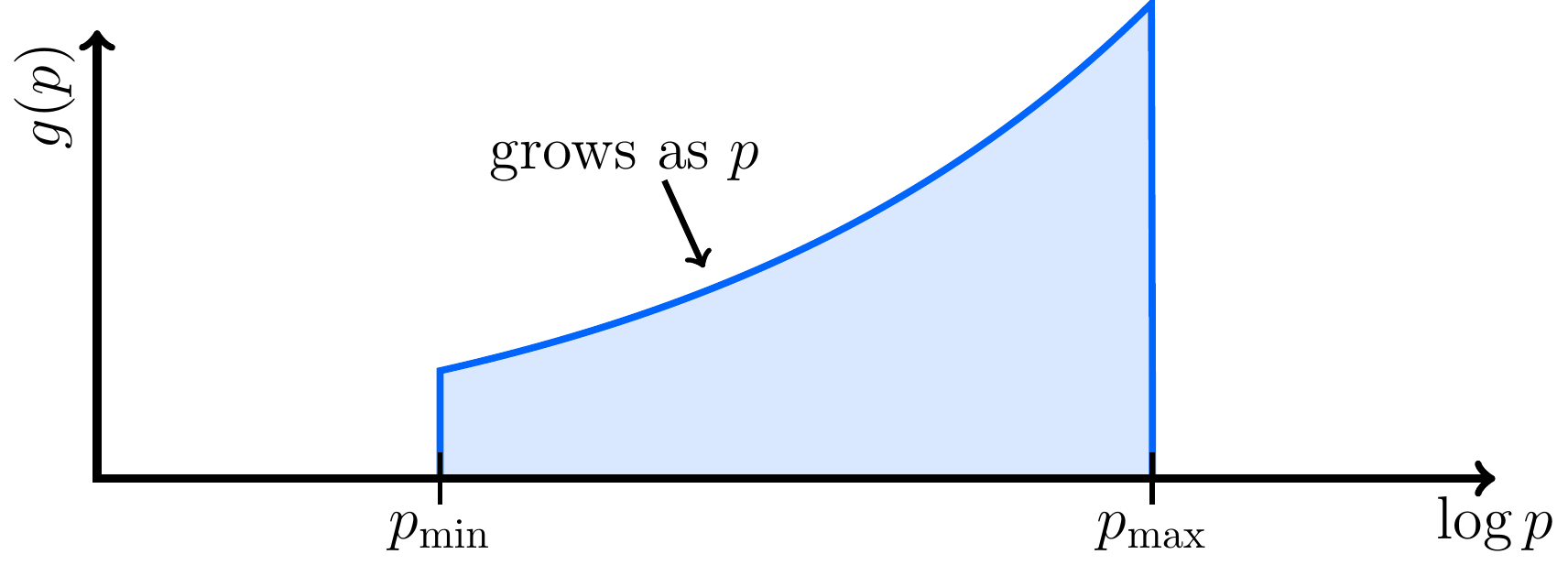} 
\caption{
The universal form for the dark-matter phase-space distribution $g(p)$ corresponding to Cases~H and K1,
under the assumptions inherent in Table~\ref{stacking}.~
This distribution exhibits
sharp left and right edges at $p_{\rm min}$ and $p_{\rm max}$, respectively,
and exhibits growth scaling linearly with $p$
for {\il{p_{\rm min}\leq p\leq p_{\rm max}}}.} 
\label{fig:boxcartoon}
\end{figure}

Finally, when combined with previous observations,
the results in Table~\ref{stacking}
even occasionally allow us to predict the full functional form for the resulting daughter packet.
For example, under the assumption of a two-body decay into identical daughters,
we have already seen that our deposits have flat profiles 
when plotted as functions of the daughter energy $E$.
In other words, when plotted in $E$-space,
each deposit is ``brick''-shaped.
However, we now see from Table~\ref{stacking} that Cases~H and K1 have the unique 
property that all of the deposits land directly on top of each other 
while sharing the same comoving edge momenta $p_{\rm min}$ and $p_{\rm max}$
throughout the decay process. 
Furthermore, we see from Table~\ref{inversepacket} that the daughters for Cases~H and K1 
are both highly relativistic, which means that a flat profile in $E$-space is equivalently a
flat profile in $p$-space.    Such a profile therefore grows linearly with $p$ when plotted versus $\log p$.
This results in the daughter packet shown Fig.~\ref{fig:boxcartoon}.

\begin{figure}[t]
\centering
\includegraphics[keepaspectratio, width=0.49\textwidth]{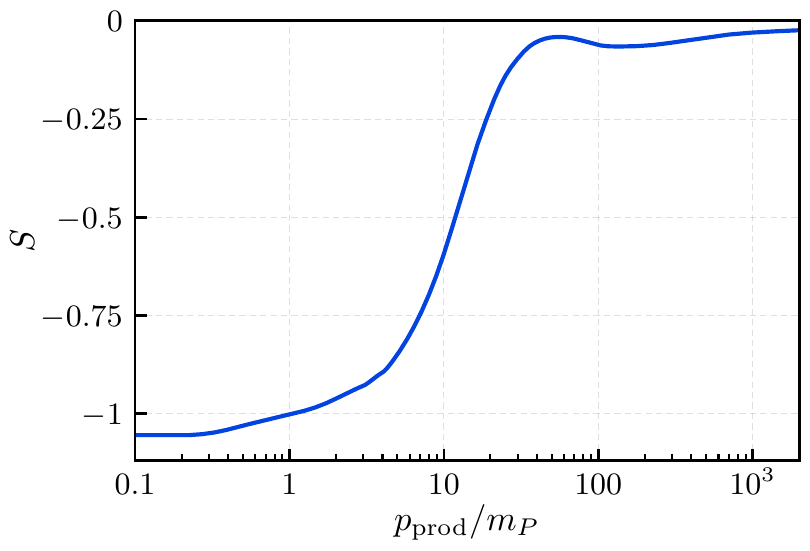} 
\caption{The skewness $S$ of the daughter packet resulting from the decay of a (skewless) log-Gaussian parent packet,
plotted as a function of the central momentum $p_{\rm prod}$ of the parent at production.
In all cases the width of the parent packet is held constant,
stretching
across two orders of magnitude in momentum, and we have
assumed  a two-body decay with identical daughters for which
{\il{m_D/m_P\approx 0.4}}.
We have also taken {\il{\Gamma/H(t_0)=10^{-3}}} and {\il{\kappa=3/2}}.
As the parent becomes increasingly relativistic, we see that the skewness of the daughter increases
dramatically from the universal asymptotic non-relativistic limit {\il{S\approx -1.06}} (in agreement with the results
in Fig.~\ref{fig:Snonrel}) towards {\il{S\approx 0}}.
This confirms that our expectations from Eq.~(\ref{tiltrules}) survive a full Boltzmann analysis.}
\label{fig:skewplot}
\end{figure}

We can also determine the extent to which
our previous 
expectation from Eq.~(\ref{tiltrules})
concerning the tilt of the daughter packet 
actually survives a full Boltzmann analysis, with all relativistic, redshifting, and exponential-decay effects included.
In Fig.~\ref{fig:skewplot} we plot the skewness $S$
of the daughter packet that emerges from the decay of a log-Gaussian
parent packet (\ie, a parent packet which has a Gaussian shape when plotted on a logarithmic axis)
as a function of the average parent momentum
at the time the parent is produced, holding the width of the parent packet fixed
and taking the reference value {\il{\Gamma/H(t_0)=10^{-3}}}.
We see that this skewness indeed shows a strong dependence
on the average momentum of the parent at production,
rising from {\il{S\approx -1.06}} 
in the non-relativistic limit
[in agreement with Fig.~\ref{fig:Snonrel}
for this value of $\Gamma/H(t_0)$]
towards {\il{S\approx 0}} as the parent becomes increasingly relativistic
at production.
Thus we see that our observations regarding skewness survive a full Boltzmann analysis.
Indeed, the maximum value of $S$ in the ultra-relativistic limit 
depends on the skewness of the original parent packet.

These observations concerning skewness
provide another tool 
which can help us determine the properties of the parent from 
the properties of the daughter.
For example,
we have already seen  in Table~\ref{inversepacket}
that
a relativistic daughter packet which is narrow, with {\il{\Delta p \ll m}} as
well as {\il{\Delta p \ll \langle p\rangle}}, could in general be the result of
either 
a relativistic parent experiencing a near-marginal decay (Case~E) or a non-relativistic
parent experiencing a decay which is farther from marginality  (Case~C).
The skewness of the daughter packet may therefore be a useful tool in helping to distinguish between
these two cases.

It is interesting that the curve in Fig.~\ref{fig:skewplot} is non-monotonic.
Indeed, even though the skewness $S$ generally increases as the parent becomes increasingly
relativistic at production, we see that there is a region {\il{p_{\rm prod}/m_P\sim \calO(50-500)}}
for the parent momentum at production
within which the skewness $S$ experiences a slight dip.
Given the parameters chosen for this plot, this is the region
in which the average parent momentum $p_P$ {\it at decay}\/  is approximately equal
to the momentum $p_D^{\rm rest}$ of the daughters in the rest frame of the parent.
Indeed, in the language of Tables~\ref{inversedeposit} and \ref{inversepacket},
this is the region in which our decay process transitions from being {\it near}\/ 
relative marginality to being {\it far}\/ therefrom.
When the decay process is near relative marginality,
the width $\Delta p$ of the daughter packet is largely set by $p_D^{\rm rest}$ while
its average momentum $\langle p\rangle$ is set by $p_P$. 
By contrast, when the
decay process is far from relative marginality,
the reverse is true.
Thus the region with
{\il{p_D^{\rm rest}/p_{\rm decay}\approx \calO(1)}}
[which for this figure translates to the region
{\il{p_{\rm prod}/m_P\sim \calO(50-500)}}]
marks the transition between these 
two behaviors, with the slight dip in the skewness emerging as a transitional effect.

\subsection{Non-minimal dark-sectors:  Overlapping decay chains and multi-modal dark-matter distributions\label{sec:overlapping}}

\begin{figure}[t]
\centering
\includegraphics[keepaspectratio, width=0.45\textwidth]{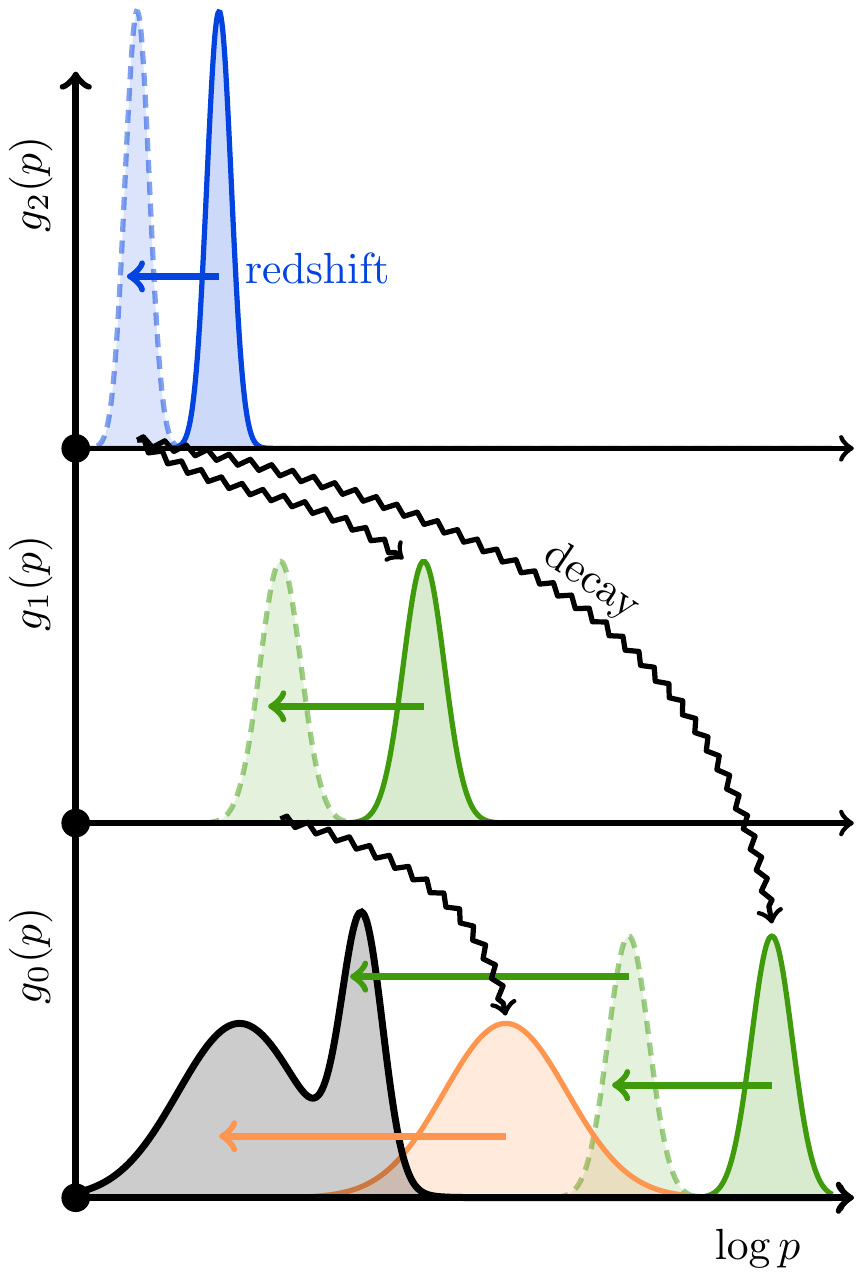} 
\caption{One possible scenario leading to a non-trivial phase-space distribution $g(p)$ at late times through a sequence of successive ``deposits'' of the sort sketched in Fig.~\ref{fig:conveyor1}.~
An excited state is initially created with a simple unimodal (and possibly even thermal) distribution at level {\il{\ell=2}} (blue) and redshifts towards smaller momenta before undergoing a two-body decay {\il{2 \rightarrow 1+0}}.
Each of these daughter distributions then redshifts until the {\il{\ell=1}} daughter undergoes its
own decay {\il{1 \rightarrow 0+0}}.
The resulting distributions combine and continue to redshift, ultimately producing a final phase-space distribution 
 (black).  Thus, even though we began with a simple unimodal distribution, a complex bi-modal distribution eventually emerges
as the result of a superposition of results from two competing decay chains.}
\label{fig:conveyor2}
\end{figure}

Having understood the individual parent-to-daughter decay process, let us now examine how intra-ensemble decays
within a non-minimal, multi-component dark sector
can yield a multi-modal dark-matter phase-space distribution.
To frame our discussion, let us consider the three-state system
illustrated in Fig.~\ref{fig:conveyor2}.~
Here the three states are labeled {\il{\ell=0,1,2}} in order of increasing mass, 
and we shall assume that only the heaviest state is initially populated.
For simplicity, we shall assume that this state
corresponds to a single non-relativistic 
unimodal packet (sketched in blue in Fig.~\ref{fig:conveyor2}) --- indeed,
this packet can even be thermal.
This packet then redshifts until it decays.
For simplicity, let us assume that each decay is a two-body decay,
and in this case we shall assume that the {\il{\ell=2}} state undergoes a decay of the form {\il{2\to 1+0}} in order to produce
two non-relativistic daughters (green), one with {\il{\ell=1}} and the other with {\il{\ell=0}}.
Each of these new daughter packets then redshifts until the {\il{\ell=1}} daughter undergoes its own decay 
into two kinematically identical non-relativistic {\il{\ell=0}} granddaughters (orange). 
As shown in Fig.~\ref{fig:conveyor2}, these granddaughters happen to have a non-negligible overlap 
with the redshifted {\il{\ell=0}} daughter (dashed green). 
They therefore superpose, and the resulting distributions continue to redshift, ultimately producing a final $g(p)$ distribution (black).

Note that each decay, whether from parent to daughter or from daughter to granddaughter,
exhibits the characteristics discussed above.
In particular,
because all of our packets are assumed non-relativistic,
each decay produces offspring packets  
whose widths in momentum space are larger than those of the packet from the preceding generation.
These offspring packets also have higher momenta 
than those of the corresponding parent packet, as discussed above.
Each of the offspring packets nevertheless has the same total area as the parent packet
from which it emerged, since each daughter is descended from a single parent.
Indeed, all of the other daughter properties discussed in Sect.~\ref{parent_to_daughter} also continue to apply.
 
As anticipated, this system furnishes us with an explicit example in which the deposits 
onto the {\il{\ell=0}} conveyor belt  exactly mirror those of packets $A$ and $B_1$ 
in Fig.~\ref{fig:conveyor1}.~ It is for this reason that  both processes produce 
the same sort of bi-modal distribution.
Thus, we see from this example that the $A/B_1$ portion of the deposit pattern in Fig.~\ref{fig:conveyor1} 
can indeed emerge naturally --- 
these deposits can simply arise as the end-products of separate decay chains from a single 
heavy source.  
(Likewise, the $B_2$ portion of Fig.~\ref{fig:conveyor1} can also emerge through an additional independent decay chain.)
We therefore conclude that 
a complex, multi-modal dark-matter distribution function can easily be produced,
even when our original parent state is unimodal (and potentially even thermal).

We may also study the deposit rate functions $\Delta(p,t)$ arising from such scenarios.
To do this, we shall work backwards and begin by focusing on those states  whose decays directly produce
ground-state daughter particles.
In particular, we begin by considering the 
set of parent packets labeled by $\ell$ which 
decay at times $t_{\ell}$
and produce daughter packets directly on the {\il{\ell=0}} dark-matter conveyor belt.
Eq.~(\ref{discrete}) then becomes
\beq
  \Delta (p,t) ~=~ \sum_\ell \Delta_\ell(p) \,\delta(t-t_\ell)~,~
\label{preprobs}
\eeq
where $\Delta_\ell(p)$ is the profile of the dark-matter phase-space contribution 
coming from the decay of parent $\ell$.
Moreover, we can further write
\beq
    \Delta_\ell(p) ~=~ \int dp'\, g_\ell(p')\, P_{\ell\to 0} (p',p)~,
\label{probs}
\eeq
where $g_\ell(p')$ is the phase-space distribution of the parent $\ell$
and where {\il{P_{\ell\to 0}}} denotes the probability per unit momentum that
a parent $\ell$ with momentum $p'$ produces an {\il{\ell=0}} daughter with momentum $p$, normalized
by the multiplicity of the decay.
Indeed, it is within these probability functions that much of 
the kinematic information embedded in Table~\ref{inversepacket} resides.
Substituting Eqs.~(\ref{preprobs}) and (\ref{probs}) into Eq.~(\ref{genexp}) then yields
\beq
  g(p)~=~\sum_{\ell} \int d p'\,
       g_{\ell}(p') \,
        P_{\ell\rightarrow 0}\left(p', p\frac{a(t)}{a(t_{\ell})}   \right)~.
\label{parentsonly}
\eeq

This result is complete as is.   However, we could push this further by recognizing that the
parents $\ell$ are most likely to be descendants of grandparents $k$, which in turn are likely
to be descendants of great-grandparents $j$,  and so forth.
We can therefore write the parent distribution $g_{\ell}(p')$ in Eq.~(\ref{parentsonly})  
in terms of deposits from grandparent decays, and recursively iterate this process 
back to a primordial initial state $a$. 
We then obtain 
\beqn
  && g(p)~=\!\!\! \sum_{\{a,b,\dots,k,\ell\}}\int d p_{\ell}\int d p_{k} \dots 
                  \int d p_{b} \int d p_{a}  \,g_a(p_a)\nonumber\\ 
    && ~~\times P_{a\rightarrow b}\left(p_a,p_b\frac{a(t_b)}{a(t_a)}\right)
              P_{b\rightarrow c}\left(p_b,p_c\frac{a(t_c)}{a(t_b)}\right) \cdots \nonumber\\
    && ~~\times P_{k\rightarrow \ell}\left(p_k,p_{\ell}\frac{a(t_{\ell})}{a(t_k)}\right)
              P_{\ell\rightarrow 0}\left(p_{\ell},p\frac{a(t)}{a(t_{\ell})}\right)~.~~~~~~~~~~\nonumber\\
\eeqn
Here the summation over $\{a,b,\dots,k,\ell\}$ represents the summation over all possible {\it decay chains}\/ 
that start from the primordial state $a$;
note, in this context, that not all decay chains are of equal length, 
as this depends on how many intermediate steps are taken during the decay process.
Likewise, each $t_i$ ({\il{i=a,b,...,\ell}})  denotes the time at which  
the state $i$ decays, where we are 
implicitly assuming that each state decays instantaneously
once its lifetime $\tau_i$ is reached.
Thus we are implicitly taking 
{\il{t_{i}\equiv t_0^{(a)}+ \tau_{a}+\tau_{b}+\dots+\tau_i}}, where
$t_0^{(a)}$ is the time at which the primordial ancestor $a$ is produced.

Thus far, we have studied the manner in which a final phase-space distribution $g(p)$ is constructed from the results
of individual decay chains, and we have seen that multi-modality can emerge naturally through the superposition
of the results from separate chains.
However, even the analyses 
we have performed 
miss certain features.
For example, we have been assuming in our discussions until this point that all decays happen instantaneously
at {\il{t=1/\Gamma}}, where $\Gamma$ is the associated decay width.   In reality, however, decays occur continuously
both before and after $1/\Gamma$.     
In a similar vein, we also implicitly assumed that
each momentum slice of a given parent
is created at the same time and therefore feels the
same ``clock''.   However, this also is not 
necessarily true.  For example, we assumed that the {\il{\ell=1}} state  
in Fig.~\ref{fig:conveyor2} 
is produced, redshifts, and then decays.   However, because of the
continuous nature of the decay process, some parts of the {\il{\ell=1}} 
packet might still remain to be created while other parts of the packet might have
already begun to decay.   This is a continuous process in which the {\il{\ell=1}} state
is essentially an intermediate resonance.  

For accurate results, features such as these must also be taken into account.
It is then natural ask whether effects such as these
might ``wash out'' the features (such
as multi-modality)
that we have been discussing --- thereby restoring
a traditional packet shape --- or whether such features survive.

To study this, we shall perform a full numerical analysis of the Boltzmann equations
corresponding to a physical system involving multiple independent decay chains.
(The Boltzmann equations for a general multi-state system are presented in Appendix~\ref{app:boltzmann}.)
In particular, we shall begin with the same setup as in Fig.~\ref{fig:conveyor2}, namely
a three-state system which begins with only the heaviest state populated.   For concreteness
we shall even imagine that this state has a non-relativistic thermal distribution  
{\il{g_2\sim p^3 \exp(- p^2/2m_2 T)}} where {\il{T=m_0/20}} is chosen as a reference value.
We shall also take the masses in the ratio {\il{m_2/7=m_1/3=m_0}},
and let $\Gamma^\ell_{ij}$ denote the width for the decay {\il{\ell\to i+j}}.  For simplicity,
we shall restrict our attention to symmetric decays with {\il{i=j}}.

\begin{figure}[t]
\includegraphics[keepaspectratio, width=0.48\textwidth]{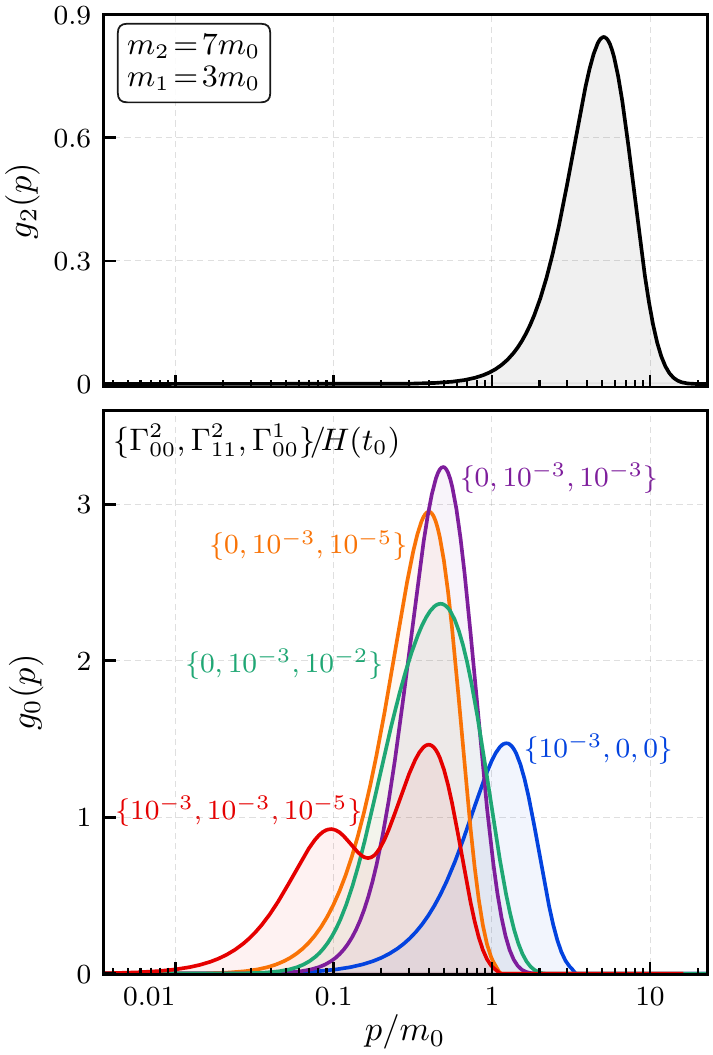}
\centering
\caption{Decays within a three-state system with mass ratios 7:3:1.
We begin with an initial thermal distribution (top panel, black curve) and 
for each of five different combinations of decay widths $\lbrace \Gamma^2_{00}, \Gamma^2_{11}, \Gamma^1_{00}\rbrace$,
we calculate the resulting phase-space distribution of the ground state (bottom panel, colored curves)
at the moment when all decays are complete.
This final time can therefore be different for each curve.
Note that our chosen decay widths are expressed 
as fractions of the Hubble parameter $H(t_0)$ at the time $t_0$ when the initial distribution is established.
We have also taken {\il{\kappa=3/2}}.
Within these plots the vertical axes are normalized such that the parent packet has unit area.}
\label{jade}
\end{figure}

Once the highest level is populated, a decay cascade is initiated and proceeds until the total
resulting abundance eventually collects in the ground state.
However, the ground-state phase-space distribution that results 
depends on the specific choice of decay widths $\Gamma^\ell_{ij}$, as each choice
corresponds to a different decay pattern.
In Fig.~\ref{jade} we plot the resulting phase-space distributions of the ground state 
for five different combinations of values for
$\lbrace \Gamma^2_{00}, \Gamma^2_{11}, \Gamma^1_{00}\rbrace$,
with each value expressed in Fig.~\ref{jade} as a fraction of the
Hubble parameter $H$ at the time when the phase space distribution $g_2(p)$ is
initially established.
In each case, we show this final distribution at the moment when the intra-ensemble decay processes
has concluded (defined as the  moment at which this distribution has accumulated 99.5\% of its final
asymptotic expected abundance).

Each of the cases shown in Fig.~\ref{jade} has a direct physical interpretation.
\begin{itemize}
\item 
{\il{\lbrace \Gamma^2_{00}, \Gamma^2_{11},  \Gamma^1_{00}\rbrace/H(t_0) = \lbrace 10^{-3},0,0\rbrace}} (blue):\\
       In this case, the heaviest state decays directly into the ground state. This is therefore effectively a two-component system. 
       Because the rest mass of the daughters is significantly less than the rest mass of the parent, the daughter
       packets emerge with considerable average momentum  and are in fact relativistic.
\item
{\il{\lbrace \Gamma^2_{00}, \Gamma^2_{11},  \Gamma^1_{00}\rbrace/H(t_0) = \lbrace 0,10^{-3},10^{-5}\rbrace}} (orange):\\
    In this case, the direct decay into the ground state is forbidden. 
    Instead, in order to reach the ground state,
       the decays must go through two steps, {\il{2\to 1+1}} followed by {\il{1\to 0+0}}.
     However, as compared with the first step, the second step has a much smaller decay rate.
    As a result, the intermediate state {\il{\ell=1}} is able to fully develop and carry significant abundance 
        before it begins decaying to the ground state. 
\item
{\il{\lbrace \Gamma^2_{00}, \Gamma^2_{11},  \Gamma^1_{00}\rbrace/H(t_0) = \lbrace 0,10^{-3},10^{-3}\rbrace}} (purple):\\
    This case is similar to the previous case, but the decay rate from {\il{\ell=1}} to {\il{\ell=0}} is the same as that from {\il{\ell=2}} to {\il{\ell=1}}.
    Therefore, the intermediate state begins decaying {\it while}\/ is still forming, thereby putting an effective
     upper limit on the abundance this state can carry.
\item
{\il{\lbrace \Gamma^2_{00}, \Gamma^2_{11},  \Gamma^1_{00}\rbrace/H(t_0) = \lbrace 0,10^{-3},10^{-2}\rbrace}} (jade):\\
    This case is similar to the previous two cases, but with a much faster second step. 
    Therefore, as soon as any portion of the intermediate state begins to form, it almost immediately 
        decays into the ground state.    As a result this state never accumulates appreciable abundance.
\item
{\il{\lbrace \Gamma^2_{00}, \Gamma^2_{11},  \Gamma^1_{00}\rbrace/H(t_0) = \lbrace 10^{-3},10^{-3},10^{-5}\rbrace}} (red):\\
    Here the {\il{\ell=2}} state has an equal probability (branching ratio) to decay into the {\il{\ell=1}} state
    or the {\il{\ell=0}} state, with the former then eventually decaying into the {\il{\ell=0}} state.
    Thus, unlike each of the previous cases, this case exhibits two independent decay chains.
    Indeed, this case represents a superposition of the blue and orange cases listed above,
    and one may easily verify that the (red) curve for this case is nothing but a weighted combination of the
    orange and (redshifted) blue curves.  
\end{itemize}

We see, then, that  even though the initial phase-space distribution of the parent state is the same for all cases (and
is even chosen to be thermal),
the resulting phase-space  distributions that emerge for the ground state differ rather substantially from each other.
For example, they differ
in their total areas (comoving number densities), with
the parent distribution and our five different ground-state distributions having total areas in the ratios
1:2:4:4:4:3.   This is precisely in accordance with our expectations,
given that each decay produces two daughters and that the final case involves two independent decay chains,
each occurring with $50\%$ probability.
However, these curves also differ in their overall shapes,
yielding results which are either unimodal or bi-modal,
the latter occurring in the single case that exhibits two independent decay chains. 
Indeed, the case shown in red in Fig.~\ref{jade} is essentially the same as that sketched in Fig.~\ref{fig:conveyor2}, except that in Fig.~\ref{jade}
so much time elapses between the two deposits onto the ground state that the first deposit has had time to redshift {\it beyond}\/ the eventual location
of the second deposit.
The fact that the red curve in Fig.~\ref{jade} is nothing but a weighted sum of the orange and (redshifted) blue curves
ultimately rests on the approximate linearity of the underlying Boltzmann equations, a feature which  allows us to treat each decay chain independently and then 
simply add the results.  It is in this final addition process that the multi-modality emerges,
and thus we see that multi-modality is a robust phenomenon, even when
a full numerical Boltzmann analysis is performed.
Moreover, each ``mode'' of the resulting distribution will individually exhibit the
properties already described in Sect.~\ref{parent_to_daughter} for individual daughters.

Note that the Boltzmann equations are generally non-linear.
However, these equations are approximately linear under certain assumptions.
One necessary assumption is that the phase-space distribution $f_i(p,t)$ for each 
species be relatively small, {\it i.e.}\/, {\il{f_i(p,t) \ll 1}},
so that any Bose enhancement (or Pauli blocking in the case of a fermionic species)
can be safely neglected.
However, the Boltzmann equations will also be non-linear
if there exist any significant processes amongst the different species
for which the initial state includes two or more species.
For example, if our system includes two-body decay processes of the form {\il{\phi_\ell \to \phi_i \phi_j}},
then our system necessarily also includes the inverse-decay processes
{\il{\phi_i \phi_j\to \phi_\ell}}
as well as scattering processes of the form
{\il{\phi_i\phi_j\to\phi_\ell\to\phi_{i'}\phi_{j'}}}.
Both kinds of processes will render the Boltzmann equations non-linear,
and consequently have the power to weaken or even potentially eliminate the  multi-modality (or even the 
non-thermality) that would have otherwise
arisen from independent decay chains.
However, as we shall see in Sect.~\ref{sec:toy_model}, it is often 
the case that the amplitudes for two-body decays
can be significant while the amplitudes for the corresponding  
inverse-decay and scattering processes remain small.
In such cases, we expect our multi-modality to be robust.

\section{No longer void and without form:
  From dark-matter phase-space distributions to matter power spectra
\label{sec:perturb}}

We now turn to the second connection indicated in Eq.~(\ref{onepointone}), 
namely that between the dark-matter phase-space distribution $f(p)$ --- or equivalently $g(p)$ --- and the matter power spectrum $P(k)$.
As in Sect.~\ref{sec:phase_space_evolution}, our goal is 
to develop intuition for how different $g(p)$ distributions affect the ultimate shape of $P(k)$.  In this way we can potentially
begin to address the ``inverse'' question of reconstructing certain rough characteristics of $g(p)$ given only the information
associated with $P(k)$.  Indeed, we shall ultimately present a closed-form conjecture which will enable us
to ``resurrect'' many features of $g(p)$, given only the information in $P(k)$.
This is important because it is ultimately only through quantities such as $P(k)$ that the dark-matter phase-space distribution
$f(p)$ makes contact with  observational data.

\subsection{Non-minimal dark sectors and the matter power spectrum}

The basic idea underpinning the connection between $f(p)$ and $P(k)$ is that
dark matter helps to promote the growth of structure
in the early universe.
According to the inflationary paradigm,
inflation gives rise to
a spectrum of primordial curvature perturbations in the early universe.
These perturbations serve as seeds for cosmological structure, allowing
overdense and underdense regions to develop across a spectrum of physical scales.
However, the manner in which this structure evolves
is sensitive to the equation of state of the different species present.
For example, in a universe composed entirely of cold dark matter (CDM),
the primordial perturbations grow linearly with the scale factor, generating
significant structure on even the smallest scales.
By contrast, if some of the dark matter is not purely cold (\ie, if the
dark matter has some non-zero pressure due a non-zero momentum), then dark-matter particles
with a sufficient speed may escape the developing gravitational wells, thereby smoothing
out the structure over the corresponding scales.
As a result, one finds that only perturbations of a certain minimum size are able to remain stable and/or grow with time.

All of this information is encoded within the matter power spectrum $P(k)$.
To define this quantity, we consider the spatial perturbations of the
energy density $\rho(\vec x,t)$ relative to the
zeroth-order, unperturbed, spatially homogeneous  energy density $\rhobar(t)$:
\beq
\delta(\vec{x},t)~\equiv~\frac{\delta \rho(\vec{x}, t)}{\bar{\rho}(t)}~=~\frac{\rho(\vec{x}, t)-\bar{\rho}(t)}{\bar{\rho}(t)}~,
\eeq
where
\beq
     \rho(\vec x, t) ~\equiv~ {1\over (2\pi)^3} \sum_i \, g_i \int d^3 p \, E_i(p) \, f_i(\vec x, \vec p, t)
\label{rhosum}
\eeq
and where
\beq
     \rhobar(t) ~\equiv~ {1\over V} \int_V d^3 x \, \rho(\vec x, t)~
\eeq
for a suitably large fiducial volume $V$.
Note that in Eq.~(\ref{rhosum}) we are explicitly summing over all species that contribute to the total energy density
of the universe, including the dark matter.
Our main interest is then in the two-point
correlation function {\il{\xi(\vec x,t) \equiv \langle \delta(\vec x + \vec y, t) \delta(\vec y,t) \rangle}},
which, under the assumptions of homogeneity and isotropy,  can be Fourier-decomposed in the form
\beq
  \xi(\vec x,t) ~\equiv~ {1\over (2\pi)^3} \int d^3 k ~ P(k,t)~ e^{-i \vec k\cdot \vec x}~,~
\label{FTone}
\eeq
where $k\equiv |\vec k|$ and where $P(k,t)$ is the time-dependent matter power spectrum.
Since $\xi(\vec x,t)$ depends only on $r\equiv |\vec x|$, we can invert Eq.~(\ref{FTone}) and perform
the angular integration to obtain
\beq
         P(k,t) ~=~ 4\pi \int_0^\infty dr \, r^2 \,{\sin kr\over kr} \, \xi(r,t)~.
\eeq
This quantity therefore describes the degree to which the two-point correlation function of
energy-density fluctuations shows power at the scale $k$,
and for convenience we shall henceforth
define {\il{P(k) \equiv P(k,t_{\rm now})}}.
Moreover, in this work we shall mostly be interested in the so-called {\it transfer function}\/ $T(k)$,
defined 
as
\beq
            T(k) ~\equiv~  \sqrt{ { P(k) \over P_{\rm CDM}(k) }}~,
\eeq
where $P_{\rm CDM}(k)$ is the matter power spectrum that would emerge in the CDM limit in 
which all of the dark matter is assumed cold.
Thus, 
for any $f(p)$, the transfer function $T(k)$ indicates the degree to which the corresponding matter power spectrum $P(k)$ deviates
from our CDM-based expectations as a function of the scale $k$.
Conveniently, $T(k)$ is independent of the particular choice of normalizations for $P(k)$,
many of  which exist in the literature.

As a dominant contributor to this process, dark matter plays a leading role in structure formation.
In order to determine which scales $k$ within the matter power spectrum $P(k)$ might be affected  
by dark matter of a given momentum $p$,  we can perform a 
straightforward horizon calculation.
Given that dark matter of momentum $p$ has a velocity
{\il{v= p/E = p/\sqrt{p^2 + m^2}}},   
the corresponding horizon size $d_{\rm hor}$ 
in a static flat universe 
would simply be $vt$, whereupon  we could identify a corresponding wavenumber {\il{k_{\rm hor}\sim 1/d_{\rm hor}}}. 
However, in an expanding universe,
the momentum is continually redshifting towards lower values.
Thus we must actually integrate over the cosmological history between
the time of dark-matter production and today.  This then yields the result
\beqn
   k_{\rm hor}(p) &\equiv & \xi \left[ \int_{t_{\rm prod}}^{t_{\rm now}}     { p/a(t) \over
       \sqrt{ p^2 / a(t)^2 + m^2 } }   {dt\over a(t) } \right]^{-1} \nonumber\\
    &=&   \xi \left[ \int_ {a_{\rm prod}}^1  {da\over H a^2 } \, {p \over \sqrt{p^2 + m^2 a^2 }}\right]^{-1}~. 
\label{kFSHdef}
\eeqn
where the quantity $p$ in this expression signifies the dark-matter momentum {\it today}\/. 
We emphasize that 
this expression for 
$k_{\rm hor}$ is merely an order-of-magnitude 
estimate, one which holds only up to an multiplicative 
factor of order $\calO(1)$.  
Indeed, while the horizon length within the square brackets in
Eq.~(\ref{kFSHdef}) is a precisely defined quantity, there are many different ways of
extracting a corresponding wavenumber, depending on the particular conventions adopted 
and on the particular system under study.
The quantity $\xi$ in our 
definition of $k_{\rm hor}(p)$ 
in Eq.~(\ref{kFSHdef}) is therefore inserted in order to represent this $\calO(1)$ factor.
Note that 
the integral in Eq.~(\ref{kFSHdef}) generally receives contributions from the radiation-dominated era
as well as later matter-dominated era, and these scale differently with time.   For this reason it is 
convenient to treat these separately, leading to the approximate total result
\beqn
    k_{\rm hor}(p) \!&\approx &\!  \xi \Biggl[
{p\over m (a^2 H)_{\rm MRE}} 
       \biggl\lbrace  2 
             + \tanh^{-1}\biggl( {m\over \sqrt{m^2 + p_{\rm MRE}^2}}\biggr) \nonumber\\ 
         && ~~ 
             - \tanh^{-1}\biggl( {m\over {\textstyle \sqrt{m^2 + p_{\rm prod}^2}}}\biggr) \biggr\rbrace \Biggr]^{-1}~,
\label{kFSHdef2}
\eeqn
where `MRE' indicates the time of matter-radiation equality
and where $p_X$ indicates the value of the momentum redshifted back to time $t_X$.

Any dark matter with momentum $p$ today has the potential
to suppress structure at all scales {\il{k \geq k_{\rm hor}(p)}}.
{\it In other words, for any $p$, the quantity $k_{\rm hor}(p)$ defines the
minimum value of $k$ for which the matter power spectrum $P(k)$ could be affected.}
However, our interest is not merely in situations involving a single momentum slice --- we
are interested in understanding situations involving an entire non-trivial distribution $f(p)$.
It is here that subtleties arise.

One standard approach that is commonly employed in the literature is to average over the different momenta
within the distribution,
essentially defining a free-streaming horizon scale $k_{\rm FSH}$ 
in terms of an average packet velocity $\langle v(t)\rangle$
via a relation of the form
\beq
   k_{\rm FSH} ~\sim~ \left[  
    \int_{t_{\rm prod}}^{t_{\rm now}}  {dt \over a(t)} \langle v(t)\rangle 
         \right]^{-1}~.
\label{kFSHavgd}
\eeq
However, while such an analysis is robust for  most thermal or unimodal momentum distributions in which averaged
quantities faithfully represent the full distribution, this is not generally the case for non-thermal
and/or multi-modal distributions~\cite{Konig:2016dzg}.
Indeed, there are implicit assumptions buried within the definition in Eq.~(\ref{kFSHavgd})
that may cause it to fail for more general momentum distributions
of the sort we have discussed in Sect.~\ref{sec:phase_space_evolution}.~
For example, if we have a bi-modal $f(p)$, it is not necessarily even the case that the average velocity $\langle v(t)\rangle$
represents the speed of {\it any}\/ particle that populates the distribution.

In order to understand these complications at a more fundamental level,
one can consider 
the Jeans equation which governs the time-evolution of energy-density perturbations on different length scales.
For dark matter with a given sound speed $c_s$,
the form of this equation typically allows one to read off a characteristic Jeans wavenumber $k_{\rm Jeans}$ which separates gravitationally stable (\ie, growing) and unstable (\ie, non-growing) modes.
This wavenumber  thereby signals the onset of suppression for the matter power spectrum $P(k)$.
Essentially $k_{\rm Jeans}$ indicates
the critical perturbation size 
at which the tendency towards gravitational collapse is precisely balanced
against the increased dark-matter pressure resisting collapse. 

However, this analysis assumes that the 
pressure perturbations propagate through the dark matter with a single well-defined sound speed $c_s$,
much as we would expect for dark matter of a single species.
For dark-matter particles of mass $m$ with a single momentum $p$,
a well-defined sound speed $c_s$ always exists.
However, it is easy to see that $c_s$ depends on $p$.
For example, we know that {\il{c_s=0}} for {\il{p=0}} (signifying the case of infinitely cold dark matter), 
while {\il{c_s=1/\sqrt{3}}} for {\il{p\gg m}} (signifying the case of highly-relativistic matter). 
Indeed, an explicit expression for $c_s$ as a function of $p$ is derived in Appendix~\ref{app:sound_speed}.~
Thus, dark matter with a non-trivial distribution $f(p)$ does not have a single well-defined sound speed.
From the perspective of the Jeans equation, such dark matter therefore behaves as if it were composed
of different ``species'' with different sound speeds, one for each momentum slice.

If the dark-matter phase-space distribution $f(p)$ is highly peaked around a central averaged momentum 
$\langle p\rangle$, it may of course be possible to work with a single averaged sound speed $\langle c_s\rangle$
and thereby extract a single Jeans scale,
in the same spirit as Eq.~(\ref{kFSHavgd}).
However, for more complicated distributions, this need no longer be the case.  
In general, the presence of multiple ``species'' is problematic in the presence of gravity because 
gravity feels all species simultaneously, thereby causing the 
evolution of the density perturbations for each species 
to be affected by the density perturbations for all of the other species.
In other words, the existence of dark matter across a non-trivial momentum distribution $f(p)$ leads to 
important cross-correlations that eliminate the existence of critical Jeans scales that
cleanly separate growing from non-growing (suppressed) perturbation modes.
This is true even if the dark matter exhibits only two distinct momenta.

In addition to these difficulties,
we are not merely interested in the range of $k$-values for which the matter power spectrum $P(k)$ might be affected 
---   we also wish to evaluate $P(k)$ itself.   
Though relatively straightforward, the calculation of $P(k)$ is also highly non-trivial.
One begins with certain initial conditions for the primordial perturbations arising from inflation.
One then propagates these forward in time through the linearized Einstein
equations, assuming a dark-matter component with a given phase-space distribution $f(p,t)$.
This also requires taking into account the evolution of the background cosmology.
Evolving to the present time and explicitly calculating the 
two-point correlation function, one thereby obtains the matter power spectrum $P(k)$ corresponding to a given $f(p,t)$ 
at {\il{t=t_{\rm now}}}.

In this work, we shall perform these calculations numerically using the 
\texttt{CLASS} 
software package~\cite{Lesgourgues:2011re, Blas:2011rf, Lesgourgues:2011rg, Lesgourgues:2011rh}.
However, for all the reasons discussed above,
we do not expect any simple relationship to exist between a given $f(p)$ distribution and the corresponding
matter power spectrum $P(k)$.
That said, we nevertheless wish to develop a rough, phenomenological way of understanding how the phase-space distribution $f(p)$ 
affects the matter power spectrum $P(k)$.
Even more ambitiously, we wish to have a way of ``inverting'' this mapping from $f(p)$ to $P(k)$, so
that we can use our knowledge of $P(k)$ in order to ``resurrect'' the most important features of $f(p)$.

\subsection{Defining a $k$-space dark-matter profile $\widetilde g(k)$}

Our approach to this problem is somewhat unorthodox.
First, we shall 
consider momentum slices through our dark-matter distribution packet $g(p)$, 
relating each slice of momentum $p$ to a corresponding value $k_{\rm hor}(p)$.
Indeed, we shall do this without any assumptions concerning the overall shape of $g(p)$.
Normally, as indicated above, 
$k_{\rm  hor}$ would be interpreted as defining the {\it minimum}\/ value
of $k$ within $P(k)$ which could potentially be affected by dark matter in that slice.
However, we shall instead take the defining relation for $k_{\rm hor}(p)$ in Eq.~(\ref{kFSHdef}) as
{\it defining a mapping}\/ between the $p$-variable
of $g(p)$ and the $k$-variable
of $P(k)$.  In other words, we shall identify $k_{\rm hor}$ with $k$ and thereby
consider $g(p)$ as having a corresponding profile in $k$-space:
\beq
        \widetilde g(k) ~\equiv~  g(k_{\rm hor}^{-1}(k) ) \,\left| {\cal J}(k) \right|~.
\label{tildegdef}
\eeq
Here 
$k_{\rm hor}^{-1}$ denotes the inverse function which relates $k$ back to $p$ [\ie, the inverse
of the mapping in Eq.~(\ref{kFSHdef})],
while {\il{{\cal J}(k)\equiv d\log p/d\log k}} is the Jacobian
for the change of variables from $\log p$ to $\log k$ [again as defined through the $k_{\rm hor}(p)$ function
in Eq.~(\ref{kFSHdef})].
It then follows that
\beq
     \calN(t) ~=~ \int  d\log p~ g(p) ~=~  \int d\log k ~ \widetilde g(k)~.
\eeq
Thus, just as the $p$-profile $g(p)$ describes the dark-matter distribution in $p$-space,
 the $k$-profile $\widetilde g(k)$ describes the dark-matter distribution in $k$-space.
Moreover, because $\widetilde g(k)$ lives in the same space as $P(k)$,
these two functions can even be plotted together along the same axis.
{\it We shall therefore approach the question of understanding the relationship between $g(p)$ and $P(k)$
by instead seeking to understand the relationship between $\widetilde g(k)$ and $P(k)$.} 

Strictly speaking,
our identification of $k_{\rm hor}$ with $k$ is potentially 
    subject to the same sorts of ${\cal O}(1)$ 
    ambiguities which entered into our original definition of $k_{\rm hor}$, 
    as represented by the ${\cal O}(1)$ prefactor $\xi$ in Eq.~(\ref{kFSHdef}).
    However, for our purposes in this paper, any further ${\cal O}(1)$ factor
    involved in relating $k_{\rm hor}$ to $k$ can be absorbed into an effective
     change in the value of $\xi$.  
     Thus, for simplicity, we shall consider $\xi$ to incorporate all such ${\cal O}(1)$ 
       effects.   We shall therefore identify $k_{\rm hor}$ directly with $k$,
as stated above, and proceed  to investigate the relationship between $\widetilde g(k)$ and $P(k)$. 

\subsection{Connecting $\widetilde g(k)$ to $P(k)$:  General phenomenological observations\label{sec:phenobs}}

In order to develop an intuition regarding this relationship, 
we shall examine the response of $P(k)$ to a series of idealized forms for $g(p)$ and $\widetilde g(k)$.
For concreteness and simplicity,
each of these idealized forms for $g(p)$ will be taken to consist of one or more log-normal distributions of
the form
\beq
 g(p)  	~=~ \frac{A}{\sqrt{2\pi}\,\sigma}
   \exp\Bigg\{-\frac{1}{2\sigma^2}\bigg[\log\left(\frac{p}{\expt{p}}\right) + \frac{1}{2}\sigma^2\bigg]^2\Bigg\}~,
\eeq
where $\sigma$ and $\langle p\rangle$ are respectively the width and average momentum of the distribution
and where $A$ is an overall normalization (or dark-matter abundance).
These distribution functions are essentially Gaussians in $(\log p)$-space.
Indeed, with this form, we can easily verify that
{\il{\int g(p)\, d\log p = A}} and {\il{\int p\, g(p)\, d\log p = \langle p \rangle}}, as claimed. 
Thus either $\sigma$ or $\langle p \rangle$ may be altered independently.
However, it is important to note that 
{\il{\langle \log p\rangle \equiv \int \log p \, g(p)\, d\log p = \log\langle p\rangle - \sigma^2/2}}.
In other words, this Gaussian is centered (with maximum height) 
at $\log \langle p \rangle  - \sigma^2/2$ in $(\log\, p)$-space,
and thus any change in the value of $\sigma$ 
will cause the location of the maximum
height to shift even though $\langle p \rangle$ is kept fixed.

\begin{figure}[t]
\centering
\includegraphics[keepaspectratio, width=.48\textwidth]{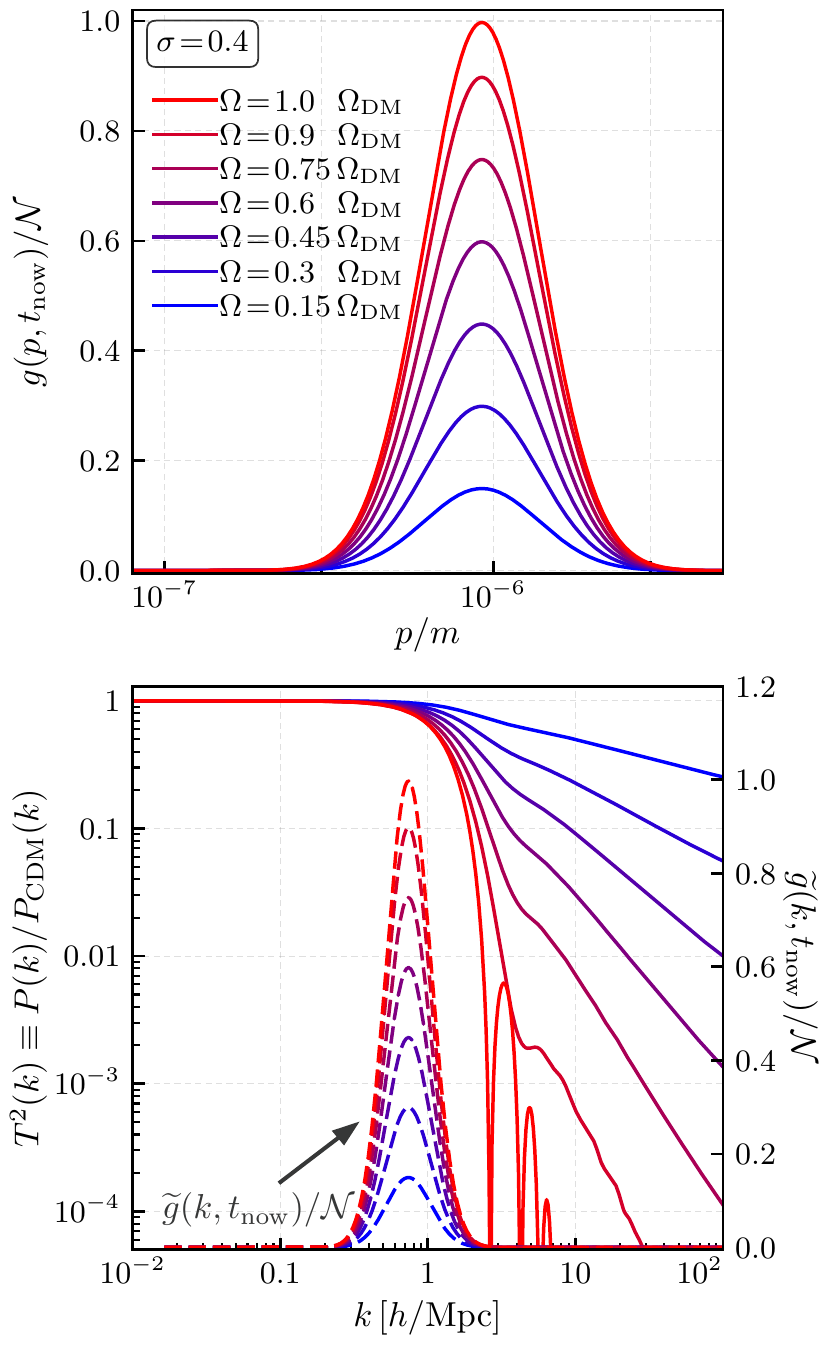}
\caption{Dark-matter distributions $g(p)$ which differ
only in their overall normalizations (\ie, their associated abundances $\Omega$), and their corresponding
transfer functions {\il{T^2(k)\equiv P(k)/P_{\rm CDM}(k)}}. 
{\it Upper panel:}\/
A variety of dark-matter distributions $g(p)$ which share a common width $\sigma$
and average momentum $\langle p\rangle$.  
In each case we imagine the existence of an additional cold dark-matter component 
(not shown) 
such that the total dark-matter abundance is held fixed at $\Omega_{\rm DM}$.
{\it Lower panel:}\/
The corresponding $k$-space dark-matter distributions $\widetilde g(k)$, calculated for {\il{\xi=5/3}},
 and the corresponding transfer functions $T^2(k)$, calculated via  
the \texttt{CLASS} software package~{\mbox{\cite{Lesgourgues:2011re,Blas:2011rf,Lesgourgues:2011rg,Lesgourgues:2011rh}}}.
We see that the greater the total dark-matter abundance associated with $g(p)$, 
the stronger
the {\it suppression}\/ for $T^2(k)$ 
and the steeper the slope $d\log T^2(k)/d\log k$ at large values of $k$. }  
\label{fig:vary_abundance}
\end{figure}

Let us consider the case in which $g(p)$ consists of a single Gaussian peak in $(\log p)$-space, and let us
study the effects of changing the overall dark-matter abundance $A$ associated with this peak, keeping $\langle p\rangle$
and $\sigma$ fixed.
Such distributions are shown in the upper panel of Fig.~\ref{fig:vary_abundance},
where we have chosen values of $A$ such that the total abundance associated with this peak in each case
is given by
{\il{\Omega= r \Omega_{\rm DM}}} where $\Omega_{\rm DM}$ is the total dark-matter abundance
and where the ratio $r$ ranges over the set 
of values specified within Fig.~\ref{fig:vary_abundance}.~   For each value of {\il{r<1}}, we are of course
implicitly assuming the existence of an additional infinitely cold 
dark-matter component (not shown) with abundance {\il{\Omega_{\rm cold} = (1-r) \Omega_{\rm DM}}} 
such that the total dark-matter abundance remains fixed
at $\Omega_{\rm DM}$.
For each value of $r$,
the corresponding $k$-space distributions $\widetilde g(k)$ and 
transfer functions {\il{T^2(k)\equiv P(k)/P_{\rm CDM}(k)}} are shown together in the lower panel
of Fig.~\ref{fig:vary_abundance}.~ 
For concreteness in defining $\widetilde g(k)$, we have taken
{\il{\xi=5/3}} within Eq.~(\ref{kFSHdef}), as this value tends to horizontally align 
the peaks $\widetilde g(k)$ with the onset of the suppression of structure within $T^2(k)$,
as shown in Fig.~\ref{fig:vary_abundance}.~
Indeed, we will find that taking {\il{\xi\approx 5/3}} has a similar alignment effect for every case
to be considered in this paper, not only here but also for the highly non-trivial cases to be examined
in Sect.~\ref{sec:toy_model}.~
We shall therefore adopt {\il{\xi=5/3}} as a universal reference value for the rest of this paper.

For values of $k$ below the location of $\widetilde g(k)$,
we see that the transfer function remains close to $1$, indicating that the development of 
structure at these scales is insensitive to the fact that a fraction $r$ 
of the dark matter is no longer being treated as infinitely cold.
In other words, for the large length scales corresponding to such values of $k$, 
even the dark matter carrying the non-zero 
momenta within $g(p)$ is effectively cold.
However, as $k$ increases and approaches the scales associated with non-zero $\widetilde g(k)$, our transfer
function begins to peel away from unity and fall rather dramatically.
This signifies the onset of the suppression of structure at these scales, as anticipated. 
In the case with the greatest suppression,
we see from Fig.~\ref{fig:vary_abundance} that the curve for $T^2(k)$  
actually begins to exhibit sharp wiggles as a function of $k$.
These are ultimately the effects of dark-matter acoustic oscillations;
such effects are generally subleading, appearing within the transfer function
only when the formation of structure is significantly suppressed,
and they shall not be our focus in this paper.   We shall therefore disregard such wiggles in our subsequent discussion.
However, we also note  
the degree of suppression is greater when $\widetilde g(k)$ carries
greater abundance.
This immediately tells us, as expected, that there is a correlation
between the abundance associated with 
$\widetilde g(k)$ and the degree of structure suppression induced
at larger values of $k$.
Likewise, we also note from Fig.~\ref{fig:vary_abundance} 
that as $k$ passes beyond the range with non-zero $\widetilde g(k)$,
the evolution of our transfer function $T^2(k)$ on this log-log plot
in each case ultimately seems to 
develop a negative slope 
which remains essentially constant for the values of $k$ shown.
This slope also appears to be correlated with the abundance
associated with $\widetilde g(k)$, with steeper (more negative) slopes corresponding
to larger abundances.
Thus, at this stage, we conclude that a greater abundance for $\widetilde g(k)$
appears to correspond not only to a stronger {\it suppression}\/ for $T^2(k)$ at larger values of $k$,
but also to a steeper {\it slope}\/ for $T^2(k)$.

\begin{figure}[t]
\includegraphics[keepaspectratio, width=.48\textwidth]{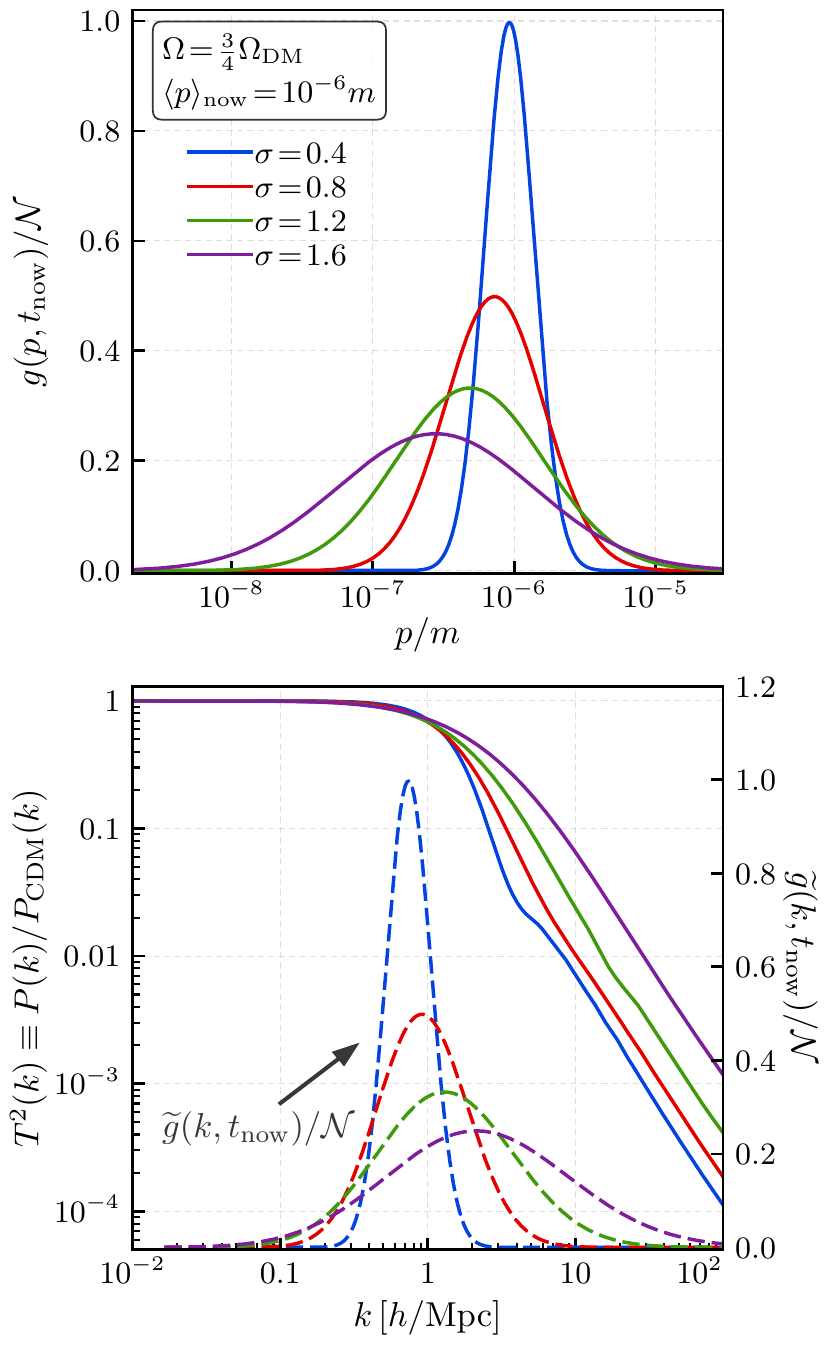}
\caption{Similar to Fig.~\ref{fig:vary_abundance}, but for dark-matter distributions in which only the width $\sigma$ is varied 
while the total abundance and average momentum $\langle p\rangle$ are held fixed.
We see that varying the width
affects the net suppression of the transfer function, {\it but 
does not affect its eventual slope $d\log T^2/d\log k$ at large $k$ beyond $\widetilde g(k)$}.
As discussed in the text, this suggests that
the accumulated abundance correlates not with the 
net suppression of the transfer function, but with its logarithmic slope.}
\label{fig:vary_width}
\end{figure}

In order to disentangle these two effects, we now consider
the case of a single peak where we now vary the {\it width}\/ $\sigma$ of the peak,  
holding $A$ and $\langle p\rangle$ constant.
This situation is shown in Fig.~\ref{fig:vary_width}.~
First, we confirm from the upper panel of Fig.~\ref{fig:vary_width} that holding $\langle p\rangle$ fixed
and increasing $\sigma$ indeed causes the maximum of the peak to shift towards smaller values of $\log p$,
as discussed above.
Second, as we progress from 
smaller to larger values of $k$ within
the lower panel 
of Fig.~\ref{fig:vary_width},
we observe that increasing the width
$\sigma$ of the dark-matter distribution
induces a more gradual suppression of $T^2(k)$ as a function of $k$,
ultimately resulting in less net suppression at large
values of $k$.

 {\it Even more importantly, however, we also observe
that increasing the width of $\widetilde g(k)$ appears to have no effect on the logarithmic slope of the transfer 
function $d\log T^2/d\log k$ at large $k$ beyond $\widetilde g(k)$.}
Indeed, the logarithmic slope of the transfer function
appears to be {\it fixed}\/, even though the overall suppression   
clearly varies with the width of $g(p)$.
Of course, even though we are varying the width of $g(p)$ 
in Fig.~\ref{fig:vary_width}, 
we are holding the overall {\it abundance}\/ associated with $g(p)$ fixed.
This then suggests that the total dark-matter abundance associated
with $g(p)$ correlates with the eventual logarithmic {\it slope}\/ of the transfer function
rather than with the net amount by which $T^2(k)$ is suppressed.

This behavior actually holds throughout the range of $k$-values plotted in
Fig.~\ref{fig:vary_width}, and not merely at large $k$.
Indeed, as we sweep a reference value $k^\ast$ from left to right in $k$-space
and pass through the $\widetilde g(k)$ distribution,
we accumulate an increasing abundance of dark matter for which {\il{k< k^\ast}}.
Likewise, as we sweep from left to right in $k$-space, we see that the logarithmic slope of the transfer 
function at {\il{k=k^\ast}} simultaneously becomes increasingly steep.
Indeed, it is only for the values of $k$ 
shown in Fig.~\ref{fig:vary_width} 
beyond $\widetilde g(k)$ 
that we stop accumulating abundance
and the logarithmic slope of the transfer function becomes effectively constant.
This observation thus allows us to refine our previous conclusions from
Fig.~\ref{fig:vary_abundance} and suggests that {\it the accumulated abundance correlates not with the 
net suppression of the transfer function, but with its logarithmic slope 
$d\log T^2/d\log k$.}

\begin{figure}[t]
\includegraphics[keepaspectratio, width=.48\textwidth]{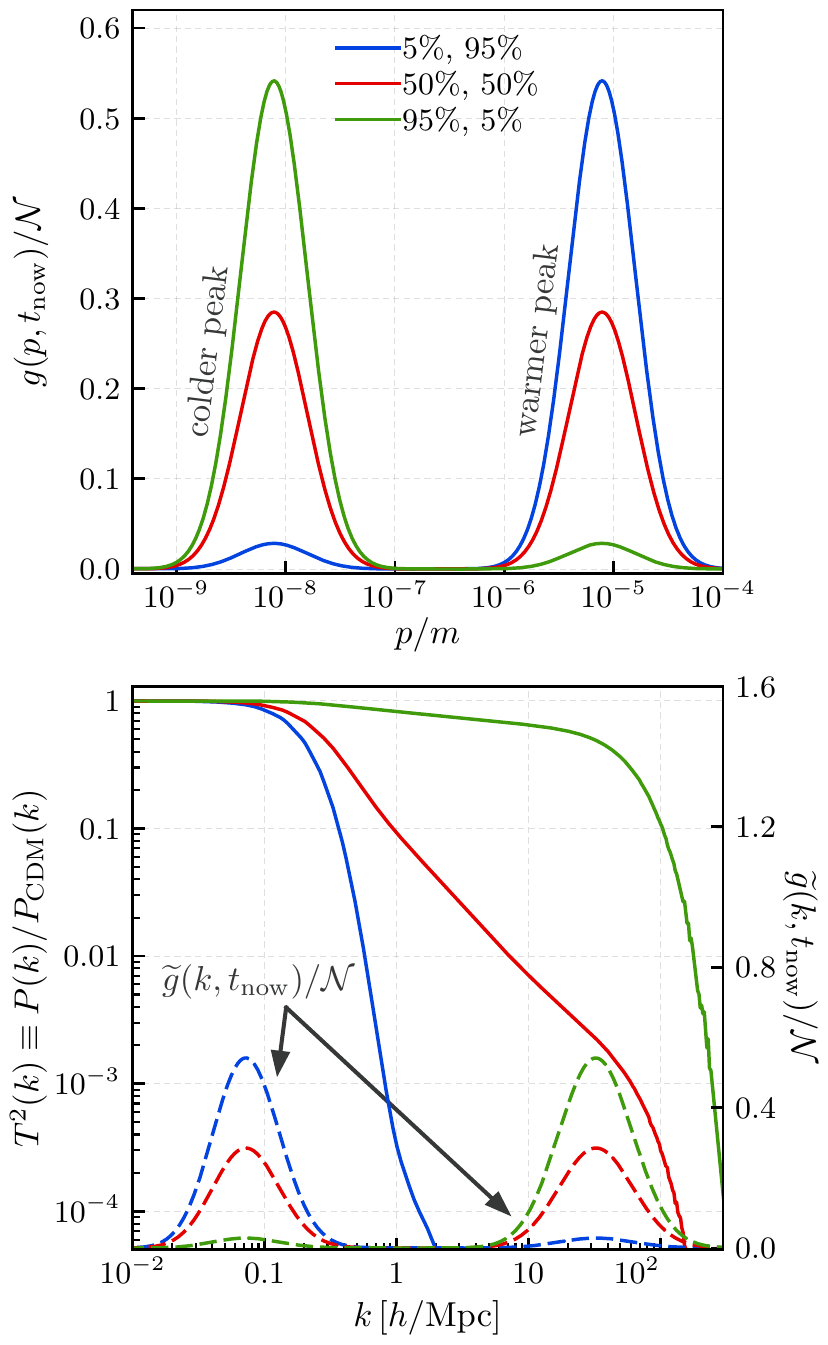}
\caption{Similar to Figs.~\ref{fig:vary_abundance} and \ref{fig:vary_width}, but for 
dark-matter distributions $g(p)$ consisting of two log-Gaussian peaks 
whose widths and average momenta are held fixed but  
amongst which the apportionment of the   total
dark-matter abundance $\Omega_{\rm DM}$ is varied.  
As we sweep from smaller to larger values of $k$ in each case, 
we see that $d\log T^2/d\log k$ 
becomes increasingly steep as our accumulated
abundance increases --- precisely as anticipated from 
the results of Fig.~\ref{fig:vary_width}.}
\label{fig:twin_peaks}
\end{figure}

To see if this behavior survives for more complex $g(p)$ distributions,
in Fig.~\ref{fig:twin_peaks} we consider the case of two disjoint peaks,
one corresponding to colder dark matter with smaller $\langle p\rangle$ and
the other to warmer dark matter with larger $\langle p\rangle$.
(Note that we refer to these peaks as colder and warmer even though they need not be thermal.)
In each case we hold the width and average momentum 
$\langle p\rangle$ of each peak fixed,
but we vary the manner in which the total dark-matter abundance $\Omega_{\rm DM}$ 
is apportioned between them.
We nevertheless find that the above behavior persists even in such cases.
Indeed, as we sweep from smaller to larger values of $k$ in the regions {\it within}\/ the peaks,
we see that the logarithmic slope of the transfer
function becomes increasingly steep as our accumulated
abundance increases --- indeed, as the
relative abundance of the warmer peak increases,
the slope of the transfer function between the peaks
also increases.
However, as we sweep from smaller to larger values of $k$ in the region {\it between}\/ the peaks,
we are no longer accumulating abundance.   Likewise, the logarithmic slope of the   
transfer function in this region is approximately constant.
Thus, we continue to find that accumulated abundance remains correlated
with the logarithmic slope $d\log T^2/d\log k$ of the transfer function.

\subsection{From $P(k)$ back to $f(p)$:  The hot fraction function $F(k)$ and a reconstruction conjecture} 

We could, of course, continue to examine further cases with more 
non-trivial $g(p)$ distributions.  However, at this stage, we are actually able to formulate a conjecture
which will incorporate all of our observations and which we believe holds quite generally.
Moreover, as we shall see, this conjecture will not only allow us to correlate features of the dark-matter distribution
$g(p)$ with those of the transfer function $T^2(k)$, but also allow us to 
invert the process and resurrect the salient features of $g(p)$ directly from $T^2(k)$.

In order to phrase our conjecture mathematically,
we first note that 
at any value of $k$, the total abundance of dark matter accumulated from even smaller $k$-values 
is given by 
\beq
  F(k) ~\equiv~ \frac
       {   \int_{-\infty}^{\log k}   ~ \widetilde g(k') ~  d\log k'}
       {   \int_{-\infty}^{+\infty}   ~ \widetilde g(k') ~d\log k'}~.
\label{Fracdef2}
\eeq
Note that this can be equivalently written in terms of $g(p)$ rather than $\widetilde g(k)$ as
\beq
  F(k) ~\equiv~ \frac
       {   \int_{\log p(k)}^\infty \, g(p')\, d\log p'}
       {   \int_{-\infty}^\infty \, g(p')\, d\log p' }~
\label{Fracdef}
\eeq
where the $p(k)$ function within the lower limit of the integral in the numerator
is the inverse of the $k_{\rm hor}(p)$ function in Eq.~(\ref{kFSHdef}).
Thus, for any value of $k$, we see that 
$F(k)$ may be interpreted physically as that fraction of the 
dark matter which may effectively be considered 
as free-streaming (or ``hot'') relative to the 
corresponding value of 
{\il{p= k_{\rm hor}^{-1}(k)}}.
We shall therefore refer to $F(k)$ as the {\it hot fraction function}\/.
Note that 
the denominators in Eqs.~(\ref{Fracdef2}) and (\ref{Fracdef}) are nothing but the overall normalizations
$\calN$ associated with the phase-space distribution, as defined in Eq.~(\ref{Ndef}).

Our conjecture, then, is 
that there is a direct approximate relationship between the hot fraction function $F(k)$ and the 
logarithmic slope $d\log T^2 / d\log k$ of the transfer function.
We shall write this in terms of an as-yet unknown function $\eta$:
\beq
            F(k) ~\approx~ \eta\left(   \left|  {d\log T^2 \over d\log k} \right| \right)~.
\label{conj1}
\eeq
Equivalently, taking the $(\log k$)-derivative of both sides, we conjecture that
\beq
             {\widetilde g(k)\over \calN} ~\approx~  \eta'\left(   \left|  {d\log T^2 \over d\log k} \right| \right) \,
         \left| {d^2 \log T^2\over (d\log k)^2} \right|~,~
\label{conj2}
\eeq
where the prime on the $\eta$-function indicates a derivative with respect to its argument.
Thus, if the first derivative of $\log T^2$ with respect to $\log k$
is related to the hot fraction function $F(k)$,
then it is a combination of the first and second derivatives that is
related to the phase-space distribution $\widetilde g(k)$.

This conjecture is quite significant, correlating 
the hot fraction function $F(k)$ not with the transfer function $T^2(k)$ but with its slope.
However, we can even push this one step further.
In Fig.~\ref{fig:vary_abundance}, we observed
that the slopes $d \log T^2/d\log k$ that emerge at the 
large values of $k$ shown in this plot  
are correlated with the abundances $\Omega$ associated
with the distributions $\widetilde g(k)$ shown.
However, within these large-$k$ regions beyond $\widetilde g(k)$,
the accumulated hot fraction functions $F(k)$ are nothing but
the fractions {\il{r\equiv \Omega/\Omega_{\rm DM}}} 
that relate these abundances $\Omega$
to the total dark-matter abundance $\Omega_{\rm DM}$.
Thus, given these ratios and the corresponding logarithmic slopes within Fig.~\ref{fig:vary_abundance},
we can investigate whether there might exist
a simple empirical relation between these two quantities, and
indeed we find that the relation
\beq
            \left| {d\log T^2\over d\log k} \right| ~\approx~  [F(k)]^2 + {3\over 2} F(k)~
\label{eq:FitFImplicit}
\eeq
holds to rather high precision.
This then fixes our unknown function $\eta$ in Eq.~(\ref{conj1}), whereupon Eq.~(\ref{conj2}) takes the form
\beq
     {\widetilde g(k) \over \calN}  ~\approx~ {1\over 2} \, 
           \left( {9\over 16} + \left| {d\log T^2\over d\log k}\right|\right)^{-1/2}
          \left|  {d^2 \log T^2 \over (d\log k)^2} \right| ~.~~~
\label{conj3}
\eeq
This, then, is the final form of our conjecture.
Indeed, if accurate, this relationship would allow us to ``resurrect'' 
$\widetilde g(k)$ from the transfer function $T^2(k)$
and thereby deduce the dark-matter distribution $g(p)$ that
produced it.

In Sect.~\ref{sec:toy_model}, starting from an explicit Lagrangian describing a complex non-minimal dark sector,
we shall explicitly test the observations we have developed thus far in this paper,
including our reconstruction conjecture in Eq.~(\ref{conj3}).
{\it As we shall find, in each case our conjecture (\ref{conj3}) is indeed remarkably successful in reproducing the salient
features of the dark-matter distribution, even when this distribution highly non-trivial and/or multi-modal.}
Our conjecture can therefore serve as a powerful tool in the archaeology toolbox,
enabling a reconstruction of many features of the early universe
starting from a present-day observable such as the matter power spectrum.

Two important caveats must be borne in mind regarding our conjecture in Eq.~(\ref{conj3}).
First, we emphasize that our conjecture is not meant to be a precise mathematical statement.
Indeed, given the rather complicated nature of the Einstein evolution equations which connect 
$g(p)$ to $T^2(k)$, we do not expect  a relation of the simple form in Eq.~(\ref{conj3}) to provide a precise
archaeological inverse (except perhaps under some 
limiting approximations and simplifications).
Rather, this conjecture is intended merely as an approximate practical guide --- 
a way of reproducing the rough characteristics of $g(p)$ given a particular transfer function $T^2(k)$.

But second --- and perhaps more importantly ---
we stress that our conjecture implicitly assumes/requires
that $\log T^2$ has a negative-semidefinite second derivative with 
respect to $\log k$, so 
that it either has a constant slope or is concave-down when
plotted versus $\log k$.
Generally, this tends to occur in situations in which
our dark-matter momentum distributions --- no matter how 
complex in shape --- are relatively ``clustered'' in $k$-space. 
By contrast, if there are widely separated clusters in the dark-matter
phase-space distribution, the transfer function can
cross an inflection point and become concave-up, potentially
even leading to a plateau~\cite{Boyarsky:2008xj}.
Indeed, this is precisely what would happen if we were to plot, {\it e.g.}\/, the transfer functions in Fig.~\ref{fig:vary_abundance} out to even larger values of $k$. 
In such cases, our conjecture is expected to hold only within
each cluster individually (just as it indeed holds for the ``clustered'' region 
of Fig.~\ref{fig:vary_abundance} shown).
As we shall see in Sect.~\ref{sec:toy_model}, this restriction to clusters is
not severe, and still allows us to resurrect the salient features
of the dark-matter phase-space distribution $g(p)$ 
for a wide variety of dynamical histories.

Finally, our conjecture in Eq.~(\ref{conj3}) also allows us to understand the limitations of certain proposed functional forms
for $T^2(k)$ which have appeared in the literature. 
For example, in Ref.~\cite{Murgia:2017lwo} it was shown that the functional form~\cite{Barkana:2001gr, Viel:2011bk, Destri:2013hha}
\beq
T(k) ~=~ \left[ 1+ (\alpha k)^\beta \right]^\gamma ~~~  {\rm with}~~ \alpha,\beta>0,~\gamma <0~ 
\label{Merle}
\eeq
is remarkably successful in fitting the results from a relatively large number of underlying models of early-universe dynamics.
(Note that more complex functional forms are also discussed in Ref.~\cite{Murgia:2017lwo}.)  
However, using our conjecture in Eq.~(\ref{conj3}), we can reconstruct the corresponding $\widetilde g(k)$, obtaining 
\beq
      \widetilde g(k) ~=~   {A x\over (1+x)^{3/2} \sqrt{ 9 + Bx}}~,
\eeq
where \il{x\equiv (\alpha k)^\beta}, \il{A\equiv  4\beta^2 |\gamma|>0}, and $B\equiv  9 + 32 \beta |\gamma| >9$.
As expected, this functional form corresponds to a localized packet, with {\il{\widetilde g(k)\sim A k/3}} as {\il{k\to 0}} and {\il{\widetilde g(k)\sim A/(\sqrt{B} k)}} as {\il{k\to \infty}}.  Moreover, it is also straightforward to verify that this packet is always unimodal, with a unique maximum located at
\beq
    x_{\rm max} ~=~ { B-9 + \sqrt{(B-9)^2 + 144 B}\over 4B}~.
\eeq
Thus we learn that the generic functional form in Eq.~(\ref{Merle}), although quite flexible, is ultimately limited to dark-matter phase-space distributions which are unimodal.
Indeed, as we have seen (and as we shall explicitly verify in Sect.~\ref{sec:toy_model}),
this is only a small part of what is possible.

\section{Deciphering the archaeological record:  An 
        illustrative end-to-end example }\label{sec:toy_model}

In Sect.~\ref{sec:phase_space_evolution}, we examined the manner in which decays within 
the dark sector can give rise to a non-trivial and often multi-modal phase-space 
distribution $f(p)$ for the lightest particle within this sector, \ie, for the particle which constitutes
the dark matter at the present time.
Likewise, in Sect.~\ref{sec:perturb}, we studied the 
relationship between a given phase-space distribution $f(p)$ and the corresponding linear matter power spectrum $P(k)$,
ultimately proposing a conjecture in Eq.~(\ref{conj3}) which would allow us to approximately reconstruct $f(p)$ from 
knowledge of $P(k)$.
However, our analysis thus far has been primarily conceptual and 
limited to examples involving only a small number of dark-sector states. 
Likewise, we have only examined each step within Eq.~(\ref{onepointone})
independently, without performing a complete end-to-end analysis.

In this section, we shall examine how these ideas play out within a more complex framework
--- one involving a relatively large number of dark-sector states and decay pathways. 
In this way, we seek to determine the extent to which our general observations from 
Sects.~\ref{sec:phase_space_evolution} and~\ref{sec:perturb} are robust, 
remaining more or less intact as the dark sector grows in complexity,
even under a full numerical Boltzmann analysis.
We also wish to see how the different steps of the process within Eq.~(\ref{onepointone}) actually
connect to each other, starting with a particular model of early-universe dynamics and 
ending with a predicted matter power spectrum $P(k)$.

Towards this end, in this section we shall construct a phenomenologically rich illustrative model 
within which our calculations will take place.
In particular, we shall begin with an explicit Lagrangian describing a relatively large number of dark-sector states.
From this we shall then proceed to study the various decay amplitudes, the patterns of possible allowed decay chains,
the resulting dark-matter phase-space distributions $f(p)$, and 
the corresponding matter power spectra $P(k)$.   
This will ultimately enable us to determine the extent to which the phenomena
discussed in Sects.~\ref{sec:phase_space_evolution} and~\ref{sec:perturb} are 
realized within a relatively complex dark sector.
This will also afford us the opportunity of explicitly testing our reconstruction conjecture in Eq.~(\ref{conj3}).
We emphasize, however, that the illustrative model we shall consider
is not meant to be a UV-complete description of an actual
fully-realized dark sector.   Rather, our purpose here is merely to establish a framework in which to study the rich set of
possible phenomenologies 
associated with intra-ensemble decays and to verify that the basic expectations we have discussed in
Sects.~\ref{sec:phase_space_evolution} and \ref{sec:perturb}
are indeed realized, including the potential for archaeological reconstruction.

\subsection{The model}

Our model consist  of an ensemble of $N+1$ real scalar fields
$\phi_\ell$, where the index {\il{\ell=0,1,\ldots, N}} labels these fields in order of increasing mass.    
We take these scalars to be singlets under the Standard-Model (SM) gauge group and assume that they couple
only negligibly to the fields of the visible sector.
Likewise, we shall assume that the behavior of these fields is governed by
the Lagrangian 
\beq
  \mathcal{L} ~=~ \sum_{\ell = 0}^{N} 
    \bigg(\frac{1}{2}\partial_\mu\phi_{\ell}\partial^\mu\phi_{\ell}
    - \frac{1}{2}m_{\ell}^2 \phi_{\ell}^2 \bigg) 
    + \mathcal{L}_{\rm int}~,
  \label{eq:FieldLagrangian}
\eeq
where $m_{\ell}$ denotes the mass of $\phi_\ell$ and where the interaction Lagrangian 
$\mathcal{L}_{\rm int}$ consists of terms involving the fields $\phi_\ell$ alone.
For simplicity, we shall consider a form for $\mathcal{L}_{\rm int}$ in which two-body
intra-ensemble decays of the form {\il{\phi_\ell \to \phi_i \phi_j}} dominate the decay width 
of each unstable $\phi_\ell$.  In particular, we consider an interaction Lagrangian which 
includes trilinear terms of the form 
\beq
  \mathcal{L}_{\rm int} ~\ni~ 
    -\sum_{\ell = 0}^{N}\sum_{i = 0}^{\ell}\sum_{j = 0}^{i}c_{\ell i j} 
    \phi_{\ell}\phi_i\phi_j~,
  \label{eq:InteractionLagrangian}
\eeq 
where the $c_{\ell i j}$ are coupling coefficients with dimensions of mass.  
We emphasize that in addition to the trilinear terms in Eq.~(\ref{eq:InteractionLagrangian}), 
$\mathcal{L}_{\rm int}$ will in general include additional quartic terms involving the
$\phi_\ell$.  Such terms play an important role in stabilizing the scalar potential;  moreover, 
they may also play a role in establishing the primordial abundances of these fields at early
times.  However, due to phase-space considerations, they typically play only a subleading role 
in the decay phenomenology of the $\phi_\ell$ at times well after those abundances have been
established.  In what follows, we shall therefore assume that the coupling coefficients associated with
such terms are sufficiently small that these terms have negligible impact on particle dynamics within 
the dark sector at late times,  and focus on the consequences of the trilinear terms in 
Eq.~(\ref{eq:InteractionLagrangian}).     

The different masses $m_\ell$ and couplings $c_{\ell ij}$ associated with our individual ensemble constituents 
will not be considered to be independent parameters of our model.  Rather,
both the masses and couplings 
will be assumed to scale across the ensemble according to a set of 
scaling relations involving only a small number of free parameters.  In particular, we shall assume that the $m_\ell$ 
scale across the ensemble according to the scaling relation
\beq
  m_\ell ~=~ m_0 + \ell^\delta\Delta m~,
  \label{eq:MassSpectrum_intra}
\eeq
where $m_0$ denotes the mass of the lightest ensemble constituent, where $\Delta m$ is a parameter 
with dimensions of mass, and where $\delta$ is a dimensionless scaling exponent which controls
how the density of states scales across the ensemble.

A variety of scaling relations are in principle possible for the coupling coefficients 
$c_{\ell i j}$.  Given that our primary aim in this paper is to investigate the extent to which 
information about the decay history of the dark sector can be gleaned from the phase-space
distribution of a single, stable particle species which plays the role of the dark matter
today, we shall adopt a scaling relation for the $c_{\ell i j}$ which is amenable to such a study.  
First, we shall assume that the coupling structure of our model is such that only 
the lightest ensemble constituent $\phi_0$ is stable and that a contribution to the decay
width of each constituent $\phi_\ell$ with {\il{\ell > 0}} arises 
from the interaction Lagrangian in Eq.~(\ref{eq:InteractionLagrangian}).  
Moreover, we also wish to examine how the interplay between the coupling structure of the 
dark sector, the decay kinematics within the ensemble, and the cosmological background
ultimately gives rise to the phase-space distribution $g_0(p,\tnow)$ for this lightest ensemble 
constituent at present time.  

There are two salient kinematic quantities which are of particular interest in 
characterizing two-body decays of the form {\il{\phi_\ell \to \phi_i\phi_j}}.  
The first of these quantities is the total energy released in the decay --- \ie, the 
difference between the mass of the decaying particle and the sum of the masses 
of the two daughter particles.  The second is the difference in mass between the two daughter 
particles.  
Motivated by these considerations, we shall therefore adopt a parametrization for the 
$c_{\ell i j}$ in which
\begin{eqnarray}
  c_{\ell i j} &=& \mu R_{\ell i j} 
    \left(\frac{m_{\ell}-m_i-m_j}{\Delta m}\right)^r    
    \left(1+\frac{\left| m_i-m_j\right|}{\Delta m}\right)^{-s} \nonumber \\   
    & & \times~\Theta(m_\ell - m_i - m_j)~.
\label{eq:clij}
\end{eqnarray}
Here $\mu$ is a parameter with dimensions of mass which sets the overall scale of the 
couplings, while $\Theta(m_{\ell}-m_i-m_j)$ is a Heaviside theta function,
$r$ and $s$ are dimensionless free parameters, and the combinatoric factor 
\beq
  R_{\ell i j} ~\equiv~ \begin{cases}
    6 & \mbox{all indices different} \\
    3 & \mbox{only two indices equal} \\
    1 & \mbox{all indices equal}~
  \end{cases}
  \label{eq:Rnpq}
\eeq
is defined such that
\beq
         \sum_{\ell=0}^N \sum_{i=0}^\ell \sum_{j=0}^i \,R_{\ell i j}\, \phi_\ell \phi_i \phi_j ~=~
         \sum_{m,n,p=0}^N \phi_m \phi_n\phi_p~.
\eeq
Note that even though {\il{i\geq j}} in Eq.~(\ref{eq:InteractionLagrangian}),
the absolute-value signs in Eq.~(\ref{eq:clij}) ensure that
{\il{c_{\ell ij}=c_{\ell ji}}}.  This property will be useful later.

The parameters $r$ and $s$ appearing in Eq.~(\ref{eq:clij}) have a straightforward
interpretation and will be critical for our analysis.  The parameter $r$ governs the manner in which 
$c_{\ell i j}$ scales with the overall kinetic energy released during the decay process
{\il{\phi_\ell \to \phi_i\phi_j}}.  Taking {\il{r > 0}} establishes a preference for highly exothermic decays involving 
a large conversion of mass energy into kinetic energy --- in other words, decays from heavy parents directly into 
relatively light daughters which therefore behave more like radiation.    By contrast, 
taking {\il{r < 0}} establishes a preference for  minimally exothermic decays in which relatively little kinetic energy is released 
and the daughter particles behave more like matter.   Likewise,      
the parameter $s$ governs the extent to which the daughter particles are close in mass to each other.
Taking {\il{s > 0}} establishes a preference for decays in which there is
a high degree of symmetry between the masses of the daughter particles.
By contrast, taking {\il{s < 0}} disfavors such decays relative to those in which such a symmetry
between the daughters is significantly broken.
Thus, while positive $r$ favors decays producing radiation,
positive $s$ favors decays whose daughters are symmetric.

Examining the phenomenology that results from different values of $r$ and $s$ therefore allows us to 
survey many different kinds of decay chains and 
their corresponding dark-matter phase-space distributions.
Of course, a fully realistic model of the dark sector is unlikely to exhibit a coupling structure 
of the specific form in Eq.~(\ref{eq:clij}).
However, as we shall see, this structure is capable of realizing many if not most of the different
decay phenomenologies that could emerge within a fully realistic multi-component dark sector,
and our goal in this section is to study the implications of these different decay phenomenologies
rather than their specific realization within a UV-complete theory.
Adopting the coupling structure in Eq.~(\ref{eq:clij}) will therefore be sufficient for our purposes.

Given the number of free parameters which govern our model and the variety of 
possible initial conditions for the $g_\ell(p,t_I)$, it will prove useful for us to adopt a few 
benchmark assumptions as we proceed.
In particular, for concreteness, we shall focus on the case of an ensemble 
with {\il{N = 9}} --- \ie, an ensemble comprising ten constituent particles with ten distinct masses.  Such a value
of $N$ is sufficiently large that a highly non-trivial pattern of intra-ensemble decays
can arise, yet sufficiently small that the evolution of the Boltzmann system is 
not computationally onerous.  We shall also fix {\il{\delta = 1}}, {\il{\Delta m = 2m_0}}, and 
{\il{\mu = m_0/10}} in what follows.  Given these benchmark assumptions, there remains one 
free parameter $m_0$ on which the mass spectrum of our model depends, as well as two 
free parameters $r$ and $s$  which govern the coupling structure of the model.    

For simplicity, we shall consider the case in which only the most massive constituent 
$\phi_9$ in the ensemble is initially populated at $t_I$, while the energy density
in all other ensemble constituents is negligible.  For sake of generality, we shall 
remain largely agnostic about the precise nature of the mechanism which establishes the
initial energy density $\rho_9(t_I)$ for this field.  Rather, we shall simply assume that this 
occurs during a production-time window $\Delta t_P$ and 
effectively ceases acting by some cosmological time {\il{t_P \leq t_I}}.  Note that in order to ensure that 
the change in $\rho_9(t)$ due to decays is not significant during the production-time window,
it is sufficient to require that     
\beq
\begin{cases}
  &  t_I - t_{P_\ell} \ll \tau_9 \\
  &  \Delta t_{P_\ell} \ll \tau_9~,
\end{cases}
\label{eq:initconds}
\eeq
where {\il{\tau_9 \equiv 1/\Gamma_9}} is the lifetime of $\phi_9$.  
Indeed, the conditions in Eq.~(\ref{eq:initconds})
are equivalent to the condition already mentioned in Eq.~(\ref{condition}) 
which ensures effectively continuous daughter packets, 
even for extremely non-relativistic parents.
We emphasize that a 
production mechanism for which {\il{\tau_9 \leq t_I}} is not in conflict with these 
conditions, provided that $\Delta t_P$ is small and provided that the time $t_I$ 
at which we begin our numerical Boltzmann evolution is sufficiently close to $t_P$. 

Depending on the nature of the production mechanism through which $\rho_9(t_I)$ is 
established, the initial phase-space distribution $g_9(p,t_I)$ for $\phi_9$ can take
a variety of forms.  In what follows, we shall assume that the
initial population of $\phi_9$ is sufficiently cold that the detailed shape of $g_9(p,t_I)$ 
has essentially no impact on the subsequent evolution of the phase-space distributions of 
all particles produced through the decay chains initiated by $\phi_9$ decay.  Indeed, 
in the regime in which $g_9(p,t)$ effectively only receives support at very small $p$, 
the phase-space packet for each daughter particle produced directly by the decay process 
{\il{\phi_9 \rightarrow \phi_i \phi_j}} will be sharply peaked around {\il{p \approx \half( m_9 - m_i - m_j)}}.
Thus, within this regime, the profiles of the phase-space packets generated for the two 
daughter particles $\phi_i$ and $\phi_j$ are not particularly sensitive to the shape
of $g_9(p,t_I)$, nor are the profiles of the packets for the particles produced via
subsequent decays.  
Indeed, we have verified numerically 
that varying the shape of $g_9(p,t_I)$ while holding the initial number density $n_9(t_I)$ 
fixed does not significantly impact our results for either $g_0(p,\tnow)$
or $P(k)$, provided that the condition {\il{f_9(p,t) \ll 1}} is satisfied for all {\il{t \geq t_I}}. 

Given these assumptions, only two initial conditions for our example scenario remain to be 
specified.  These are the initial time $t_I$ itself and the overall normalization of 
$g_9(p,t_I)$ at this initial time.  The system then evolves dynamically according the Boltzmann equations
outlined in Appendix~\ref{app:boltzmann}.~   
However, as briefly noted in Sect.~\ref{sec:phase_space_evolution},
the Boltzmann equations are greatly simplified and possess several attractive properties
if they can be approximated as linear.   We can make such an approximation
if the phase-space distributions are sufficiently small ({\it i.e.}\/, if {\il{f_\ell(p,t) \ll 1}},
so that Pauli-blocking and Bose-enhancement effects  
are negligible) and if scattering and inverse-decay processes amongst the scalars are 
also negligible.
As discussed in Appendix~\ref{app:boltzmann},
these conditions are in turn both satisfied as long as the normalization of $g_9(p,t_I)$ remains sufficiently small.
Thus, the Boltzmann evolution in our ensemble is effectively linear, and rescaling the initial 
normalization of $g_9(p,t_I)$ within this limit simply has the effect of rescaling the other phase-space distributions
by the same constant factor.
Furthermore, we have shown that our system of Boltzmann equations is also invariant under a particular
set of transformations of the dimensionful parameters in our model.  
Thus, for the sake of generality, in what follows we shall refrain from specifying 
particular values for these quantities until we examine the consequences of 
our model for the matter power spectrum and thereby make contact with observational constraints.   
Of course, due to phenomenological considerations, we shall ultimately require that these
parameters be chosen such that the energy density carried by each $\phi_n$ with {\il{n > 1}} is negligible by 
the time of Big-Bang Nucleosynthesis (BBN).

\subsection{Partial widths and decay chains}\label{sec:ic}

Given the interaction Lagrangian in Eq.~(\ref{eq:InteractionLagrangian}) and the mass
spectrum and coupling structure in Eqs.~(\ref{eq:MassSpectrum_intra}) and~(\ref{eq:clij}),
it is straightforward to determine the partial decay widths $\Gamma^\ell_{ij}$
for all kinematically-allowed intra-ensemble decay processes of the form 
{\il{\phi_\ell \to \phi_i+\phi_j}} arising in our model.  The partial widths are given by
\begin{equation}
  \Gamma^{\ell}_{ij} ~=~
  \frac{1}{1+\delta_{ij}}\frac{|\vec{p}_{\rm CM}|}{8\pi m_{\ell}^2}
      |c_{\ell i j}|^2 
\end{equation}
where   the momentum of either daughter particle in the rest frame of $\phi_\ell$ is
given by
\beq
  |\vec{p}_{\rm CM}|~=~\frac{\sqrt{\left[m_{\ell}^2 - \left(m_i+m_j\right)^2\right]
  \left[m_{\ell}^2 - \left(m_i-m_j\right)^2\right]}}{2m_{\ell}}~.
  \label{eq:3momentumcubic}
\eeq
Note that {\il{\Gamma^\ell_{ij} = \Gamma^\ell_{ji}}}.
Since no additional decay processes with non-negligible partial widths exist for any
of the ensemble constituents, the total width $\Gamma_\ell$ of $\phi_\ell$ is merely
\beq
         \Gamma_\ell ~\equiv~   \sum_{i=0}^{\ell-1} \sum_{j=0}^i  \,\Gamma_{ij}^\ell~
\eeq
where the summation is over all configurations of daughters $(i,j)$ with $j\leq i$ 
such that {\il{m_i+m_j\leq m_\ell}}. 
 
\begin{figure}[t!]
\includegraphics[width=0.48\textwidth,keepaspectratio]{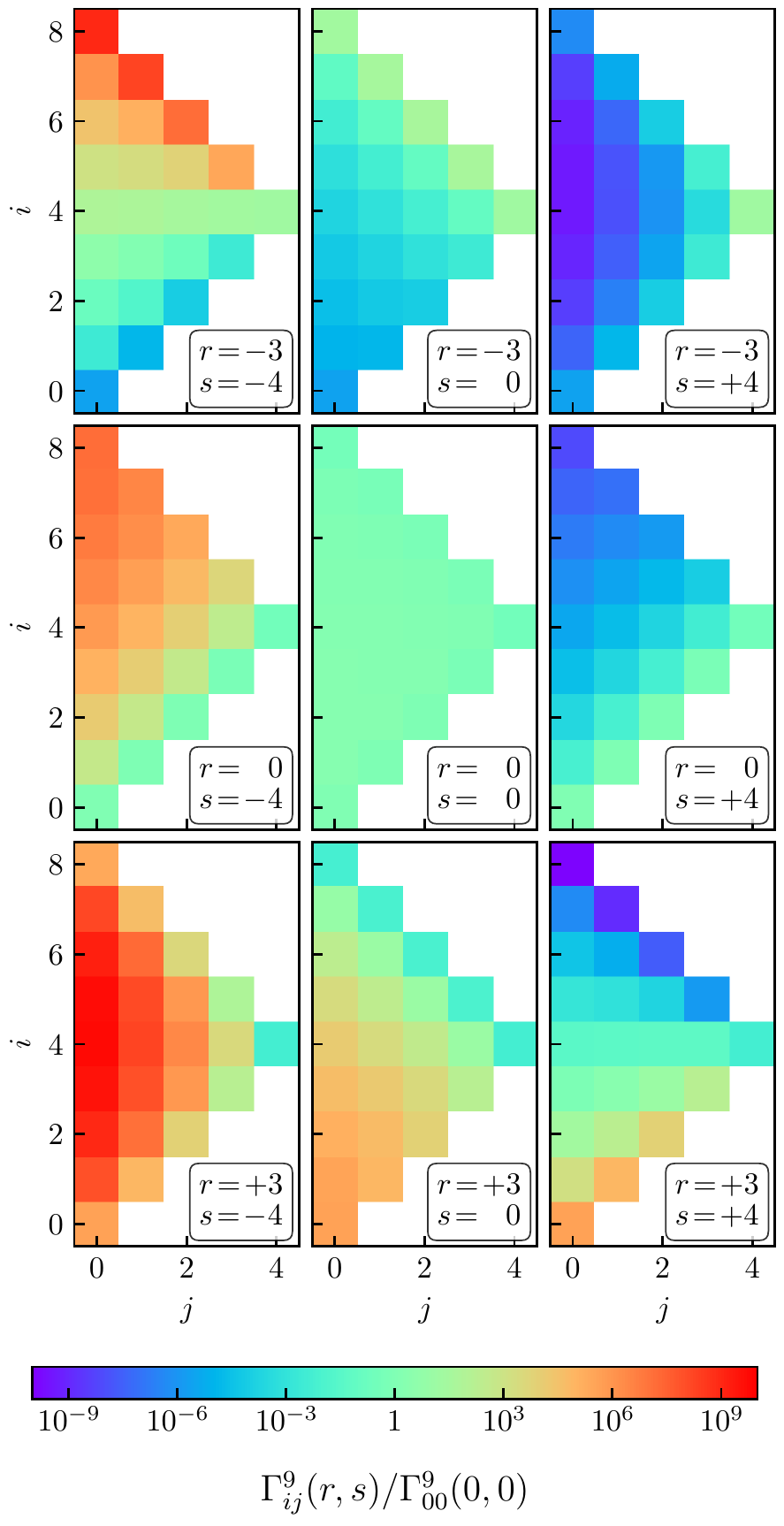}
\caption{Partial widths for the various decay processes
of the form 
{\il{\phi_9 \to \phi_i \phi_j}} with {\il{i\geq j}} through which the heaviest constituent $\phi_9$ in our
particle ensemble decays.  Each panel of this figure corresponds to a 
particular choice of $r$ and $s$, with $r$ increasing from top to bottom and $s$ increasing 
from left to right.  Within a given panel, the color of a particular square in the grid 
indicates the value of $\Gamma^{9}_{ij}$ for the corresponding assignment of 
$r$ and $s$, normalized to the partial width $\Gamma^{9}_{00}$ obtained for 
the parameter assignments {\il{r = s = 0}}.
\label{fig:decay_width_channels}}
\end{figure}

In Fig.~\ref{fig:decay_width_channels}, we show 
how the partial widths $\Gamma^{9}_{ij}$ for the decays {\il{\phi_9 \rightarrow \phi_i\phi_j}} 
which initiate these decay chains depend on the parameters $r$ and $s$ introduced in 
Eq.~(\ref{eq:clij}).  Each individual panel of the figure corresponds to a particular choice 
of the parameters $r$ and $s$, with $r$ increasing from top to bottom and $s$ increasing from 
left to right.  Within a given panel, the color of a particular square in the grid indicates 
the value of $\Gamma^{9}_{ij}$ for the corresponding assignment of $r$ and $s$, normalized
to the partial width $\Gamma^{9}_{00}$ obtained for the parameter assignments 
{\il{r = s = 0}}.
    
It is evident from Fig.~\ref{fig:decay_width_channels} that a variety of scaling behaviors
for the $\Gamma^{9}_{ij}$ as a function of the indices $i$ and $j$ can arise as a result of the 
interplay between the parameters $r$ and $s$.  For example, for {\il{r < 0}} and {\il{s < 0}}
(top left panel), decays to daughters $\phi_i$ and $\phi_j$ with vastly different values of 
$i$ and $j$ are preferred, as are decays with small mass gaps $m_9 - m_i - m_j$.  As a result, 
the partial widths are largest for final states such as {\il{\phi_9 \rightarrow \phi_8 \phi_0}} and 
{\il{\phi_9 \rightarrow \phi_7 \phi_1}}.  By contrast, for {\il{r > 0}} and {\il{s < 0}} (bottom left panel), 
the same preference for daughters with with vastly different values of $i$ and $j$ persists,
but is now combined with a preference for highly exothermic decays.
This leads to a situation in which the largest partial 
widths are those to final states such as {\il{\phi_9 \rightarrow \phi_5 \phi_1}} and 
{\il{\phi_9 \rightarrow \phi_4 \phi_0}}.  For  {\il{r < 0}} and {\il{s > 0}} (top right panel), a strong 
preference for symmetry between the daughter-particle masses, combined with a preference 
against strongly exothermic decays, leads to a situation in which the single decay channel 
{\il{\phi_9 \rightarrow \phi_4 \phi_4}} dominates.  For the case in which {\il{r = s = 0}}, the $\Gamma^{9}_{ij}$ 
for all possible decay channels are quite similar.  However, we emphasize that they are not 
precisely identical, due to the dependence of $|\vec{p}_{\rm CM}|$ in 
Eq.~(\ref{eq:3momentumcubic}).  The scaling behaviors for the $\Gamma^{9}_{ij}$ displayed
in the remaining panels of the figure can be also understood based on these same considerations. 

We also see from Fig.~\ref{fig:decay_width_channels} that the values of $\Gamma^{9}_{ij}$ for 
the processes which dominate the total width $\Gamma_9$ of $\phi_9$ are significantly 
larger for certain combinations of $r$ and $s$ than others.  
In particular, we see that $\Gamma_9$ is largest when $r$ and $s$ are positive and negative, respectively,
and smallest when $r$ and $s$ are respectively negative and positive.
Thus, we see that $\Gamma_9$ itself varies significantly as a function of $r$ and $s$.

\begin{figure*}
\centering
\includegraphics[width=0.99\textwidth,keepaspectratio]{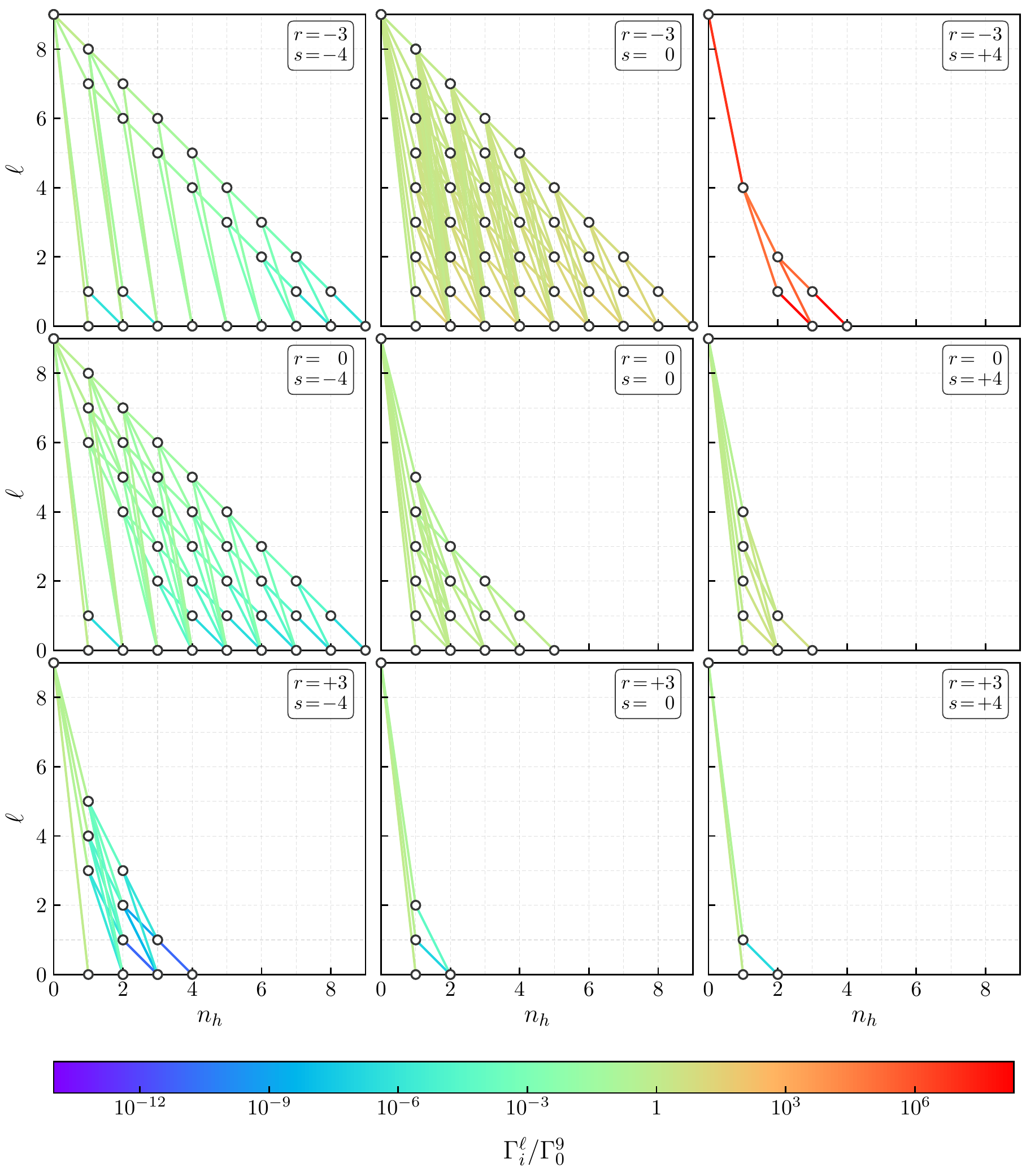}
\caption{The dominant reduced decay chains that arise within our model, starting from the heaviest
field $\phi_9$.
The different panels of this figure correspond to the
same choices of $r$ and $s$ as in Fig.~\protect\ref{fig:decay_width_channels}.~  
The vertical axis in each panel indicates the index $\ell$ associated
with a particular particle produced along the decay chain,
while the horizontal axis indicates the number $n_h$ of individual steps or ``hops''
that have occurred along the reduced decay chain in order to produce that particle.
Each line segment from parameters $(\ell_1,n_h)$ to parameters $(\ell_2,n_h+1)$ thus corresponds
to the production of a daughter particle $\phi_{\ell_2}$ directly from the decay of 
a parent particle $\phi_{\ell_1}$, with an associated production rate $\Gamma^{\ell_1}_{\ell_2}$.
For visual simplicity, 
we have indicated 
only those ``dominant'' parent-to-daughter segments
for which 
$\Gamma^{\ell_1}_{\ell_2}$ exceeds 5\% of the total
production rate $2\Gamma_{\ell_1}$ stemming from
$\phi_{\ell_1}$, 
with the corresponding colors indicating 
the value of $\Gamma^{\ell_1}_{\ell_2}$ 
normalized to 
$\Gamma^9_0$
in each panel. 
\label{fig:decay_chain}}
\end{figure*}

We now proceed to examine the decay chains which ultimately determine
the structure of $g_0(p,\tnow)$.  
Indeed, while an examination of the partial widths $\Gamma^{9}_{ij}$ provides some insight into how the decay 
phenomenology of our model depends on the choice of our model parameters $r$ and $s$, 
we must consider the full set of possible decay pathways 
through which the energy density initially stored in $\phi_9$ is ultimately transferred to
$\phi_0$ in order to characterize how the coupling
structure of our model impacts the present-day phase-space distribution $g_0(p,\tnow)$ 
of the lightest ensemble constituent $\phi_0$. 

Because we are considering two-body decays of the form {\il{\phi_\ell\to \phi_i\phi_j}} (each with
a different branching fraction),
and because each of these daughters then produces two granddaughters (whose identities 
are determined according to another set of branching fractions), the collection of particles produced from a single
ancestor proliferates quickly through each generation, as does the full set of potential decay chains and their associated net 
branching fractions.
In this paper we have nevertheless performed
this analysis exactly, and all of the results that we shall show in this paper have traced all of the possible
decay chains in this way.
However, in order to interpret our results physically and illustrate them graphically, 
it will suffice to simplify our discussion somewhat by considering
what we shall call {\it reduced}\/ decay chains.
Essentially, for any given parent $\phi_\ell$, we can ask 
which of all possible daughters $\phi_i$ with masses {\il{m_i\leq m_\ell}} 
is most likely to be produced {\it regardless of the 
identity of the other sibling}.   Indeed, the total rate for producing a given daughter $\phi_i$ 
from a parent $\phi_\ell$ is given by   
\beq
  \Gamma^{\ell}_{i} ~\equiv~ \sum_{j=0}^{N} 
    \mathcal{N}_{ij} \Gamma^{\ell}_{ij}~,
  \label{eq:prod_rate}
\eeq
where {\il{\mathcal{N}_{ij} = 1 + \delta_{ij}}} is the multiplicity of $\phi_i$ in the final state 
of any individual process {\il{\phi_\ell\to\phi_i\phi_j}}.  
We shall then construct our reduced decay chains by
stitching together sequences of such {\il{\ell \to i}} parent-to-daughter
transitions.

In Fig.~\ref{fig:decay_chain}, we show the dominant reduced decay chains which arise
within the context of our model, given our choice of initial conditions.  
Each line segment shown in the figure corresponds to the production of a daughter particle 
$\phi_i$ directly from the decay of a parent particle $\phi_\ell$,  while the color of the segment 
provides a measure of the rate $\Gamma^{\ell}_{i}$ at which daughters $\phi_i$ are produced directly 
from the decays of $\phi_\ell$.  
Thus the segments whose colors tend towards red in each panel indicate
a more rapid production of daughters than those in the same panel whose colors tend towards blue.
In order to maintain visual clarity, for each parent $\phi_\ell$ we have indicated 
only those segments whose daughters $\phi_i$ have the highest net probabilities for being produced.
These are defined as those
processes for which the corresponding production rate $\Gamma^\ell_i$
exceeds 5\% of the total particle-production rate stemming from $\phi_\ell$ decay --- \ie, processes for
which {\il{\Gamma^{\ell}_{i}/(2\Gamma_{\ell}) \geq 0.05}}.

It is straightforward to interpret the results
shown in Fig.~\ref{fig:decay_chain} in terms of deposits onto the cosmological conveyor belt
associated with the lowest state $\phi_0$, as discussed in Sect.~\ref{sec:phase_space_evolution}.
Of course, each decay chain shown within the different panels of Fig.~\ref{fig:decay_chain}
ultimately terminates when $\phi_0$ is produced.
This then is the time of the deposit onto the $\phi_0$ conveyor belt.
However, it is evident from
the different colors
associated with these decay chains that they can potentially proceed
with very different overall rates.
In general, we may associate a characteristic timescale associated with a given reduced decay chain
by considering
the sum of the inverses of the aggregate production rates $\Gamma^\ell_i$ for all of the
decay processes occurring within that decay chain.
Thus, a rough estimate of this
timescale can be obtained from the timescale associated with the slowest individual
decay step within the chain.
In Fig.~\ref{fig:decay_chain}, these are segments whose colors tend towards the blue end of the
color spectrum rather than the red.

In the four panels of the figure for which {\il{r \leq 0}} and
{\il{s \geq 0}} --- \ie, the four panels constituting the upper right corner of the figure ---
the timescales for all decay chains which contribute significantly to the production
of $\phi_0$ are quite similar.
In other words, each of these decay chains in these panels tends to make its deposit
onto the $g_0$ conveyor belt at roughly the same time.
As discussed in Sect.~\ref{sec:phase_space_evolution}, we therefore expect the corresponding $g_0(p,\tnow)$
to be essentially uni-modal for such combinations of $r$ and $s$.
By contrast, we observe that in the remaining figures in the plot --- \ie, those for which {\il{r > 0}} and/or
{\il{s < 0}} ---  many of the decay chains which contribute significantly to the production of $\phi_0$
have vastly different timescales, as indicated in Fig.~\ref{fig:decay_chain} through their significantly different colors.
Consequently, in such cases we expect that a non-trivial,
multi-modal phase-space distribution will be generated for $\phi_0$.

Our understanding of the conveyor-belt dynamics
from Sect.~\ref{sec:phase_space_evolution} also enables us to form comparative expectations
regarding $g_0(p,\tnow)$  {\it across}\/ the different panels.
For example, we have already seen that in Fig.~\ref{fig:decay_width_channels}
that the decay widths involved for {\il{r <0}} and {\il{s>0}} are significantly smaller than those involved for
{\il{r>0}} and {\il{s<0}}.   Thus, even though all of the decay chains within the first case arise might make
their deposits at approximately the same time as each other,
this time of deposit is much later than the time of the first deposits that occur in the second case.
Thus we expect the uni-modal dark-matter phase-space distribution $g_0(p,\tnow)$ in the first case
to have less time to redshift
than the contributions from the first mode of the phase-space distribution in the second case,
implying that $g_0(p,\tnow)$ should be situated at higher momenta than the
the lowest-momentum peak within the distribution $g_0(p,\tnow)$ in the second case.

\subsection{Dark-matter phase-space distributions and  matter power spectra}\label{sec:id}
 
In order to determine the extent to which these expectations are realized within our model, 
we have calculated the actual dark-matter phase-space distributions $g_0(p,\tnow)$ that result
in each case at the present time, long after all of the decays within each decay chain have terminated.
We stress that this calculation has been performed through
a full Boltzmann analysis, as outlined in Appendix~\ref{app:boltzmann},
with all source and decay terms included.
Indeed, as the decay process unfolds towards the ground state, there is considerable 
variety in the intermediate stages 
through which our dark sector passes.
This non-trivial time-evolution of the dark sector is outlined in Appendix~\ref{app:time_evol}.~
However, in this section, our main interest concerns the final results that
emerge after all of the decays are concluded.

\begin{figure*}[t]
\includegraphics[width=.99\textwidth,keepaspectratio]{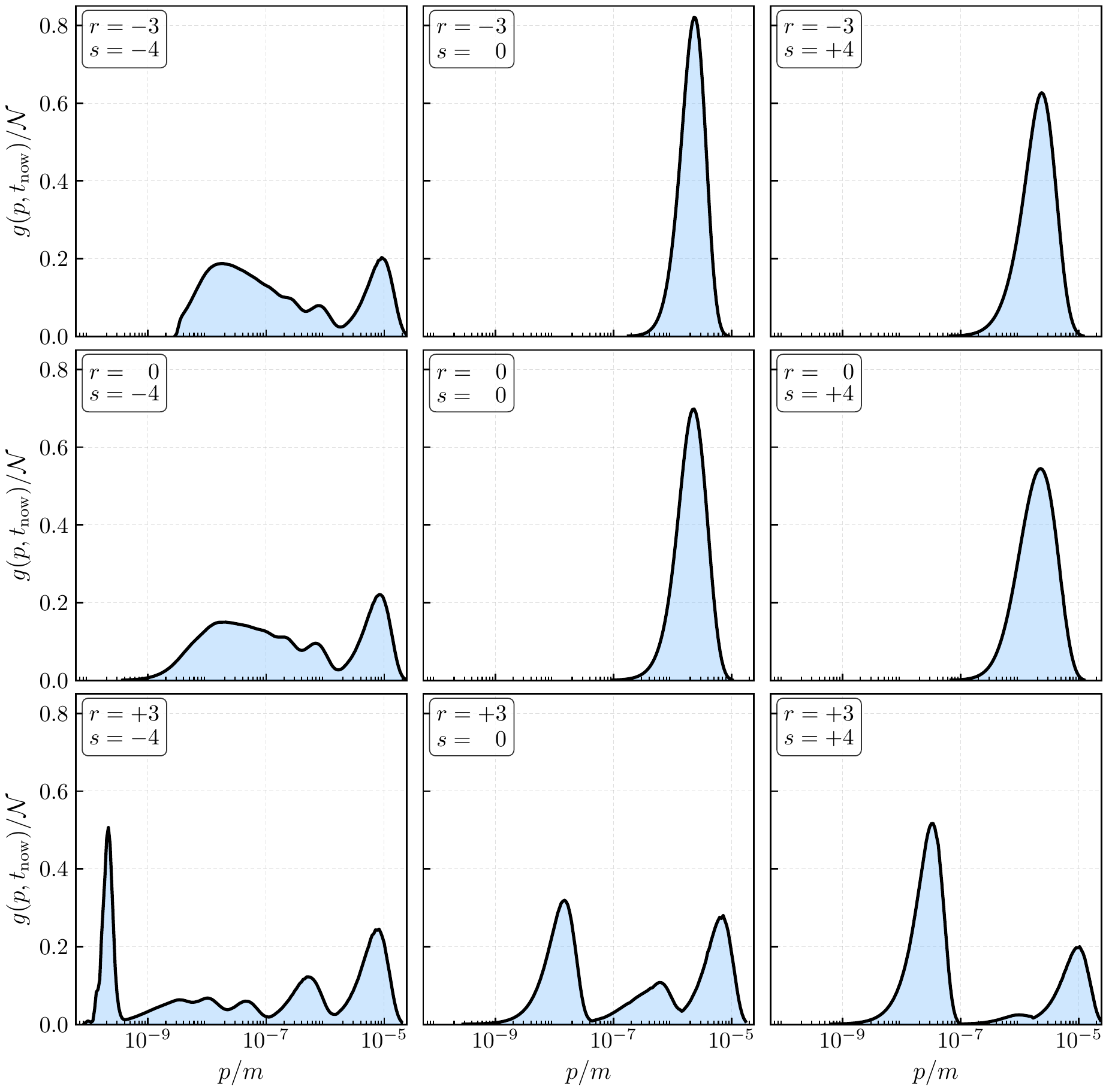}\centering
\caption{The phase-space distributions $g_0(p,\tnow)$ in our model for the lightest ensemble
constituent $\phi_0$ at the present time. 
The different panels 
correspond to the same choices of $r$ and $s$ as in 
Figs.~\protect\ref{fig:decay_width_channels}  and \ref{fig:decay_chain}.~
The initial conditions in each case
have been chosen such that these distributions share not only a common overall normalization
but also a common value for the ratio 
$\langle p\rangle_{\rm now}/m_0$, where $\langle p\rangle_{\rm now}$ is the average 
momentum of the distribution at present time.
\label{fig:f0_all}}
\end{figure*}

Our results for the  dark-matter phase-space distributions $g_0(p,\tnow)$ 
are shown in Fig.~\ref{fig:f0_all}.~
Once again, the different 
panels of Fig.~\ref{fig:f0_all} correspond to the same choices of $r$ and $s$ as in 
Figs.~\ref{fig:decay_width_channels} and \ref{fig:decay_chain}.~
In order to facilitate comparison between the 
results obtained for different values of $r$ and $s$, we have chosen the initial conditions 
in each case such that all of our final distributions $g_0(p,\tnow)$ 
share not only a common overall normalization but also a common value for the ratio 
$\langle p\rangle_{\rm now}/m_0$,
where $\langle p\rangle_{\rm now}$ is the average present-day momentum of the distribution.
The overall normalization for $g_0(p,\tnow)$ is chosen 
to correspond to the observed present-day dark-matter energy density 
{\il{\rho_{\rm DM} \approx 0.27 \times 10^{-5} \,h^2 \,\rm GeV\, cm^{-3}}}.
Likewise, 
for all of the plots in Fig.~\ref{fig:f0_all} we have chosen the value 
\begin{equation}
  \frac{\langle p \rangle_{\rm now}}{m_0} ~=~ \frac{T_{\rm CMB}}{100\mathrm{~eV}} 
  ~\approx ~ 2.3\times 10^{-6}~,
\label{chosenvalue}
\end{equation}
where {\il{T_{\rm CMB} \approx 0.23\times 10^{-3}}}~eV is the present-day temperature of
the cosmic microwave background (CMB) radiation. 
This value is motivated as follows.
In general, we are interested in situations in which a significant portion of the total dark matter
exhibits speeds lying within
the rough window 
{\il{v \approx p/m_0 \sim \mathcal{O}(10^{-8} - 10^{-5})}}.
Indeed, if a significant fraction of the dark matter has speeds above this window,
the resulting dark matter will be too ``hot'' to accord with observation.
On the other hand, the free-streaming of dark matter with speeds below this window
will eventually have effects on $P(k)$ which are evident only at large values of $k$ 
which are beyond the range at which they can reliably be probed observationally.
Motivated by these considerations, we have therefore chosen to anchor the 
overall scale of the present-day momentum $\langle p\rangle$
for our $g_0(p,\tnow)$ distribution as in Eq.~(\ref{chosenvalue}) such that the speed of a typical
dark-matter particle lies within our window of interest.  

In this connection, we observe that
the Boltzmann equations which give rise to 
$g_0(p,t)$ are invariant under the rescalings specified in Eq.~(\ref{eq:RescalingTransf}).
Likewise, our eventual power spectrum  
$P(k)$ depends only on the velocities {\il{v \approx p/m_0}} of $\phi_0$ rather than on
$p$ and $m_0$ independently.  
This result of course assumes that 
the conditions {\il{f_i(p,t) \ll 1}} are likewise satisfied for all of the 
$\phi_i$ distributions in our ensemble during all times {\il{t \geq t_I}} during which our 
Boltzmann evolution takes place.    
In particular this must be true for $f_9(p,t_I)$ --- corresponding to the heaviest state the ensemble ---
prior to the decay process.
We have verified numerically that even for $m_0$ as small as $20$~keV, which corresponds to
a present-day average momentum {\il{\langle p \rangle_{\rm now} = 200 \,T_{\rm CMB}}}, 
these consistency criteria are satisfied.  Indeed,
the required normalization for $g_9(p,t_I)$ 
is such that 
{\il{f_i(p,t) \ll 1}} for all $\phi_i$ in the ensemble at all 
{\il{t \geq t_I}} for all combinations of $r$ and $s$ considered in Fig.~\ref{fig:f0_all}.

We see from the results in Fig.~\ref{fig:f0_all} that all of our expectations from Sect.~\ref{sec:phase_space_evolution} 
are indeed borne out.
For example, as anticipated, 
the phase-space distributions $g_0(p,\tnow)$ obtained for
{\il{r \leq 0}} and {\il{s \leq 0}} are essentially unimodal, while  the distributions which
arise either for {\il{r > 0}} or for {\il{s < 0}} exhibit a complicated, multi-modal structure.
Likewise, we see that single peaks within the unimodal distributions sit at higher momenta than the lowest-momentum peaks
within the multi-modal distributions, again in line with our expectations above.

These results also 
demonstrate one of our main themes of this paper,
namely that
fairly complex and even multi-modal dark-matter phase-space distributions can easily
arise, even when the full details of 
the Boltzmann evolution are incorporated into the analysis.   Indeed, the existence of many independent levels and the proliferation
of overlapping decay chains 
do not wash out  the non-trivial structures for $g_0(p)$ that we anticipated in Sect.~\ref{sec:phase_space_evolution}.~
Instead, they only serve to enhance these structures.  

\begin{figure*}[t]
\includegraphics[width=0.99\textwidth,keepaspectratio]{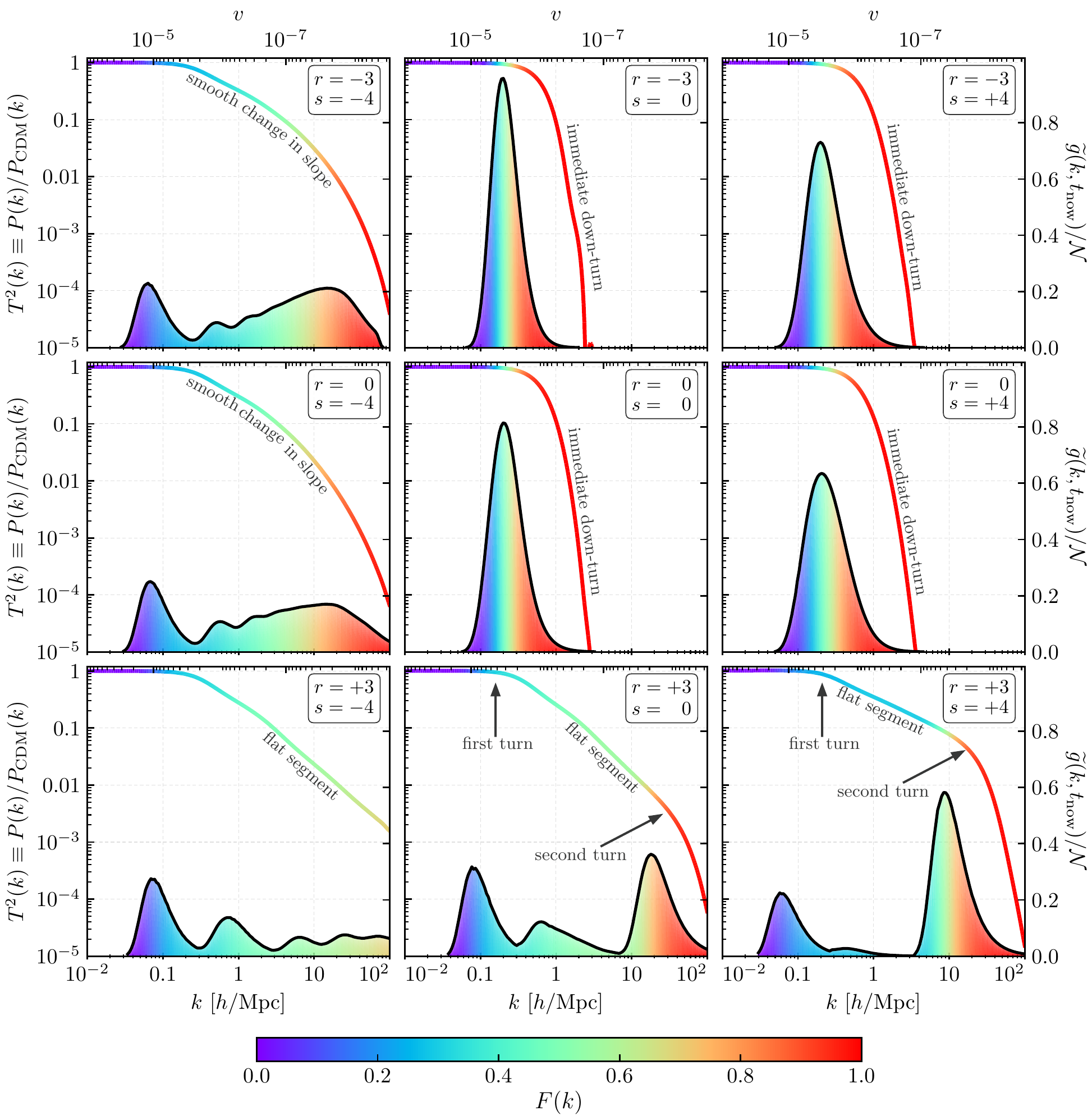}\centering
\caption{The $k$-space distributions $\widetilde g(k)$ defined in Eq.~(\ref{tildegdef})
and squared transfer functions {\il{T^2(k) \equiv P(k)/P_{\rm CDM}(k)}}  
corresponding to each of the phase-space distributions $g_0(p,\tnow)$ shown in 
Fig.~\protect\ref{fig:decay_width_channels}.  
As we sweep from smaller to larger values of $k$ within the 
$k$-space distributions $\widetilde g(k)$ (outlined in solid black curves),
we have shaded the area inside the $\widetilde g(k)$ distributions according to the accumulated hot fraction function $F(k)$
defined in Eq.~(\ref{Fracdef2}).   These same colors are then used for plotting the corresponding
transfer functions $T^2(k)$ as functions of $k$. 
The tick-marks along the bottom, top, left, and right axes of each panel correspond
respectively to the wavenumber $k$, the corresponding velocity $v$, the transfer function $T^2(k)$, and 
the $k$-space distribution $\widetilde g(k)$.
\label{fig:Pk_rainbow}}
\end{figure*}

We now turn to examine the 
$k$-space distributions $\widetilde g(k)$ 
and squared transfer functions {\il{T^2(k) \equiv P(k)/P_{\rm CDM}(k)}}  
corresponding to each of the phase-space distributions $g_0(p,\tnow)$ shown in 
Fig.~\ref{fig:f0_all}.~
Our results are shown in Fig.~\ref{fig:Pk_rainbow}.~
For each phase-space distribution $g_0(p,\tnow)$, the corresponding
$k$-space distribution $\widetilde g(k)$ is defined
in Eq.~(\ref{tildegdef}).
Likewise, we have calculated each corresponding transfer function $T^2(k)$
using the \texttt{CLASS}~{\mbox{\cite{Lesgourgues:2011re,Blas:2011rf,Lesgourgues:2011rg,Lesgourgues:2011rh}}} 
software package.   
As we sweep from smaller to larger values of $k$ within the 
$k$-space distributions $\widetilde g(k)$ 
in Fig.~\ref{fig:Pk_rainbow},
we have shaded the area inside the $\widetilde g(k)$ distributions according to the accumulated hot fraction function $F(k)$
defined in Eq.~(\ref{Fracdef2}).   These same colors are then used for plotting the corresponding
transfer functions $T^2(k)$ as functions of $k$. 

The results in Fig.~\ref{fig:Pk_rainbow} are once again in complete accordance
with our expectations from Sect.~\ref{sec:perturb}.~
In particular, we see that the logarithmic slope $d \log T^2(k)/ d\log k$ 
of the transfer function indeed appears to correlate with the hot fraction function $F(k)$.
Indeed, this slope holds steady in regions where the hot fraction function is relatively
constant (with relatively little change of color), while this slope changes more rapidly in 
regions where the hot fraction function (and thus the corresponding color) is also changing rapidly.
This is precisely the behavior anticipated in Sect.~\ref{sec:perturb}.~
However, we now see that this behavior survives robustly even for dark-matter phase-space
distributions which are fairly complicated, potentially even exhibiting many different peaks and troughs
as functions of $k$.

\begin{figure*}[t]
\includegraphics[width=0.99\textwidth,keepaspectratio]{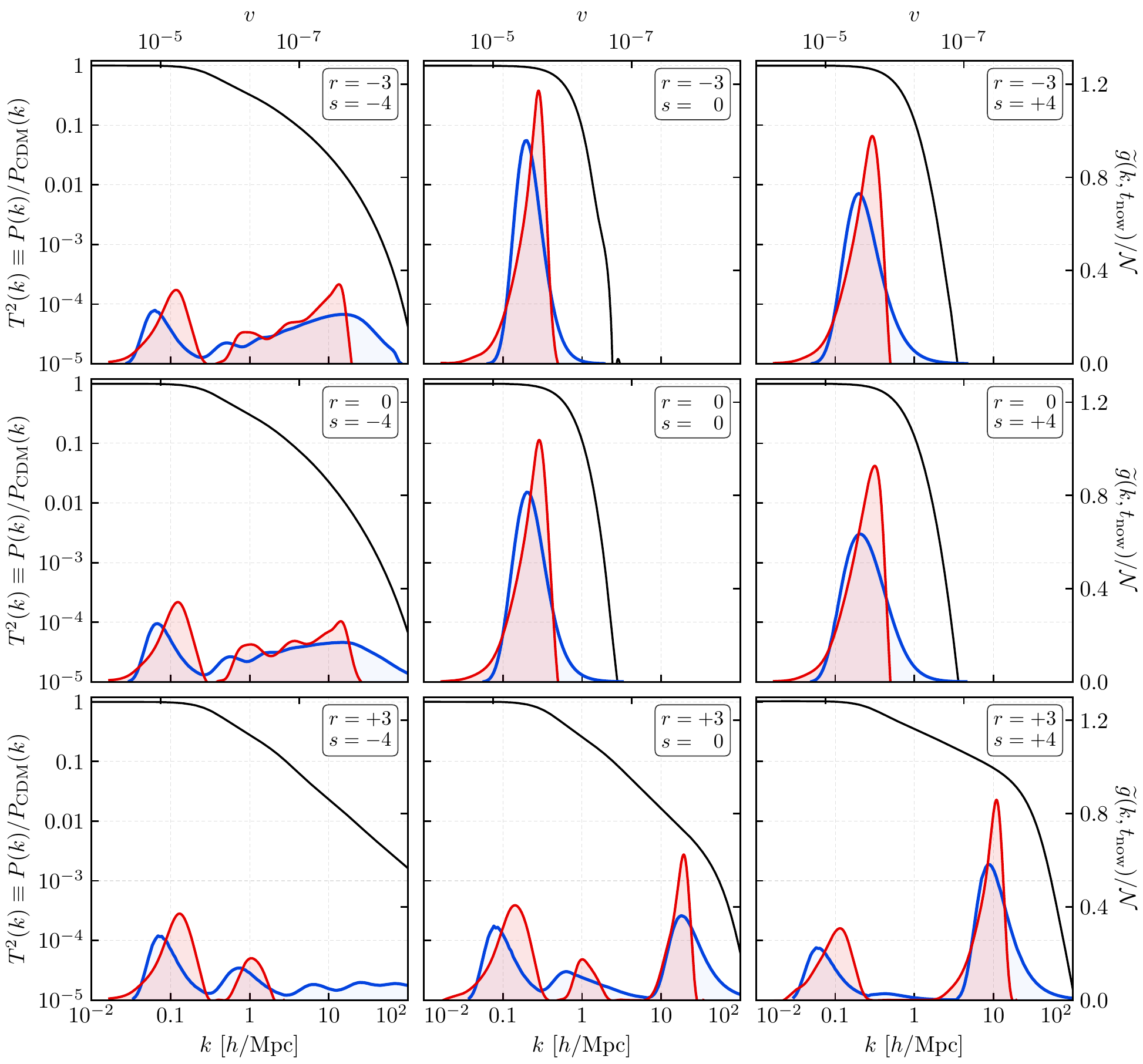}
\centering
\caption{An explicit test of our reconstruction conjecture in Eq.~(\ref{conj3}).
In each panel, we show three curves:  the original dark-matter phase-space distribution (blue),
the corresponding transfer function (black) to which it gives rise,
and the phase-space distribution (red/pink) which has been ``reconstructed'' from the transfer function via Eq.~(\ref{conj3}).
We see that in all cases --- including those with unimodal, bi-modal, and even tri-modal distributions, with
peaks of assorted heights and widths ---
our conjecture reproduces many of the critical features of the original phase-space distribution with 
remarkable accuracy.
\label{fig:Fk_fit}}
\end{figure*}

\subsection{An explicit test of our reconstruction conjecture}\label{sec:ie}

Finally, given the transfer functions in 
Fig.~\ref{fig:Pk_rainbow}, we can now perform a test 
of our conjectured relation in Eq.~(\ref{conj3}).
In particular, we can
now determine the extent to which this relation
allows us to reconstruct the dark-matter phase-space distribution $\widetilde g(k)$ directly
from the transfer function $T^2(k)$.
Our results are shown in Fig.~\ref{fig:Fk_fit}.~
In each panel of this figure,
the black curve shows the transfer function $T^2(k)$ from 
Fig.~\ref{fig:Pk_rainbow}, 
while the blue outline shows the original
underlying 
$k$-space dark-matter phase-space
distribution
$\widetilde g(k)$ from which it was derived, also from 
Fig.~\ref{fig:Pk_rainbow}.~ 
By contrast, the pink shaded regions show the reconstructed phase-space distributions
$\widetilde g(k)$ that follow directly from the transfer function $T^2(k)$ via our conjectured ``inverse'' 
relationship in Eq.~(\ref{conj3}).

Although the reconstructed phase-space distributions do not match the original phase-space
distributions exactly, they nevertheless do faithfully capture the most salient features of these distributions.
Indeed, this is true not only for the unimodal distributions that emerge for
{\il{r \leq 0}} and {\il{s \leq 0}}, but also for the
multi-modal distributions that emerge otherwise.
Likewise, this remains true even in cases for which our distributions are relatively
extended, with no sharp peaks at all, such as those which emerge for {\il{s<0}}.
We thus conclude that our conjectured relationship holds remarkably well across a variety of
possible dark-matter distribution
shapes (thermal, non-thermal,  unimodal, multi-modal, and so forth).
{\it A priori}\/, the relationship between the phase-space distribution $g(p)$ and the resulting power spectrum $P(k)$
is complex, 
involving rather non-trivial initial conditions incorporating primordial
perturbations of the inflaton and other fields which are then propagated forward in time 
using the linearized Einstein equations in the presence of these non-cold dark-matter species.
Given this situation,
we find this degree of agreement for what is essentially a relatively simple analytic ``inversion'' formula to be rather stunning.  

The results of this section thus illustrate that many features of the early-universe dynamics
associated with the non-minimal dark sector of our illustrative model
are indeed imprinted on the resulting matter power spectrum,
and that an ``archaeological'' study of the matter power spectrum is indeed capable of reconstructing many aspects of this 
cosmological history.   Indeed, we have seen that many features of the dark-matter phase-space distribution can be reconstructed
from the transfer function 
via our conjecture in Eq.~(\ref{conj3}).   Likewise, the specific pattern of peaks and valleys within this reconstructed
phase-space 
distribution reflects the integrated deposit history onto the dark-matter conveyor belt, and through the particular properties
of the 
peaks involved we may even use results from Sect.~\ref{sec:phase_space_evolution}
--- such as those in Table~\ref{inversepacket} --- in order to elucidate many aspects of the particular     
decay phenomenology involved.
Thus, while no archaeological study can reconstruct every aspect of the cosmological past, we have seen
there is much that we can indeed learn
through these sorts of analyses.

\section{Conclusions and discussion}\label{sec:intra_decay_conclusion}

Throughout this paper, our goal has been to develop
a toolbox of methods for learning about, and potentially constraining,
the features of non-minimal dark sectors and their dynamics in the early universe.
By necessity this quest has been fundamentally ``archaeological'' --- we seek to exploit present-day data in order to
learn about the past. 
In this paper our approach to this question has been centered around two somewhat independent 
linkages, as first outlined in Eq.~(\ref{onepointone}):   the first connects early-universe dynamics (such as that
associated with a non-minimal dark sector) to a resulting dark-matter phase-space distribution $f(p)$, 
and the second 
connects this phase-space distribution to a corresponding matter-power spectrum $P(k)$.   
In principle, this latter quantity is observable ---
an assertion which we shall discuss further below ---
and thus our overarching goal has been to 
examine the extent to which we can ``invert'' the flow chart in Eq.~(\ref{onepointone}) and
thereby constrain not only the corresponding dark-matter distribution $f(p)$  
but also the possible early-universe dynamics from which this $f(p)$ distribution might have arisen --- all
while starting from 
a given matter power spectrum $P(k)$.

Needless to say, a complete inversion is not possible, and we have seen numerous examples
in this paper
in which several different cosmological histories give rise to the same (or similar) present-day data.
Yet, there have also been instances in which we can draw fairly robust conclusions regarding the inverse map
with only a handful of reasonable assumptions.
Collectively, 
such results may be of critical importance 
if the dark sector turns out to interact with the visible sector
exceedingly weakly, or
perhaps even only gravitationally.
Studies focusing on inverting such quantities as the matter power spectrum
may then be the only ways 
of ever learning about the dark sector and its early-universe dynamics.

In this paper, we have examined each of the linkages within Eq.~(\ref{onepointone}) from a number 
of different directions, always with an eye towards assessing the extent to which the mappings they
represent might be inverted.
Along the way, we have generated what we believe to be a number of interesting and potentially useful results.

In Sect.~\ref{parent_to_daughter}, we began by studying the decay process from a given parent phase-space distribution
         (or ``packet'') to a corresponding daughter packet.  
              We found that this process is highly complex, as illustrated in Fig.~\ref{fig:conveyor3},
          involving a mixture of effects due to time dilation, cosmological redshifting, and the exponential 
          nature of the decay process itself.  
          Under certain assumptions that are discussed in the text,
          we believe that our chief results in this direction include the following:
\begin{itemize}
\item   First, under certain assumptions, 
             we found that 
            the width and average momentum of a given daughter packet allow us to
             greatly constrain the properties of the parent 
               from which it arose
            as well as the marginality of the corresponding decay process.
           These results are shown in Table~\ref{inversepacket}.~
        Indeed, we found that in some cases this reconstruction is unique, while in other cases 
           several possibilities exist.

\item    Second, in some cases we were able to obtain {\it universal functional forms}\/
         which describe the resulting daughter packets. 
          For example, in the limit that the parent
           is highly non-relativistic, we found that the daughter packet 
           always approaches the universal functional form 
        shown in Eq.~(\ref{gnonrel}) and illustrated in Fig.~\ref{fig:gnonrel}.  
           By contrast, for certain highly relativistic decays, we found 
           that the daughter packet will instead have the very different universal
             functional form shown in Fig.~\ref{fig:boxcartoon}.

\item    Third, by comparing the results of Tables~\ref{inversedeposit} and \ref{inversepacket},
         we were able to  explore the interplay between decay kinematics and cosmological expansion.
          This resulted in our observations, as outlined  in Table~\ref{stacking}, concerning the manner in which
          individual decay deposits onto the daughter conveyor belt accrue
            as time evolves.   These general observations allowed us to determine the extent to which
           cosmological expansion causes the daughter packets
           to be wider than would have been predicted on the basis of decay kinematics alone.
           In some cases, we found that cosmological expansion has little overall effect on the shape
        of the resulting daughter packet --- indeed, in such cases a packet of similar shape would have emerged
        even if the universe had not been expanding.
        As indicated in Table~\ref{stacking}, these packets tend to have relatively sharp edges.
          In other cases, by contrast, we found that cosmological expansion plays a significant role in shaping the
      resulting daughter packet.  In these cases the resulting packets tend not to have such sharp edges.
 
\item    Finally, we demonstrated that the overall tilt or {\it skewness}\/ of the daughter packet
        also carries information concerning the parent from which the daughter emerged.
         For example, as anticipated in Eq.~(\ref{tiltrules}) and verified explicitly in Fig.~\ref{fig:skewplot},
              we found that there is a close relationship between the skewness of the
          daughter packet and the degree to which the parent was relativistic at the time it was produced.
           In Eq.~(\ref{leftcondition}) we 
           even determined the conditions under which such relativistic skewness effects
               will be significant.
           Moreover, in a similar way, we were also able to demonstrate an entirely different skewness relationship:
         one between the skewness of the daughter packet and the {\it decay rate}\/ $\Gamma$ experienced by the parent (as
               expressed in units of the Hubble parameter $H(t_0)$ at the time $t_0$ when the parent was established).
                  These results are illustrated in Fig.~\ref{fig:Snonrel}.
           Together, these results imply that the skewness of the daughter packet
          can also serve as a useful tool in reconstructing the properties of the parent.

\end{itemize}

Next, in Sect.~\ref{sec:overlapping},
we enlarged our discussion beyond the case of a single parent
to examine 
the full set of 
decays that can occur within a non-minimal dark sector containing
many different constituents.  Within this context, our primary observations
are as follows: 
\begin{itemize}
\item    We explicitly demonstrated that such a 
          non-minimal dark sector can leave dramatic
              imprints on the phase-space distribution associated
           with the dark-sector ground state (presumed to be the dark matter that survives
           to the present time).
           We found that the resulting dark-matter phase-space distribution {\it need no longer be thermal}\/ --- in fact, it may 
        even be {\it multi-modal}\/, exhibiting a non-trivial pattern of peaks and troughs
           as a function of momentum.
           Such results therefore carry us beyond the sorts of thermal phase-space distributions
             that are traditionally assumed for dark-matter particles.
          In fact, we found that 
           multi-modal phase-space distributions emerge quite generically  as the result of
            overlapping decay chains, even if we start 
              from a single ancestor with a simple, unimodal phase-space distribution.
            This remains true even if our ancestor was produced thermally.
                {\it Thus we see that 
                   our final dark-matter phase-space distribution can be non-thermal (and potentially even multi-modal) 
                     --- even if the dark sector was initially populated through thermal means.} 
\end{itemize}

In Sect.~\ref{sec:perturb}, we 
turned to examine the second linkage in Eq.~(\ref{onepointone}) --- that connecting the
dark-matter phase-space distribution $g(p)$ to the matter power spectrum $P(k)$.
Given our results from Sect.~\ref{sec:overlapping},
we allowed our analysis to remain completely general and 
we did not assume any particular form for the dark-matter phase-space distribution.
In this context, we regard our main results to be the following:
\begin{itemize}
\item    First, given an arbitrary phase-space distribution $g(p)$, we 
                introduced a corresponding new quantity, a ``dual'' $k$-space distribution function $\widetilde g(k)$.
               We view this as an important conceptual step since the $p$ and $k$ variables are not 
               dual to each other in any ``Fourier'' sense.  Rather, we 
                utilized the defining relation for the horizon wavenumber $k_{\rm hor}(p)$ 
                  in Eq.~(\ref{kFSHdef}) in order to define a mapping between the $p$-variable
               of $g(p)$ and the $k$-variable of $P(k)$, implicitly 
                identifying $k_{\rm hor}$ with $k$ itself.
            We stress that this treatment is completely unorthodox, 
               since $k_{\rm hor}(p)$ 
               is technically the {\it minimum}\/ value
               of $k$ for which $P(k)$ could potentially be affected by dark matter of momentum $p$.
            As such, there is therefore no direct relation between $k_{\rm hor}$ and the $k$-variable of $P(k)$.
           Nevertheless, by choosing to identify these two variables with each other, 
           it becomes possible to consider the dark-matter phase-space
           distribution $g(p)$ in the same $k$-space as the matter power spectrum $P(k)$.  This ultimately proved to be
             a critical step in reconstructing the former from the latter.

\item   Second, prompted by our phenomenological observations in Sect.~\ref{sec:phenobs},
               we also defined a so-called ``hot fraction function'' $F(k)$.
           Once again, this is a fairly non-trivial step because it too rests again on our unorthodox
           mapping between the $p$-variable of $g(p)$ and the $k$-variable of $P(k)$.
         Indeed, the hot fraction function $F(k)$ is the total accumulated dark-matter abundance
          from all momenta $p$ greater than $k_{\rm hor}^{-1}(k)$.

\item    Finally, given these two new quantities $\widetilde g(k)$ and $F(k)$,
          we were able to put forth 
           what may prove to be one of the most important results
             of this paper:  our ``reconstruction'' conjecture in Eq.~(\ref{conj3}).
           In this context, it is important to emphasize that we regard this conjecture as having
          two fairly independent components. The first component of our conjecture is the assertion that
           the hot fraction function $F(k)$ is directly correlated
         {\it not}\/ with the transfer function {\il{T^2(k)\equiv P(k)/P_{\rm CDM}(k)}} itself, but
       rather with its logarithmic slope $d\log T^2 / d\log k$. 
         This assertion is written explicitly in Eq.~(\ref{conj1}) in terms of 
           an unknown correlation function $\eta$,
         and thus stands independently of any particular correlation function $\eta$. 
       Given this, the second component of our conjecture asserts
         a particular empirical form for this correlation between $F(k)$ and the logarithmic slope   
        $d\log T^2 / d\log k$.   
         This correlation is given  in Eq.~(\ref{eq:FitFImplicit}).
          This result then implicitly furnishes us with an 
             explicit form for the $\eta$-function described above, and thereby
            leads directly to our final conjecture in Eq.~(\ref{conj3}). 
\end{itemize}

One remarkable feature of our conjecture is that it relates $P(k)$ to $\widetilde g(k)$ {\it point-by-point}\/
in $k$-space.   Indeed, as long as we know the transfer function $T^2(k)$ and its derivatives at a specific
value of $k$, our conjecture allows us to reconstruct $\widetilde g(k)$ at that value of $k$.  
In this sense
our reconstruction is {\it local}\/, mapping each portion
of the matter power spectrum curve to a corresponding portion of the dark-matter phase-space distribution.
This is a very useful feature because each portion of the $\widetilde g(k)$ curve directly
maps back to a corresponding portion of the original $g(p)$ phase-space distribution,
and this in turn maps back to a sum of deposits along a particular 
``backwards FRW momentum lightcone'' [here borrowing the language below Eq.~(\ref{integralform})].
Thus we can trace our inverse map point-by-point
along these curves, and thereby potentially correlate 
specific features in the matter power spectrum
with specific deposit profiles stemming from early-universe dynamics.

In Sect.~\ref{sec:toy_model}, we then proceeded to test all of these ideas
within the context of an actual illustrative model.
We began by specifying a particular Lagrangian governing early-universe dynamics and proceeded all the way to 
a calculation of the corresponding matter power-spectrum. 
          As expected, this model was able to give rise to dark-matter phase-space distributions of 
great variety, including some which are unimodal, some which are bi- or tri-modal,
     and even some which have fairly non-trivial patterns of peaks and troughs ---
all arising as a consequence of the large number of dark-sector states 
and a corresponding  multiplicity of decay pathways
involving different characteristic timescales.
These distributions are shown in Fig.~\ref{fig:f0_all}.
Within this model we were also able to perform 
a rigorous test of our reconstruction conjecture in Eq.~(\ref{conj3}).
         In all of the cases explored,
         we obtained results 
         for the reconstructed distributions $\widetilde g(k)$ 
          which, though not exact,  successfully captured all of their salient phenomenological features. 
         This is illustrated explicitly in Fig.~\ref{fig:Fk_fit}. 
Likewise, in Appendix~\ref{app:time_evol},
        we traced the time-development of this model {\it during}\/ the intra-ensemble decay process
         and demonstrated how the total energy density and equation of state of the dark sector evolved non-trivially
         as the decay process unfolded.
All of these results confirmed our main conclusions as highlighted above.

Finally, last but not least,
we would be remiss not to mention one further unique feature of our work:
\begin{itemize}
\item  Within this paper, 
          we developed a very intuitive picture for viewing the cosmological time-development of 
         a given dark-matter phase-space distribution $g(p)$
           in terms of a cosmological ``conveyor belt'' and deposits onto it.
           Of course, as we discussed, this picture is nothing but a physical representation of the integral
           form of the standard Boltzmann equations, and as such it is not necessarily ``new''.  
            Nevertheless, throughout this paper, we repeatedly 
           found this physical picture to be a particularly useful one, especially for
          non-minimal dark sectors and the decays that can arise amongst their constituents.
         Indeed, this intuitive picture of deposits onto a cosmological conveyor belt 
           has served as the backbone of much of the discussion in this paper.
\end{itemize}

Needless to say, 
this paper has spanned considerable territory.
As such, a number of comments are in order, concerning
not only the work we have done but also possible directions for future investigation.

First and foremost, our ability to perform the kind of archaeological 
reconstruction we have described in this paper is predicated on our ability 
to measure $P(k)$ at large $k$. Understanding the extent
to which we are currently able to probe the matter power spectrum
is therefore critical for our work.  
It is also important to  assess the future prospects for probing $P(k)$ at even
higher $k$.

Measurements of the linear matter power spectrum based on data obtained at low redshifts 
are currently reliable for {\il{k \lesssim 0.05 - 0.1}}~Mpc$^{-1}$.
However, observations 
of the Lyman-$\alpha$ forest at higher redshifts up to {\il{z\sim 5}} can provide additional 
information about the linear matter power spectrum at wavenumbers up to {\il{k \sim 1}}~Mpc$^{-1}$.  
At higher $z$, the density of neutral hydrogen is so large that the hydrogen emission 
spectrum becomes difficult to measure.  However, observations of the 21-cm spectral line 
of neutral hydrogen at redshifts up to {\il{z \sim 30}} could potentially yield information 
about the linear matter power spectrum at much higher $k$.  There are of course certain practical 
considerations  which would render such a measurement challenging.
For example, any instrument capable of attaining the requisite sensitivity would need 
to have an enormous collection area, and ionospheric effects on radio signals with 
frequencies below $100$~MHz would likely render a terrestrial telescope unsuitable for this 
purpose.  Such a measurement is nevertheless possible in principle.

Another strategy for probing the dynamics of a non-minimal dark sector
during the early universe would be to perform a reconstruction of the dark-matter 
phase-space distribution similar to the one we have discussed in this paper, but 
based on the {\it non-linear}\/ matter power spectrum.  For example, 
information about the clustering of matter on very small scales
can be obtained 
when light emitted by quasars and other distant astrophysical objects 
is gravitationally lensed by small foreground objects with masses down to $\sim 10^6 M_{\odot}$,
producing arcs and other similar observable features. 

The challenges involved in extracting meaningful information concerning the underlying
dark sector from the non-linear matter power spectrum are not only observational,
but also theoretical and computational.
A detailed analysis of how structure evolves in
the non-linear regime generally requires computationally expensive $N$-body simulations such 
as \texttt{GADGET-2}~\cite{Springel:2005mi}.  That said, less
computationally intensive tools and approximation methods exist which
can provide insight into how dark-matter velocity distributions
affect the power spectrum on small scales.  Approaches along these lines include 
the application of fitting procedures such as \texttt{HALOFIT}~\cite{Smith:2002dz} 
and approximations concerning 
the collapse of dark-matter halos~\cite{Cooray:2002dia}.
While these approaches have been applied to warm-dark-matter scenarios~\cite{Smith:2011ev},
any results obtained from the highly 
non-thermal and multi-modal distributions we have discussed here
would likely be considerably different.

Despite the complications involved in extending our analysis 
to the non-linear regime, there are compelling motivations for studies along these lines.  
It has been suggested~\cite{Schneider:2013wwa} that single-component dark-matter
scenarios with a thermal-like distribution are not capable of
successfully addressing small-scale structure anomalies such as the 
``too-big-to-fail'' problem~\cite{BoylanKolchin:2011dk}
while at the same time satisfying Lyman-$\alpha$ constraints.
This incompatibility is ultimately due to the steepness of the suppression 
in the power spectra associated with thermal-like distributions.
Likewise, standard warm-dark-matter models 
have been proposed as a way to reproduce the observed density profiles 
of dark-matter halos, but suffer from 
the so-called ``Catch-22'' problem~\cite{Maccio:2012qf}. 
By contrast, the highly non-thermal phase-space distributions we 
have studied here may be able to address these issues with more success.
Indeed, the linear power spectra which arise from such distributions often
do not fall as steeply with $k$, primarily because $\widetilde{g}(k)$ 
can span a broader range of scales.  Moreover, even in the simplest such 
scenarios --- \eg, dark matter produced through the decay of a single
unstable species ---  constraints are easier to satisfy and
structure on small scales differs appreciably from the predictions
of thermal dark-matter models~{\mbox{\cite{Kaplinghat:2005sy,Wang:2014ina}}}.

We emphasize that small-scale structure in the theories we have studied
here is also likely to differ in other ways from that expected in theories of thermal warm dark matter.
For example, in thermal warm-dark-matter scenarios, a minimum mass 
{\il{M_{\rm min} \propto k_{\rm FSH}^{-3}}} (ultimately dictated by the 
free-streaming horizon $k_{\rm FSH}$) exists below which dark-matter halos do not form. 
Such a cutoff in the associated halo-mass function is corroborated by $N$-body 
simulations~\cite{Angulo:2013sza}.  By contrast, 
as we have seen,
the free-streaming horizon is generally not 
a reliable indicator of the dark-matter phase-space distribution
for non-minimal dark sectors.


As our conjecture makes clear,
some features of non-minimal dark sectors may be more easily reconstructed than others.
As noted above, our conjecture relates each point in the matter power spectrum to a corresponding
point in the dark-matter phase-space distribution.
Thus, our conjecture gives us information about the
dark-matter phase-space distribution
only for the specific momentum scales which correspond, via Eq.~(\ref{kFSHdef}), to the wavenumber scales
at which we can observe the matter power spectrum.
If, as discussed above, one can eventually extend the linear regime towards even higher values of $k$,
our conjecture may then allow us to reconstruct portions of the dark-matter phase-space distribution
at even lower velocity.
Further details might also be accessible depending 
 on the precision with which such observations can be made and perhaps even
 the extent to which our theoretical conjecture might be refined.
This provides additional motivation for extending the reach of the linear regime.

There are many possible dark-matter scenarios for which our
reconstruction techniques could be useful. 
Of course, in this paper we have demonstrated the utility of these techniques 
within the context of an explicit model presented in Sect.~\ref{sec:toy_model}.~
This model also demonstrated how competing decay chains in particular
yield highly non-thermal dark-matter momentum distributions.
As already noted, this model was chosen for its illustrative power and 
was not meant to describe a UV-complete description of the dark sector. 
However, models with many of the same qualitative features emerge 
naturally within a number of UV-complete scenarios for new physics.  

For example, such features generically arise in
theories involving extra spacetime dimensions.  
Indeed, in scenarios where the SM is localized on a brane within the 
higher-dimensional bulk, the Kaluza-Klein (KK) excitations of bulk fields 
are necessarily neutral under the SM gauge group.  From a four-dimensional 
perspective, such excitations therefore manifest themselves as towers of ``dark'' particles.    
The self-interactions of bulk fields can therefore give rise to intra-ensemble 
decays among the corresponding KK modes, as can interactions between 
bulk fields and fields on the brane~{\mbox{\cite{Dienes:2011ja,Dienes:2011sa,Dienes:2012jb}}}.
In either case, the decay properties of the KK modes are determined both by the 
couplings involved and by the geometry of the extra dimensions.  
For example, in scenarios involving warped extra dimensions, the pattern of 
decays --- and therefore the shape of the resulting dark-matter phase-space distributions --- depends
sensitively on the degree of warping~\cite{Buyukdag:2019lhh}.

Clockwork scenarios~{\mbox{\cite{Choi:2015fiu,Kaplan:2015fuy,Giudice:2016yja}}}
also involve large numbers of dark particles.
The structure of the dark sector in clockwork models effectively
consists of a single light weakly coupled state, along with a compressed mass spectrum of heavier states.
In a manner similar to dimensional deconstruction~\cite{ArkaniHamed:2001ca},
these models can be constructed as discretizations of extra-dimensional 
theories which are similar to 
the linear-dilaton model~{\mbox{\cite{Giudice:2016yja,Craig:2017cda,Giudice:2017suc,Choi:2017ncj,Giudice:2017fmj}}}, 
an approximate holographic dual to little string theory~\cite{Antoniadis:2011qw}.
The interactions among the fields which drive the clockwork mechanism can 
naturally involve intra-ensemble decays among these states.
The particular spectrum of masses and couplings in clockwork models implies 
a pattern of intra-ensemble decays that would differ from those in the other scenarios described above.

There also exist other frameworks in which the dark sector
generally comprises large numbers of dark particles.
For example, such dark particles may emerge as the ``hadrons''  
associated with the confining phase of a strongly-coupled dark sector~\cite{Dienes:2016vei,Buyukdag:2019lhh}.
Likewise, the dark particles may also arise naturally as the gauge-neutral bulk states of Type~I string 
theories~\cite{Dienes:2016vei}.
Such frameworks are likely to have very different phenomenologies from those discussed above,
since the density of dark states in such theories grows {\it exponentially}\/ with mass and the states themselves
lie along linear Regge trajectories.
It would be interesting 
to explore the phenomenology of the intra-ensemble decays and resulting dark-matter phase-space distributions 
that might emerge in such scenarios.

There also exist many possible generalizations and extensions of our work.
One aspect of our work that may be broadly extended is the production mechanism 
for our dark sector.  To a large extent, the production mechanism is dependent 
on how our ensemble couples to external fields.
In this work, we have generally assumed that our ensemble has only gravitational
couplings to the visible sector.
Additionally, within our example model in Sect.~\ref{sec:toy_model},
we assumed that our ensemble has
``top-heavy'' initial conditions --- \ie, conditions
in which the heavier states acquire larger initial energy densities.

Although the latter assumption was not critical for our main results, 
there exist many production mechanisms that yield such initial conditions.
Perhaps the most obvious of these is gravitational production.
For example, dark particles can be generated 
due to the changing spacetime metric at the end of inflation~\cite{Ford:1986sy}.   
Indeed, this process is often invoked as a potential abundance-generation mechanism 
for ultra-heavy dark-matter candidates~{\mbox{\cite{Chung:1998zb,Chung:2001cb}}}.  
Furthermore, as shown in 
\mbox{Refs.~{\mbox{\cite{Bassett:1997az,Ema:2015dka,Markkanen:2015xuw,Ema:2016hlw,Ema:2018ucl,Fairbairn:2018bsw,Markkanen:2018gcw,Ema:2019yrd}}}}, 
oscillations of background quantities induced by inflaton dynamics can also 
contribute to particle production in the early universe.  This also constitutes
an abundance-generation mechanism for dark matter.
In general, depending on the details of the inflationary model and 
how the dark-sector particles interact with one another,
gravitational production can make substantial contributions to the cosmological abundances 
of particles with a potentially broad range of mass scales.  
Moreover, in many cases
the energy density bestowed upon a given dark-sector species in such scenarios generally increases as a function of its mass.  
Thus, we see that gravitational production often furnishes a ``top-heavy'' distribution of energy densities. 
Indeed, gravitational production is particularly appealing within the context of this paper because
it is not predicated on 
the presence of any non-gravitational interactions between the fields of the dark 
and visible sectors.

Beyond gravitational production, our dark sector can also be 
populated through interactions with the inflaton field.  
For example, the initial population could simply be produced through the perturbative 
decay of the inflaton as part of the reheating process.  
As with gravitational production, this production mechanism can
also yield a ``top-heavy'' distribution of energy densities, 
provided that the branching ratios for inflaton decays 
are larger for the more massive states in our ensemble.  
Furthermore, the distribution of dark-sector energy densities 
can also be sensitive to the shape of the inflaton potential during reheating, 
particularly if the curvature of this potential changes significantly over the 
range of field values through which the inflaton oscillates~\cite{lucien}.

Interactions with the inflaton can also populate our ensemble through non-perturbative
processes.  In particular, once the inflaton 
begins oscillating coherently, parametric instabilities develop which lead to the explosive 
production of bosonic dark-sector particles within certain  
momentum bands~{\mbox{\cite{Traschen:1990sw,Kofman:1994rk,Shtanov:1994ce,Kofman:1997yn}}}.
The resulting dark-sector energy densities and momentum distributions depend on a variety 
of factors, including their couplings to the inflaton
and the shape of the inflaton potential.  Interactions among the ensemble constituents 
can have an impact on the outcome as well, since these affect the modulation 
of the effective masses of the dark-sector fields, a possibility
we shall discuss in more detail below.
Additionally, we note that the process of thermalization is generally
complicated in these scenarios --- with phases of non-linear dynamics, turbulence, {\it etc.}\ ---
and these could play an important role in establishing the relative initial
abundances of the different dark-sector states.

Additional production mechanisms are also available if the ensemble couples to 
other external states.  For example, production can occur through the 
freeze-in mechanism~{\mbox{\cite{McDonald:2001vt,Kusenko:2006rh,Petraki:2007gq,Hall:2009bx}}},
which is of particular interest in this regard because the couplings involved need not
be large.  The freeze-in production of our ensemble can occur through interactions with either
visible-sector particles, or particles in some other separate dark sector.
Similarly, depending on the strength of our intra-ensemble couplings, the freeze-in 
and freeze-out mechanisms can also occur through intra-ensemble interactions.  
In particular, these processes could play an important role in the thermalization of our
phase-space distributions at early times.

Another possible extension of our work relates to visible-sector couplings. 
In this paper we have assumed for simplicity that our dark sector is essentially decoupled
from the visible sector.  We therefore assumed that the decays of our dark-sector states do not 
produce SM particles.  While this is certainly a viable possibility, it would also be interesting to
examine the consequences of relaxing this assumption.  For example, as discussed above,
the presence of non-gravitational interactions between these sectors opens 
up new possible dark-matter production mechanisms.
Moreover, such interactions generically give rise to scattering processes involving 
both the dark and visible sectors.  These processes, which bring the
dark-sector states towards kinetic equilibrium, could potentially distort the phase-space 
distributions of these states and thereby wash out the associated imprints in the matter power spectrum.  
However, the presence of these interactions also opens up the possibility that the 
dark sector could be probed through other, complementary means.  
For example, these interactions can introduce dark-matter decay channels involving SM particles 
in the final state.  Not only do such decay channels transfer energy from the 
dark to the visible sector, but they can also have other observational consequences.
Thus, it might be possible to simultaneously observe both a signal of dark-sector 
dynamics within the matter-power spectrum and a complementary 
signal of interactions between the dark and visible sectors. 
It would be interesting to investigate how large these couplings could be 
without erasing information which would otherwise have been imprinted within 
the matter power spectrum.

Another assumption we have made that can potentially be relaxed concerns the lifetimes
of our dark-sector states. In particular, we have focused on the regime in which the lifetimes 
of all unstable dark-sector states are sufficiently short that all of their energy 
density has been transferred to the lightest species by the beginning of the BBN epoch.  
We have made this assumption since the decays of unstable particles at subsequent times ---
even decays solely to other, lighter states within the dark sector --- are constrained 
by their impact of the cosmic expansion rate and its effect on various
\mbox{observables~\cite{
    Gong:2008gi,
    Blackadder:2014wpa,
    Blackadder:2015uta,
    Desai:2019pvs}.}
These observables include the spatial distribution of CMB anisotropies, 
baryon-acoustic-oscillation data, and the relationship between the redshifts and luminosity 
distances of Type~Ia supernovae.
Moreover, if these particles can decay into visible-sector states,
even more stringent bounds apply.  To allow for dark-matter decays
during or after BBN, one must ensure that observational limits on the production
of such states are satisfied.  The corresponding bounds on the lifetime and abundance
of a single decaying particle species were derived in 
\mbox{Refs.~\cite{
    Hu:1992dc,
    Hu:1993gc,
    Kawasaki:1994sc,
    Sarkar:1995dd,
    Adams:1998nr,
    Cyburt:2002uv,
    Chen:2003gz,
    Slatyer:2009yq,
    Cyburt:2009pg,
    Finkbeiner:2011dx,
    Slatyer:2012yq,
    Kawasaki:2017bqm}.}
Likewise, the constraints on decays to 
electromagnetically-interacting final states for an entire ensemble of 
decaying particles have been formulated in a model-independent 
way in Ref.~\cite{Dienes:2018yoq}.

Of course, if one or more of these unstable dark-sector states is extremely long-lived, 
with {\il{\tau_\ell \gtrsim \tnow}}, these states contribute to the present-day dark-matter abundance.  
Such dark-matter scenarios then fall within the purview of the Dynamical 
Dark Matter framework~{\mbox{\cite{Dienes:2011ja,Dienes:2011sa,Dienes:2012jb}}}.
The phase-space distributions associated with all of these dark-matter components 
will generically have non-thermal profiles as a consequence of intra-ensemble
decays.  As a result, any such component which retains a non-negligible energy density 
into the matter-dominated epoch contributes to the growth (or suppression) of matter 
perturbations, rendering the process of structure formation far more complicated.
Furthermore, these dark-matter components presumably continue to decay throughout the epoch 
during which these perturbations become non-linear, thereby altering the dynamics of 
dark-matter halo formation and modifying the resulting spatial mass distribution of 
the halos.

Another assumption that may be relaxed concerns the effects of Bose enhancement or
Pauli blocking.  Throughout this paper, for simplicity we focused exclusively 
on the {\il{f_\ell(p,t)\ll 1}} regime for each of our dark-sector species.
However, there are several reasons to extend our analysis by relaxing this assumption.
For example, in scenarios in which the dark-sector states are scalar fields --- \eg, moduli 
or axion-like particles --- these fields can acquire vacuum 
expectation values (VEVs) which are misaligned from the minimum of the scalar potential.
Indeed, such a misaligned VEV is associated with a zero-momentum condensate, 
severely violating the condition {\il{f_\ell(p,t) \ll 1}}~\cite{Kolb:1990vq}.
A complete description of the scalar ensemble then necessarily involves this condensate, 
in addition to any contributions to $f_\ell(p,t)$ from other production mechanisms.

There are several considerations that can impact the dynamics of the ensemble when 
the criterion {\il{f_\ell(p,t)\ll 1}} does not hold and quantum effects 
come into play.  One of these is that the 
contributions to the collision operator $C[f]$ in Eq.~(\ref{eq:CollisionTermBreakdown}) 
associated with scattering, decay, and inverse-decay processes all involve factors 
of $(1+f_\ell)$ --- factors which cannot be approximated as unity in the presence of a 
condensate.  The impact of these effects on the phenomenology of scalar fields has been 
investigated, for example, in the context of asymmetric 
\mbox{reheating~{\mbox{\cite{Adshead:2016xxj,Hardy:2017wkr}}}}.  Other considerations associated with
the VEVs of dark-sector fields can alter the phenomenology in significant ways.  For example, within the context of the illustrative model 
presented in Sect.~\ref{sec:toy_model}, the presence of a set of VEVs 
$\langle\phi_\ell\rangle$ for the $\phi_\ell$ implies that each dark-sector species 
acquires a field-dependent contribution to its mass as a consequence of the 
trilinear coupling in Eq.~(\ref{eq:InteractionLagrangian}).  In particular, the 
effective mass of each such species in this case is given by 
{\il{(m_\ell^2)_{\rm eff} = m_\ell^2 + 2\sum_{i=0}^N c_{i \ell \ell}\langle\phi_i\rangle}}.

While this is interesting in and of itself, even more interesting is the possibility
that the $\langle\phi_\ell\rangle$ --- and therefore the effective mass of each
dark-sector species --- could be time-dependent.  This occurs naturally if any of the
$\langle\phi_\ell\rangle$ are displaced from the minimum of the scalar potential at early times.
Then, once the Hubble parameter falls to \il{H \sim m_{\ell}}, the corresponding field VEV will begin to 
oscillate coherently.  As a result, parametric instabilities can develop --- instabilities which lead to an enhanced
production of scalars within particular momentum bands.  
Alternatively, a time-dependence for the $(m_{\ell}^2)_{\rm eff}$ can also arise
directly as a consequence of dynamical processes which do not involve modulation of the VEVs of other dark-sector
fields --- \eg, from a cosmological phase transition.  
Even considering only the effect on the VEVs, multiple fields receiving such dynamical mass contributions
exhibit an array of possible \mbox{behaviors~{\mbox{\cite{Dienes:2015bka,Dienes:2016zfr,Dienes:2019chq}.}}}
Furthermore, when the $(m_{\ell}^2)_{\rm eff}$ evolve in this way, the pattern of intra-ensemble decays could
be drastically altered, thereby modifying the ultimate shape of the resulting dark-matter phase-space distributions.


In this paper we have shown that multi-modality of the dark-matter phase-space distribution
can emerge quite generically within a multi-component dark sector when there are overlapping
decay chains with different decay rates.
However, strictly speaking, one does not require 
independent decay chains in order to produce multi-modality.
Indeed, even a single decay pathway can produce a multi-modal distribution
for the daughter if the parent itself experiences production while the decay
is proceeding.   
For example, a grandparent might decay and thereby replenish the parent after the parent
has already decayed.
Alternatively, the parent might be in thermal contact with an external source (such as
might occur if the parent is experiencing thermal freezeout), and thus
its abundance might be 
continually replenished as it decays.
In fact, a model exhibiting the latter phenomenon 
already exists within the context of 
sterile-neutrino dark matter~{\mbox{\cite{Merle:2015oja,Konig:2016dzg}}}.  

In general, as we have discussed in Sect.~\ref{sec:phase_space_evolution}, 
a multi-modal dark-matter phase-space distribution will emerge
whenever there are widely separated dark-matter ``deposits'' onto the cosmological conveyor belt. 
Of course, such deposits need not all be the results of decays from more massive states --- 
any sequence of production episodes separated in time and/or momentum can realize the same
end result.
An example of such a phenomenon can be found in Ref.~\cite{Lovell:2015psz}.

One of the most important results of our paper is our reconstruction
conjecture in Eq.~(\ref{conj3}).
As emphasized above, we regard this conjecture as having two distinct components:
the first is the assertion that the hot fraction function $F(k)$ is connected to the
 {\it slope}\/ of the transfer function
\il{d\log T^2/d\log k},
as indicated in Eq.~(\ref{conj1}),
and the second is the assertion of a particular function $\eta$ which describes this
connection, as indicated implicitly through Eq.~(\ref{eq:FitFImplicit}).
Indeed, these two assertions together yield our final conjecture in Eq.~(\ref{conj3}).
Although our conjecture is remarkably successful in reproducing the salient features of $\widetilde{g}(k)$,
we regard this conjecture as at best purely empirical.
It is therefore possible that one or both aspects of this conjecture 
might be further refined.
For example, it is possible that the hot fraction function $F(k)$ might also carry
a weak dependence on other (higher) derivatives of the transfer function, or on the value of
the transfer function itself.
Likewise, even with the assumption given in Eq.~(\ref{conj1}), it is possible
that the $\eta$-function might carry higher-order corrections beyond those in Eq.~(\ref{conj3}).
Although it is not possible to rigorously invert the mathematical procedure through 
which a given dark-matter phase-space distribution $g(p)$ produces a corresponding matter power spectrum $P(k)$,
it may be possible to trace through such a calculation algebraically to leading order in order to learn
which features of $g(p)$ might dominate the resulting $P(k)$, 
and vice versa.  In this way, one might hope to eventually derive our conjecture analytically, 
along with possible correction terms.

Finally, it is interesting to consider how machine-learning techniques could be 
applied to the archaeological inverse problem of deciphering the properties of an underlying 
dark sector from the matter power spectrum.  Indeed, there has recently been considerable 
interest in how machine-learning techniques, such as the implementation of neural networks,
can be applied to various aspects of early-universe cosmology.  
For example, emulators trained on Einstein-Boltzmann solvers have been used to
generate estimates for observables such as the linear matter power spectrum and the
CMB directly from either standard cosmological 
parameters~{\mbox{\cite{Fendt:2006uh,Fendt:2007uu,Auld:2006pm,Auld:2007qz,Manrique-Yus:2019hqc}}}
or the parameters associated with specific models~\cite{Bae:2019sby}. 
Neural networks have also been used to
eliminate computational bottlenecks involving the most time-intensive or least-parallelizable 
steps in the calculations performed by these solvers~\cite{Albers:2019rzt}.
There are several ways in which machine learning might likewise be applied to
the work  in this paper.  For example, one could potentially employ these techniques in order to learn
solutions to the Boltzmann evolution of our dark-matter phase-space distributions or 
the resulting cosmological perturbations.  

As is abundantly clear from this discussion,
the work we have presented here represents but a first foray in the general
direction of
the archaeological 
reconstruction of the dark sector based on the matter power spectrum.
As such, many avenues remain open for future research.
Some constitute potential refinements or generalizations of the work 
we have presented here, while others extend our results
in a number of new directions.
As further observational data accumulates concerning the properties of the matter power spectrum, 
many different ideas along these lines will be needed in order to
exploit this data in pursuit of our overall archaeological goals. 
We therefore hope that our results can play a significant role in this endeavor.


\begin{acknowledgments}
We would like to thank 
K.~Boddy, D.~Curtin, P.~Draper, A.~Erickcek, L.~Leskovec, S.~McDermott, C.~Miller, 
Y.~Park, M.~Peloso, E.~Rozo, G.~Shiu, T.~Tenkanen, K.~Zurek, and especially J.~Shelton and S.~White for discussions.  
The research activities of KRD, FH, and SS were supported in part by the Department of Energy 
under Grant DE-FG02-13ER41976 / DE-SC0009913,
while those of JK were supported in part by the IBS under project code   IBS-R018-D1
and those of BT were supported in part by the National Science Foundation under Grant PHY-1720430.
The research activities of KRD were also supported in part by
the National Science Foundation through its employee IR/D program.
This work was completed in part at the Aspen Center for Physics, which is supported 
by the National Science Foundation under Grant PHY-1607611. 
KRD and BT would also like to acknowledge the hospitality of the 
Institute for Basic Science in Daejeon, South Korea,
as well as the hospitality of the 
Institut Pascal at the Universit\'e Paris-Saclay, which 
is supported by the IPhT, the P2I and SPU research departments, and
the P2IO Laboratory of Excellence (Programs ANR-11-IDEX-0003-01 Paris-Saclay and ANR-10-LABX-0038).
The opinions and conclusions
expressed herein are those of the authors, and do not represent any funding agencies.
\end{acknowledgments}

\appendix

\section{Boltzmann equations}\label{app:boltzmann}

In this Appendix, we describe the Boltzmann equations which govern the time-evolution 
of the phase-space distributions $f_\ell(p_\ell,t)$ for the dark-sector fields $\phi_\ell$ 
in the model introduced in Sect.~\ref{sec:toy_model}.~       
In general, the Boltzmann equation which governs the evolution of $f_\ell(p_\ell,t)$ for
a given dark-sector species $\phi_\ell$ may be written in the form 
\begin{equation}
  \frac{\partial f_\ell(p_\ell,t)}{\partial t} ~=~ H(t) \,p_\ell\, 
  \frac{\partial f_\ell(p_\ell,t)}{\partial p_\ell}
   + \frac{C[f]}{E_\ell}~,
 \label{eq:BoltzmannEqn}
\end{equation}
where $H(t)$ is the Hubble parameter, where {\il{E_\ell=\sqrt{p_\ell^2+m_\ell^2}}} is the energy 
of a particle of this species with momentum $p_\ell$, and where $C[f]$ is the collision 
operator.  The collision operator for our model can be written as a sum of three terms
\begin{equation}
  C[f] ~=~ C_D^{(-)}[f] + C_D^{(+)}[f] + C_S[f]~.
  \label{eq:CollisionTermBreakdown}
\end{equation} 
The first of these terms represents the contribution 
from decay processes of the form {\il{\phi_\ell \rightarrow \phi_i\phi_j}}, which serve as a sink for 
$\phi_\ell$, along with the corresponding inverse-decay processes.  The second 
represents the contribution from decay processes of the form 
{\il{\phi_i \rightarrow \phi_j\phi_\ell}}, which serve as a source for $\phi_\ell$, along
with the corresponding inverse-decay processes.  The third term represents the 
contribution from {\il{2\rightarrow 2}} scattering processes which involve one or more
particles of species $\phi_\ell$ in the initial state, the final state, or both. 
  
The first term on the right side of Eq.~(\ref{eq:CollisionTermBreakdown}) is given by 
\begin{eqnarray}
  C_D^{(-)}[f] &=& -\sum_{i,j}
    \int d\Pi_id\Pi_j \bigg\{ |\mathcal{M}^\ell_{ij}|^2 \nonumber \\ 
    &&  \times 
    (2\pi)^4\delta^4(p_\ell  - p_i - p_j ) \nonumber \\ 
    && \times  \Big[f_\ell(1 +\! f_i)(1 +\! f_j) - \! f_if_j(1+\! f_\ell)\Big] \bigg\}~,
  \label{eq:CDplusf}~~~
\end{eqnarray}
where the phase-space measure $d\Pi_i$ for the real scalar $\phi_i$ is
given by
\begin{equation}
  d\Pi_i ~=~ \frac{1}{(2\pi)^3} \frac{d^3p_i}{2E_i}~.
\end{equation}     
The squared matrix element $|\mathcal{M}^\ell_{ij}|^2$ for the decay process
{\il{\phi_\ell \rightarrow \phi_i\phi_j}} in our model is simply
\begin{equation}
  |\mathcal{M}^\ell_{ij}|^2 ~=~ \frac{1}{\mathcal{N}_{ij}}c_{\ell i j}^2~,
\end{equation}
where $\mathcal{N}_{ij} = 1 + \delta_{ij}$ is the multiplicity of $\phi_i$ 
in the final state of the decay process.  We note that $|\mathcal{M}^\ell_{ij}|^2$ is 
defined here such that it incorporates the symmetry factor which arises 
for combinations of the indices $i$ and $j$ for which multiple 
identical particles appear in the final state of the decay process (or in
the initial state of the inverse-decay process). 
Likewise, the second term on the right side of Eq.~(\ref{eq:CollisionTermBreakdown}) 
is given by
\begin{eqnarray}
  C_D^{(+)}[f] &=& -\sum_{i,j}\mathcal{N}_{\ell j}
    \int d\Pi_id\Pi_j \bigg\{ |\mathcal{M}^i_{\ell j}|^2 \nonumber \\ 
     &&  \times 
    (2\pi)^4\delta^4(p_i  - p_j - p_\ell ) \nonumber \\ 
    && \times  \Big[f_i(1 +\! f_j)(1 +\! f_\ell) -\! f_jf_\ell(1+\! f_i)\Big] \bigg\}~.~~~
    \nonumber \\
  \label{eq:CDminusf}
\end{eqnarray}
The squared matrix element in this case is
\begin{equation}
  |\mathcal{M}^i_{\ell j}|^2 ~=~ \frac{1}{\mathcal{N}_{\ell j}}c_{i \ell j}^2~,
\end{equation}   
where once again the squared matrix element has been defined such that it incorporates 
the relevant symmetry factor.  
Finally, the third term on the right side of Eq.~(\ref{eq:CollisionTermBreakdown} is given by 
\begin{eqnarray}
  C_S[f] &=& -\sum_{i,j,k}\mathcal{N}_{\ell i}\int d\Pi_id\Pi_jd\Pi_k 
    \bigg\{ |\mathcal{M}^{\ell i}_{jk}|^2 \nonumber \\ 
       &&  \times  
    (2\pi)^4\delta^4(p_\ell + p_i - p_j - p_k)  \nonumber \\ 
        && \times 
    \Big[f_\ell f_i(1 +\! f_j)(1 +\! f_k) 
     -\! f_jf_k(1+\! f_i)(1+\! f_\ell) \Big] \bigg\}~. \nonumber\\
 \label{eq:CSf}
\end{eqnarray}
Provided that the coupling coefficients associated with any quartic terms in the interaction
Lagrangian for the $\phi_\ell$ are sufficiently small that they can be safely neglected,
the squared matrix element for the scattering process 
{\il{\phi_\ell \phi_i \rightarrow \phi_j\phi_k}} is  
\begin{eqnarray}
  |\mathcal{M}^{\ell i}_{jk}|^2 &=& \frac{1}{\mathcal{N}_{\ell i}\mathcal{N}_{jk}}\Bigg|
    \sum_n \bigg[ 
    \frac{c_{\ell i n}c_{n j k}}{(p_\ell + p_i)^2 - m_n^2 + im_n\Gamma_n} \nonumber \\
    &&+ \frac{c_{\ell j n}c_{n i k}}{(p_\ell - p_j)^2 - m_n^2 + im_n\Gamma_n} \nonumber \\
    &&+ \frac{c_{\ell k n}c_{n i j}}{(p_\ell - p_k)^2 - m_n^2 + im_n\Gamma_n} 
    \bigg]\Bigg|^2~,
\end{eqnarray}
where once again the squared matrix element has been defined such that it incorporates 
the relevant symmetry factor.  

The expressions in Eqs.~(\ref{eq:CDplusf}), (\ref{eq:CDminusf}), and~(\ref{eq:CSf})
are completely general and applicable across the full parameter space of our model.  
However, within our region of interest within this parameter space, two conditions are 
satisfied which enable us to simplify these equations considerably.  The first of these 
conditions, which we explicitly enforce throughout our numerical study, is that
{\il{f_\ell(p_\ell,t) \ll 1}} for all dark-sector species $\phi_\ell$ at all times {\il{t \geq t_I}}.   
Within this regime, all Bose-enhancement factors can be neglected in 
Eqs.~(\ref{eq:CDplusf}), (\ref{eq:CDminusf}), and~(\ref{eq:CSf}). These individual
contributions to the collision operator therefore reduce to 
\begin{eqnarray}
  C_D^{(-)}[f] &\approx& -\sum_{i,j}
    \int \frac{d^3p_id^3p_j}{4(2\pi)^2 E_iE_j} \Big[ |\mathcal{M}^\ell_{ij}|^2 
    \nonumber \\ && \times 
    \delta^4(p_\ell  - p_i - p_j ) 
    (f_\ell - \! f_if_j)\Big]  \nonumber \\
  C_D^{(+)}[f] &\approx& -\sum_{i,j}\mathcal{N}_{j\ell}
    \int \frac{d^3p_id^3p_j}{4(2\pi)^2E_iE_j} \Big[ |\mathcal{M}^i_{j\ell}|^2  
    \nonumber \\ && \times 
    \delta^4(p_i  - p_j - p_\ell ) 
    (f_i -\! f_jf_\ell)\Big]  \nonumber \\ 
  C_S[f] &\approx& -\sum_{i,j,k}\mathcal{N}_{i\ell}\int 
    \frac{d^3p_id^3p_jd^3p_k}{8(2\pi)^5E_iE_jE_k} \Big[ 
     |\mathcal{M}^{\ell i}_{jk}|^2 \nonumber \\ && \times  
    \delta^4(p_\ell + p_i - p_j - p_k)  
    (f_\ell f_i 
    -\! f_jf_k) \Big]~.  \nonumber \\
  \label{eq:CDfCSfSimplified}
\end{eqnarray}

The second condition which is satisfied within our parameter-space region of interest
is that the overall scale of the couplings among the dark-sector fields $\phi_\ell$ --- a 
scale set by the value of the parameter $\mu$ --- be sufficiently small that the terms in 
$C[f]$ associated with scattering and inverse-decay processes have a negligible effect 
on the evolution of the phase-space distributions $f_\ell(p_\ell,t)$ for these fields.  
In the regime in which these terms can be neglected, the expressions in 
Eq.~(\ref{eq:CDfCSfSimplified}) reduce to
\begin{eqnarray}
  C_D^{(-)}[f] &\approx& -\sum_{i,j}
    \int \frac{d^3p_id^3p_j}{4(2\pi)^2 E_iE_j} \Big[ |\mathcal{M}^\ell_{ij}|^2 
    \nonumber \\ && \times 
    \delta^4(p_\ell  - p_i - p_j ) f_\ell \Big]   \nonumber \\
  C_D^{(+)}[f] &\approx& -\sum_{i,j}\mathcal{N}_{j\ell}
    \int \frac{d^3p_id^3p_j}{4(2\pi)^2E_iE_j} \Big[|\mathcal{M}^i_{j\ell}|^2  
    \nonumber \\ && \times 
    \delta^4(p_i  - p_j - p_\ell ) f_i \Big]  \nonumber \\ 
  C_S[f] &\approx& 0~.~~~  
  \label{eq:CDfCSfSimplifiedMore}
\end{eqnarray}
Indeed, we have verified numerically that for the parameter choices specified in 
Sect.~\ref{sec:toy_model}, the interaction rates for these processes are all much 
smaller than $H(t)$ for all {\il{t\geq t_I}}. 

One useful property of the expressions in Eq.~(\ref{eq:CDfCSfSimplifiedMore})
is that they are linear in the phase-space densities $f_\ell(p_\ell,t)$.  It therefore
follows that the Boltzmann equations for the $\phi_\ell$ are linear in the $f_\ell(p_\ell,t)$ 
within our parameter-space region of interest as well.  This in turn implies that the effect of 
rescaling the initial normalization $f_9(p_9,t_I)$ by an overall constant factor 
is simply to rescale the normalizations of all of the $f_\ell(p_\ell,t)$ at all subsequent times 
by the same factor --- provided, of course, that the condition {\il{f_\ell(p_\ell,t) \ll 1}} continues to be 
satisfied for all $\phi_\ell$ at all times {\il{t \geq t_I}}.  

Another useful property of the Boltzmann equations which arises within our parameter-space 
region of interest is an invariance under a certain class of transformations involving 
an arbitrary dimensionless scaling parameter $\alpha$.  In particular, for any {\il{\alpha>0}}, 
it can be shown that the Boltzmann equations are invariant under the algebraic replacements
\beq
  \begin{cases}
  m_0 ~\rightarrow~ \alpha m_0 \\ 
  \Delta m ~\rightarrow~ \alpha \Delta m  \\
  \mu ~\rightarrow~ \alpha \mu  \\
  f_\ell(p_\ell,t) ~\rightarrow~ f_\ell\left(p_\ell/\alpha,\alpha t\right)  \\
  t_I ~\rightarrow ~  t_I/\alpha \\ 
  \end{cases}
\label{eq:RescalingTransf}
\eeq
for all of the $\phi_\ell$.  Physically speaking, the transformation listed
in Eq.~(\ref{eq:RescalingTransf}) 
for the phase-space distribution $f_\ell(p_\ell,t)$ represents a uniform shift 
such that the occupation density in phase space at momentum 
$p_\ell$ becomes the occupation density at momentum $\alpha p_\ell$.  

The invariance of the Boltzmann equations under the transformations in 
Eq.~(\ref{eq:RescalingTransf}) has important implications for the phenomenology of
our model.  In particular, the phase-space distribution $f_\ell(p_\ell,t)$ for any of 
our dark-sector fields $\phi_\ell$, expressed as a function of the dimensionless ratio 
$p_\ell/m_0$, 
is identical to the phase-space 
distribution obtained for any other choice of model parameters for which the
values of the dimensionless quantities $\Delta m/m_0$, $\mu/m_0$, and $m_0 t_I$ are 
the same.

\section{Decay from parent to daughter:  An explicit numerical example}\label{app:decay_example}

\begin{table*}[t]
\begin{center}
\begin{tabular}{|| c || c | c || c| c || c | c | c ||}
\hline
\hline
 ~~~& \multicolumn{2}{c||}{Parent} 
 & \multicolumn{2}{c||}{Decay}
 & \multicolumn{3}{c||}{~~Daughter distribution~~}\\ \cline{2-8}
\rule[-0.15 truein]{-0.06 truein}{0.37 truein}
~Case~
 & ~$\displaystyle{\frac{\langle p_{\rm prod}\rangle }{m_P}}$~
 & ~$\displaystyle{\frac{\langle p_{\rm decay}\rangle }{m_P}}$~
 & ~$\displaystyle{\frac{m_D}{m_P}}$~
 & ~$\displaystyle{\frac{\langle p_D^{\rm rest}\rangle }{m_P}}$~
 & ~$\displaystyle{\frac{\langle p_D\rangle}{m_D}}$~
 & ~$\displaystyle{\frac{\Delta p_D}{m_D}}$~ 
 & ~$\displaystyle{\frac{\Delta p_D}{\langle p_D\rangle}}$ ~\\ 
\hline
\hline
\rule[0 truein ]{-0.06 truein}{0.14 truein}
~A~  &{\il{  ~~4.6\times 10^{-5}~~ }}&$  10^{-6} $&  \multirow{4}{*}{~~$\half-10^{-6}$~~} & \multirow{4}{*}{$10^{-3}$}   &{\il{ ~~6.4\times 10^{-4}~~  }}&{\il{ ~~3.3\times 10^{-4}~~   }}&$  0.52           $\\ 
~B~  &{\il{  4.5\times 10^{-2} }}&$  10^{-3} $&$                                 $&$                              $&{\il{  7.2\times 10^{-4} }}&{\il{ 3.3\times 10^{-4}   }}&$ 0.46            $\\ 
~D~  &$  53 $&$  1       $&$                                 $&$                            $&$ 0.38  $&{\il{ 6.7\times 10^{-4}   }}&{\il{  1.8\times 10^{-3}             }}\\
~H~  &{\il{  1.4\times 10^{+6} }}&$  10^{+3} $&$                                 $&$                              $&$   405 $&$   0.55 $&{\il{  1.4\times 10^{-3}              }}\\  
\hline
\rule[0 truein ]{-0.06 truein}{0.14 truein}
~C~  &{\il{ 4.6\times 10^{-2}  }}&$  10^{-3} $&  \multirow{3}{*}{$0.4333$}   & \multirow{3}{*}{$1/4$}   &$ 0.19  $&$ 0.1  $&$  0.52             $\\
~E~  &$  12 $&$  1/4     $&$                            $&$                        $&$ 0.22  $&$ 0.1   $&$  0.45             $\\
~K1~ &{\il{  1.4\times 10^{+6} }}&$  10^{+3} $&$                            $&$                        $&$  472 $&$  136   $&$  0.29             $\\
\hline 
\rule[0 truein ]{-0.06 truein}{0.14 truein}
~F~  &{\il{  4.5\times 10^{-5} }}&$  10^{-6} $&  \multirow{5}{*}{$10^{-4}$}  & \multirow{5}{*}{$1/2$}   &{\il{ 1.6\times 10^{+3}  }}&$ 832   $&$   0.52             $\\
~G~  &{\il{  4.7\times 10^{-3} }}&$  10^{-4} $&$                            $&$                        $&{\il{  1.6\times 10^{+3} }}&$ 831   $&$  0.52             $\\
~I~  &$  0.47 $&$  10^{-2} $&$                            $&$                        $&{\il{  1.5\times 10^{+3} }}&$  803   $&$  0.52             $\\
~J~  &$  25 $&$  1/2     $&$                            $&$                        $&{\il{  2.0\times 10^{+3} }}&$  848    $&$  0.43            $\\
~K2~ &{\il{  1.5\times 10^{+6} }}&$  10^{+3} $&$                            $&$                        $&{\il{  2.1\times 10^{+6} }}&{\il{ 1.2\times 10^{+6}   }}&$  0.58             $\\
\hline
\hline 
\end{tabular}
\end{center}
\caption{An explicit numerical example of the decay of a parent packet into a daughter packet, shown for a variety of
cases corresponding to the cases in Table~\ref{inversepacket}.~     
In obtaining this data we have assumed a parent of mass $m_P$ undergoing a two-body decay into identical daughters
of mass $m_D$, and we have taken {\il{\Gamma/H(t_0)=10^{-3}}} and {\il{\kappa=3/2}}, where $t_0$ is the time at which
the parent is originally produced.
The data in this table was generated through a full numerical Boltzmann analysis including all relativistic, redshifting,
and exponential-decay effects.   
The different cases in this table are labeled according to the same labeling scheme as in Table~\ref{inversepacket}, with
K1 and K2 corresponding to the two possibilities for Case~K within Table~\ref{inversepacket} with
absolute marginalities which are either $\calO(1)$ or far, respectively. 
In all cases we find that the results of this explicit example
agree with the general properties outlined in Table~\ref{inversepacket}.}  
\label{Boltzmann_data}
\end{table*} 

In Table~\ref{inversepacket}, we described certain general properties of the daughter packets which
result from relatively narrow parent packets undergoing two-body decays into identical daughters.
In this Appendix, we provide an explicit set of numerical examples 
which may further elucidate these general results.

These examples are shown in Table~\ref{Boltzmann_data}.~
In particular, the data in this table was generated through a full numerical Boltzmann analysis including all relativistic, redshifting,
and exponential-decay effects.
For this analysis, we have assumed {\il{\Gamma/H(t_0)= 10^{-3}}} where $t_0$ is the time at which the parent packet is produced,
and we have taken {\il{\kappa=3/2}}.
Within this table, $m_P$ and $m_D$ are respectively the parent and daughter masses;
$\langle p_{\rm prod}\rangle$ 
is the average
momentum of the parent packet at production;
and $\langle p_{\rm decay}\rangle$ 
is the average momentum of the parent packet when the proper time elapsed since production reaches {\il{\tau\equiv \Gamma^{-1}}}.  
Likewise $p_D^{\rm rest}$ is the momentum of the daughter in the rest frame of the parent, a quantity which depends
on the masses alone and which is given in this case by {\il{p_D^{\rm rest}=\sqrt{(m_P/2)^2 - m_D^2}}}. 
Each different case within Table~\ref{Boltzmann_data} can be specified by choosing values of $m_P$, $m_D$, and either $p_{\rm prod}$ or $p_{\rm decay}$ (we shall select $p_{\rm decay}$ for this purpose);
the remaining entries in this table are then calculated accordingly.

Note that we have chosen the different cases within Table~\ref{Boltzmann_data} in such a way as to correspond to the 
different cases in Table~\ref{inversepacket}.~ 
These different cases are therefore listed in Table~\ref{Boltzmann_data} using the same labeling scheme as 
in Table~\ref{inversepacket}, 
with K1 and K2 in Table~\ref{Boltzmann_data} corresponding to the two possibilities for Case~K within Table~\ref{inversepacket} with
absolute marginalities which are either $\calO(1)$ or far, respectively. 
In this connection, recall that the {\it absolute}\/ marginality in Table~\ref{inversepacket} is determined by the ratio $p_D^{\rm rest}/m_P$, while
the {\it relative}\/ marginality in Table~\ref{inversepacket} is determined by the ratio $p_D^{\rm rest}/p_{\rm decay}$.  
The different cases shown in Table~\ref{Boltzmann_data} therefore correspond to situations
in which 
{\il{p_D^{\rm rest} \ll p_{\rm decay}}},
{\il{p_D^{\rm rest} \sim p_{\rm decay}}},
or
{\il{p_D^{\rm rest} \gg p_{\rm decay}}}.
Likewise, for each choice, we have then considered 
only those choices for which 
{\il{p_{\rm decay}\ll m_P}},
{\il{p_{\rm decay}\sim m_P}},
or 
{\il{p_{\rm decay}\gg m_P}}.
As a result, Cases~A, B, D, and H are near absolute marginality (with {\il{p_D^{\rm rest}/m_P \ll 1}}), while 
Cases~F, G, I, J, and K2 are far from absolute marginality (with $p_D^{\rm rest}/m_P$ approaching its maximum kinematically-allowed value, which
in this case is $1/2$).
Likewise, Cases~D, H, K1, and K2 are near relative marginality (with {\il{p_D^{\rm rest} \ll p_{\rm decay})}},
while Cases~A, C, F, G, and I are far from relative marginality (with {\il{p_D^{\rm rest}\gg p_{\rm decay})}}.
Indeed, Case~F is the farthest from relative marginality, as already anticipated below Table~\ref{inversepacket}.

It is clear from Table~\ref{Boltzmann_data}
that our numerical results conform 
quite well to our general expectations in Table~\ref{inversepacket}.~
In making this assessment,
we note that the relative hierarchies 
between our input parameters in Table~\ref{Boltzmann_data} have usually been chosen to be approximately $10^{3}$ or larger.
We therefore use
this same hierarchical scale when deciding, for example,
whether a given daughter width $\Delta p_D$ is to be considered
much smaller than, of the same order of magnitude as, or  
much larger than $m_D$ or $\langle p_D\rangle$.
The agreement between these two tables thus provides further confirmation
of the results in Table~\ref{inversepacket}, with 
Table~\ref{Boltzmann_data}
serving as an explicit example
of the physics underlying Table~\ref{inversepacket}.

\FloatBarrier
\section{Adiabatic sound speed}\label{app:sound_speed}

As background for a technical point discussed in Sect.~\ref{sec:perturb},
in this Appendix 
we provide a short derivation of the adiabatic sound speed $c_s$ associated with dark matter of a given momentum $p$.

In general, $c_s$ describes the response of the pressure $P$ due to a change in the energy density $\rho$. 
Any pressure perturbation $\delta P$ can be written in terms of an entropy perturbation $\delta s$ and an energy-density
perturbation $\delta \rho$ as
\beqn
  \delta P &=& \left.\frac{\partial P}{\partial s} \right\vert_{\rho} \delta s 
                 + \left.\frac{\partial P}{\partial \rho} \right\vert_s \delta\rho \nn\\
           &=&\delta P_{\rm nad} + c_s^2 \delta\rho~,
\eeqn
where $P_{\rm nad}$ is a non-adiabatic contribution which will not concern us and where 
\beq
       c_s^2~\equiv~\left.\frac{\partial P}{\partial \rho} \right\vert_s~.
\eeq
As such, the adiabatic sound speed depends only on background quantities, and 
these do not vary in space but only in time.
We can therefore re-express $c_s$ in the form
\beq
    c_s^2~=~\dot{\bar{P}}/\dot{\bar{\rho}}~
\eeq
where $\bar{\rho}$ and $\bar{P}$ are average background values.

Some dark sectors have dark matter exhibiting relatively simple phase-space distributions $f(p)$.
In such cases, as discussed in Sect.~\ref{sec:perturb},
one might consider calculating $c_s$ 
by averaging across all momenta in order to
consider the time variations of a momentum-averaged pressure and a momentum-averaged energy density.
This would then hopefully provide a characteristic value of the associated sound speed. 
In this paper, by contrast, we are interested in situations in which the dark-matter phase-space
distribution is relatively complex and potentially even multi-modal.   In such cases,
a momentum average might
then fail to capture all of the relevant information. 

For this reason, we shall proceed by viewing each momentum slice through the $f(p)$ distribution
as its own effective ``species'' having its own sound speed $c_s(p)$.
Indeed,
as we shall demonstrate, the sound speed varies non-trivially with $p$, thereby justifying this approach.
In order to calculate $c_s(p)$, let us assume that $n_p$ is the number density of the dark-matter
particles with momentum $p$.
We then have
\beq
   \rho_p = \bar{\rho}_p = n_p E_p~, ~~~~ P_p = \bar{P}_p= n_p \frac{p^2}{3E_p}~,
\eeq
where {\il{E_p\equiv \sqrt{p^2+m^2}}}.  Given that {\il{n_p\sim a^{-3}}} and {\il{p\sim a^{-1}}},
we find that {\il{dE_p/da= -p^2/(E_pa)}}, whereupon it follows that 
\beq
    \frac{d\rho_p}{da}= - \frac{\rho_p}{a} \left( 3 + \frac{p^2}{E_p^2}\right)~,~~~~ 
    \frac{d P_p}{da}= - \frac{P_p}{a} \left( 5 - \frac{p^2}{E_p^2}\right)~.
\eeq
We therefore find that the sound speed
is given by
\beq
    c_s^2(p) ~=~ \frac{ dP_p/da}{d\rho_p/da} ~=~ {1\over 3}{p^2\over E_p^2} \left( { 5-p^2/E_p^2\over 3+ p^2/E_p^2} \right)~.
\eeq
From this result we easily verify that {\il{c_s\to 0}} for cold (non-relativistic) matter with {\il{p/E_p\to 0}},
while {\il{c_s\to  1/\sqrt{3}}} for highly relativistic matter (or radiation) with {\il{p/E_p\to 1}}.
This accords with our usual expectation that {\il{c_s=\sqrt{w}}} for matter with equation-of-state parameter $w$.

\section{Time-evolution of the dark sector \label{app:time_evol} }

In Sect.~\ref{sec:toy_model}, 
we introduced an illustrative model involving a collection of decaying fields $\phi_\ell$ and focused on the 
properties of the resulting present-day phase-space distribution $g_0(p,\tnow)$ associated with
the stable, lightest constituent $\phi_0$.
We also analyzed the impact of this phase-space distribution on the matter power spectrum $P(k)$.
While the properties of $g_0(p,\tnow)$ are an important focus of this paper,
it is nevertheless
also interesting to examine how the properties of the dark sector 
evolve {\it while}\/ the individual constituent fields within the dark sector are actively evolving and decaying.
These properties include not only 
the phase-space 
distributions and energy densities of the individual $\phi_\ell$, but also the aggregate 
equation of state for the dark sector as a whole.

These quantities are defined and calculated as follows.
Given the particular decay patterns that arise in our model,
solving the associated Boltzmann equations provides us with 
the phase-space distribution $f_\ell(p,t)$ for each species
$\phi_\ell$ as the decays proceed. 
Using
the expressions in Eq.~(\ref{fdefs})   
it is then straightforward to calculate the
corresponding number densities $n_{\ell}(t)$, energy densities $\rho_{\ell}(t)$, and 
pressures $P_{\ell}(t)$ which are also functions of time as the decays proceed.
Likewise, 
the total number density $n_{\rm tot}(t)$, total energy density $\rho_{\rm tot}(t)$, and 
total pressure $P_{\rm tot}(t)$ of the dark sector as a whole are simply sums of these
individual contributions:
\beq
  X_{\rm tot}(t)~\equiv~ \sum\limits_{\ell=0}^{N} X_\ell(t)~~~~{\rm for}~ X= n, \rho, P~.
\label{eq:ntotrhototPtot}
\eeq

We shall also  
define the individual
species equation-of-state parameters {\il{w_\ell(t) \equiv P_\ell(t)/\rho_\ell(t)}}
as well as an aggregate equation-of-state parameter {\il{w_{\rm tot}(t) \equiv P_{\rm tot}(t)/\rho_{\rm tot}(t)}} for the dark
sector as a whole.  
Note that these quantities are time-dependent as a result of the decays which occur within the dark sector.
It is also important to note that unlike
the aggregate quantities appearing in 
Eq.~(\ref{eq:ntotrhototPtot}), 
$w_{\rm tot}(t)$ is actually a {\it weighted}\/ sum of the individual $w_\ell(t)$ contributions:
\begin{equation}
  w_{\rm tot}(t) ~\equiv~ \frac{P_{\rm tot}(t)}{\rho_{\rm tot}(t)} 
    ~=~ \sum_{\ell=0}^{N}\left[\frac{\rho_\ell(t)}{\rho_{\rm tot}(t)}\right]w_\ell(t)~,
  \label{eq:EoS_tot}
\end{equation}
where the energy-density ratios
$\rho_\ell(t)/\rho_{\rm tot}(t)$ 
serve as the appropriate weighting factors.
Finally, we observe that $w_{\rm tot}(t)$ 
may  equivalently be expressed as~\cite{Dienes:2011ja}
\begin{equation}
  w_{\rm tot}(t) ~=~ 
    -\left(\frac{1}{3H}\frac{d\log\rho_{\rm tot}}{dt} + 1\right)~.
  \label{eq:wtot_energy}
\end{equation}
In general, a relation of this sort which directly connects $w(t)$ to $\rho(t)$ 
follows from the fundamental definition {\il{w(t)\equiv P(t)/\rho(t)}} 
under the additional assumption that the only changes to the total energy density 
$\rho(t)$ are those due to Hubble expansion, so that {\il{\delta \rho = P \delta V}}
where {\il{V\sim a^3}}.
In other words, the expression in Eq.~(\ref{eq:wtot_energy}) assumes that our dark-matter ensemble 
couples negligibly to any other non-gravitational sector.
However, the total energy density of our dark sector as a whole
is unaffected by decays that occur purely within the dark sector (such as we are considering in our model).
We therefore obtain Eq.~(\ref{eq:wtot_energy}) for $w_{\rm tot}$. 
By contrast, a similar equation would not hold for each $w_\ell$ separately
during any time period in which $\phi_\ell$ particles are being created or lost through decays.

We begin our analysis of the time-evolution of the dark sector in our model 
by examining how the energy densities $\rho_\ell(t)$ evolve in time.   
For reasons to be discussed below, we shall consider the time-dependence of 
the quantities $a^3 \rho_\ell$ rather than $\rho_\ell$ alone.
Note that there are generally three different effects that can cause
a given $a^3\rho_\ell$ to vary with time.   First, this quantity can increase if $\phi_\ell$ particles
are created via the decays of heavier ensemble constituents.
Second, this quantity can decrease if the $\phi_\ell$ particles themselves decay
into lighter ensemble constituents.
Finally, however, $a^3 \rho_\ell$ can also scale non-trivially with Hubble expansion.
In the absence of $\phi_\ell$ production or decay, $a^3\rho_\ell$ would be constant
only if the $\phi_\ell$ particles were all non-relativistic.    Indeed, in this case
we could even interpret $a^3\rho_\ell$ as a {\it comoving}\/ energy density.
By contrast, if the $\phi_\ell$ particles 
carry a significant momentum, this momentum will experience
a gravitational redshift towards smaller values, and this will also cause $a^3 \rho_\ell$ to drop
as a function of time.
For example, if $\phi_\ell$ is highly relativistic (\eg, radiation), we know that
$\phi_\ell$ scales as $a^{-4}$ rather than $a^{-3}$.
Thus $a^3\phi_\ell$ will scale as $a^{-1}$, even if no $\phi_\ell$ particles are being
created or lost through decay.

This last effect also applies to the total energy density $a^3\rho_{\rm tot}$.
Because each decay process within the dark sector necessarily conserves energy,
$a^3 \rho_{\rm tot}$ is unaffected by decays within the dark sector.
However, Hubble expansion continues its inexorable degradation of the overall
kinetic energy associated the dark sector. 
As a result, if our dark sector consists of highly relativistic matter,
its energy density $a^3 \rho_{\rm tot}$ 
will initially scale as $a^{-1}$.
By contrast, at late times,
$a^3 \rho_{\rm tot}$ will become effectively constant as our dark sector 
becomes increasingly cold.
Thus, a significant variation of $a^3\rho_{\rm tot}$ with time
indicates that
a significant fraction of the total dark-sector energy density
is still carried by relativistic particles.


\begin{figure*}
\includegraphics[width=.99\textwidth,keepaspectratio]{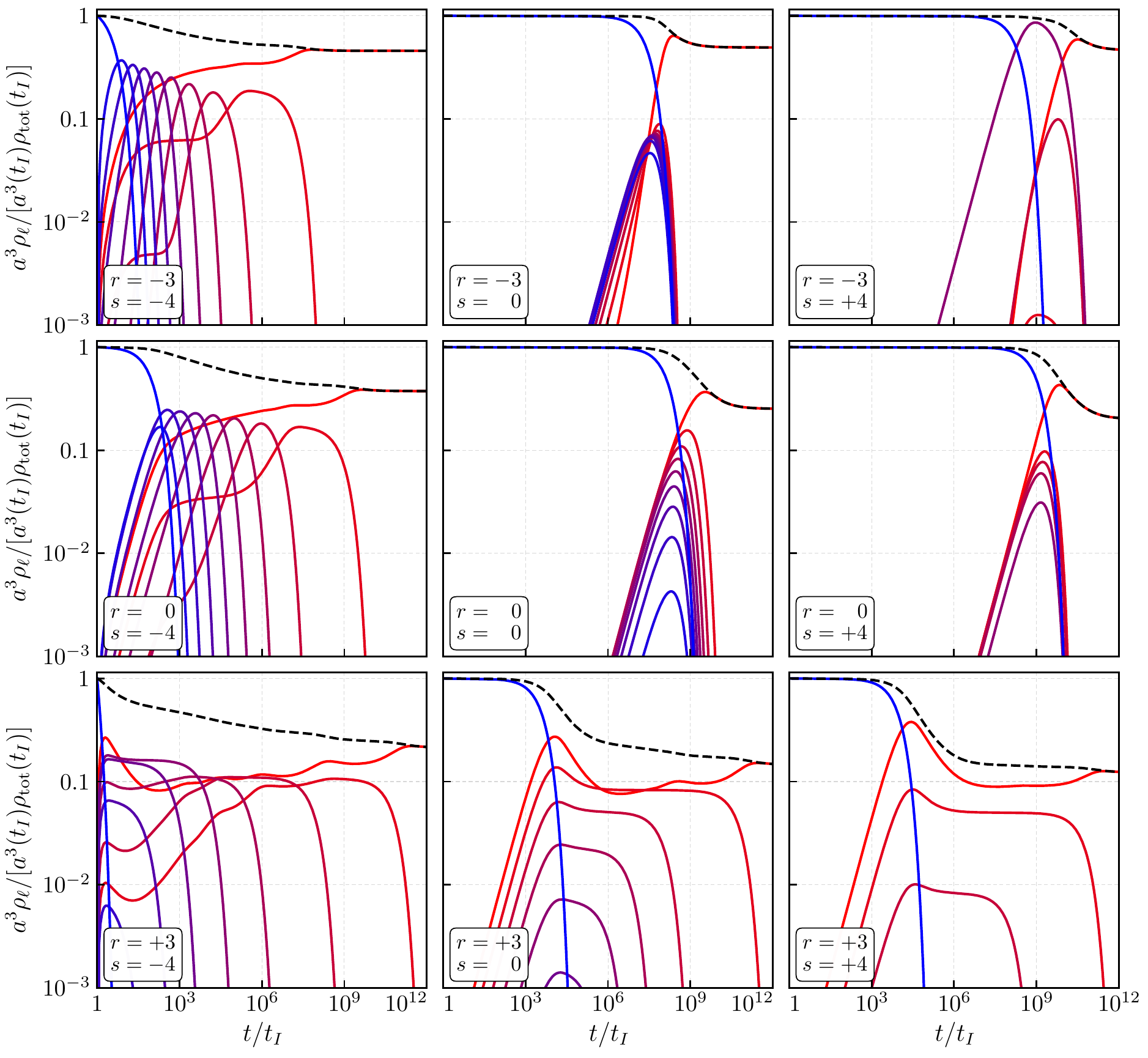}\centering
\caption{The individual constituent energy densities $a^3 \rho_\ell$ 
for {\il{\ell=0,...,9}} (solid colored curves ranging from red to blue, respectively),
each normalized to the total energy density
$a^3 \rho_{\rm tot}$ evaluated at the initial time {\il{t=t_I}}
and plotted as a function of 
the dimensionless ratio $t/t_I$.
The total energy density $a^3 \rho_{\rm tot}$, similarly normalized 
to its initial value, is also plotted (black dashed curve).}
\label{fig:density}
\end{figure*}

In Fig.~\ref{fig:density} we plot the time-evolution of $a^3\rho_\ell$ for each species $\phi_\ell$ in our model, where
the different panels in this figure correspond to the same choices for the parameters $r$ and $s$ as in 
Figs.~\ref{fig:decay_width_channels} through \ref{fig:Fk_fit}.~
Note that the ten solid curves 
(ranging from red to blue)
shown within each panel correspond to the individual $a^3 \rho_\ell$ for 
{\il{\ell=0,1,...,9}}, respectively,
with each normalized to 
the total $a^3\rho_{\rm tot}$ evaluated at the initial time {\il{t=t_I}}.
Likewise, the black dashed curve shows the evolution of
the total energy density
$a^3\rho_{\rm tot}$, similarly normalized to its initial value.
 
It is straightforward to understand the curves shown in these panels.
Let us first consider the 
regime in which {\il{r \leq 0}} and {\il{s \geq 0}} --- \ie, the four panels in 
the upper right corner of Fig.~\ref{fig:density}.~
As indicated in Fig.~\ref{fig:decay_chain}, the 
characteristic timescales associated with all relevant decay chains are quite similar
in this regime.  As a result, 
even if 
the intermediate states in these decay chains acquire significant 
energy densities --- as indeed occurs in most of these figure panels --- 
these individual energies $\rho_\ell$ are ultimately transferred to $\phi_0$ 
in each case on essentially the same timescale. 
Moreover, within this regime, decays in which a significant amount of kinetic energy is 
transferred to the daughter particles are not preferred --- and for {\il{r < 0}}
these decays are actively suppressed.  As a result, the fraction of 
the initial energy density $\rho_9(t_I)$
which is 
eventually dissipated by the redshifting of daughter particles
is small compared to the fraction which is converted into the mass energy of $\phi_0$ particles.
This implies that the overall decrease in $a^3\rho_{\rm tot}$ is comparatively
small within this regime, and markedly smaller for {\il{r = -3}} than for {\il{r = 0}}.               
 
By contrast, in the {\il{s < 0}} and {\il{r \leq 0}} regime, the preference for decays which are highly 
asymmetric but yet not highly exothermic leads to a decay-chain structure involving 
multiple subsequent steps of the form {\il{\phi_\ell \rightarrow \phi_i\phi_j}}, each of which 
yields one very light particle $\phi_i$ and another particle $\phi_j$ with $j$ only slightly 
below $\ell$.  In this regime, the fraction of the initial mass energy of $\phi_9$ particles 
converted to kinetic energy along any given decay chain is likewise small --- and likewise 
markedly smaller for {\il{r = -3}} than for {\il{r = 0}}.  However, the timescales on which the
individual intermediate $\phi_\ell$ 
acquire significant energy densities are also quite different in this regime. 
In particular, 
successively lighter states acquire a significant $\rho_\ell$ at successively later 
times 
due to an injection of energy density from the decays of the states just above them.  
The fact that a significant population of $\phi_0$ particles is generated at each step along 
these decay chains gives rise to the complex multi-modal phase-distributions which appear
in the corresponding panels of Fig.~\ref{fig:f0_all}.

Finally, in the {\il{r > 0}} regime, the preference for decays which are highly 
exothermic leads to a decay-chain structure in which $\phi_9$ particles decay directly to
final states comprising much lighter $\phi_\ell$.  Since a considerable fraction of the initial 
mass energy of $\phi_9$ particles is converted into kinetic energy in the process, the 
decrease in $a^3\rho_{\rm tot}$ is comparatively large within this regime.  
Moreover, the timescales on which $\phi_0$ particles are produced from the direct decays
of $\phi_9$ are significantly shorter than the timescales associated with the decays of
other light $\phi_\ell$ particles which might also be produced from the decays of $\phi_9$ 
down to $\phi_0$.  Thus, as we see in the panels along the bottom row of Fig.~\ref{fig:density}, 
the timescales on which the energy densities $\rho_\ell$ associated with these other light 
species are transferred to $\phi_0$ are quite long.  The fact that one population of $\phi_0$
particles is generated from the direct decays of $\phi_9$ while another population of $\phi_0$ particles
is produced
from the decays of these lighter longer-lived states is ultimately responsible for the
multi-modal structure of the phase-space distributions appearing in the corresponding
panels of Fig.~\ref{fig:f0_all}.

\begin{figure*}
\includegraphics[width=.99\textwidth]{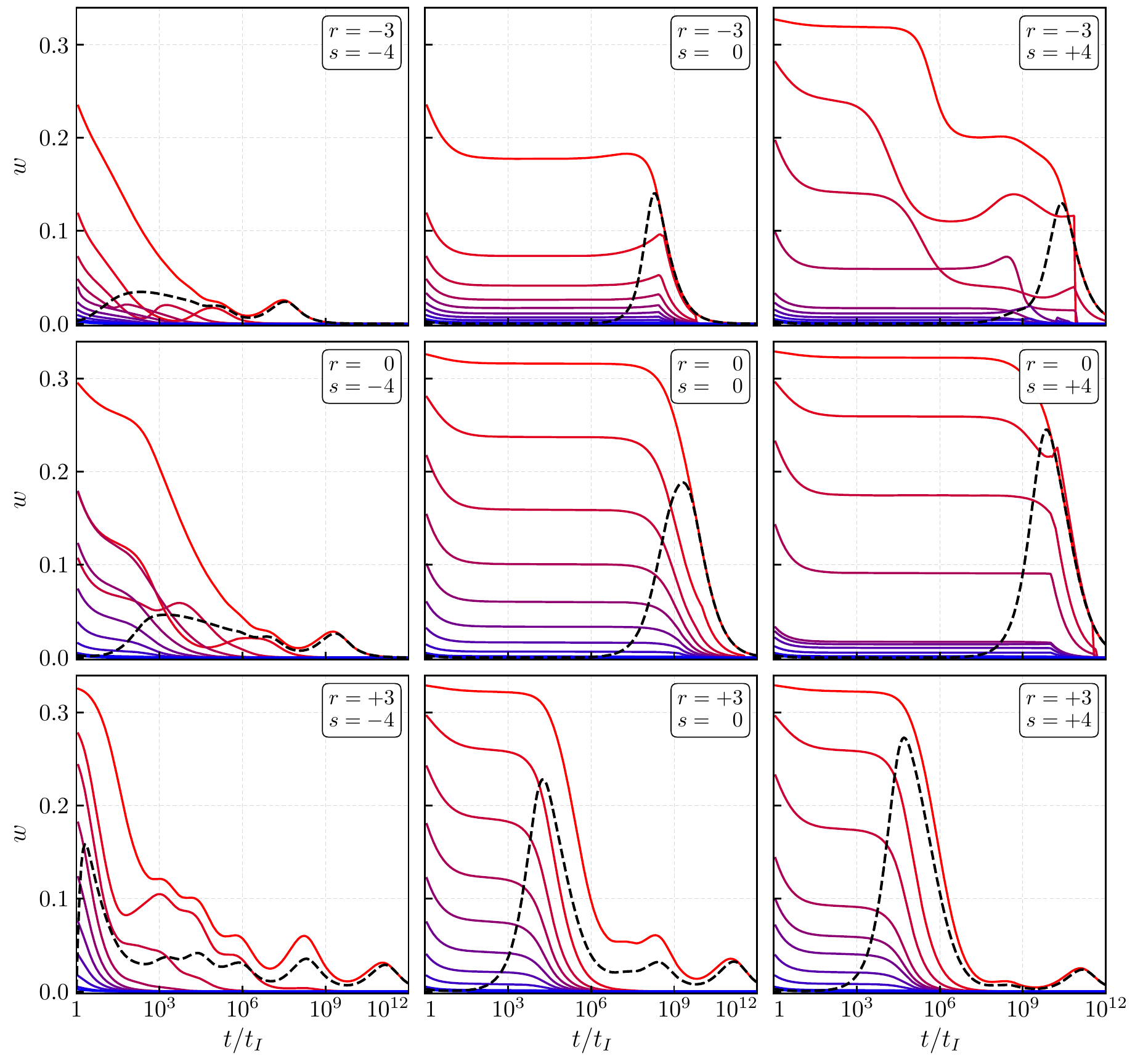}\centering
\caption{The equation-of-state parameters $w_\ell(t)$ for each dark-sector
species $\phi_\ell$, 
plotted as functions of the dimensionless ratio $t/t_I$ for {\il{\ell=0,...,9}} (red to blue, respectively).
By contrast, the black dashed curve represents the aggregate equation-of-state parameter $w_{\rm tot}(t)$ 
of the dark sector as a  whole.}
\label{fig:w_all}
\end{figure*}

A similar analysis can also be performed for our time-dependent
equation-of-state parameters $w_\ell(t)$ and $w_{\rm tot}(t)$, 
as defined in and above Eqs.~(\ref{eq:EoS_tot}) and (\ref{eq:wtot_energy}).
In Fig.~\ref{fig:w_all}, we display $w_\ell(t)$ for 
{\il{\ell=0,1,...,9}} 
as functions of the ratio $t/t_I$, with the same color and panel configurations
as in Fig.~\ref{fig:density}.~
We also display the 
aggregate equation-of-state parameter $w_{\rm tot}(t)$ for the dark sector as a whole
(dashed black curve).

Once again, just as with Fig.~\ref{fig:density}, 
the behavior of each individual equation-of-state parameter $w_\ell(t)$ ultimately reflects the 
populating and depopulating of the $\phi_\ell$ state as the decay process unfolds.
Moreover, the relation in Eq.~(\ref{eq:wtot_energy}) implies that we may view $w_{\rm tot}(t)$ at any given time as measure of how 
rapidly $\rho_{\rm tot}(t)$ is changing with time.  Indeed, we observe that 
when $w_{\rm tot}(t)$ differs significantly from zero in any 
given panel of Fig.~\ref{fig:w_all},  
the slope of the black dashed curve in the corresponding panel of Fig.~\ref{fig:density} 
also differs significantly from zero.  Combinations of $r$ and $s$ for which there exist multiple 
decay pathways producing $\phi_0$ particles on significantly different characteristic 
timescales therefore typically give rise to $w_{\rm tot}(t)$ curves with multiple peaks.
Thus, if the $w_{\rm tot}(t)$ curve exhibits such non-monotonicities as a function of time,
there is a good chance that the corresponding 
$g_0(p,\tnow)$ will turn out to be multi-modal as a function of $p$.

\begin{figure*}
\includegraphics[width=.99\textwidth]{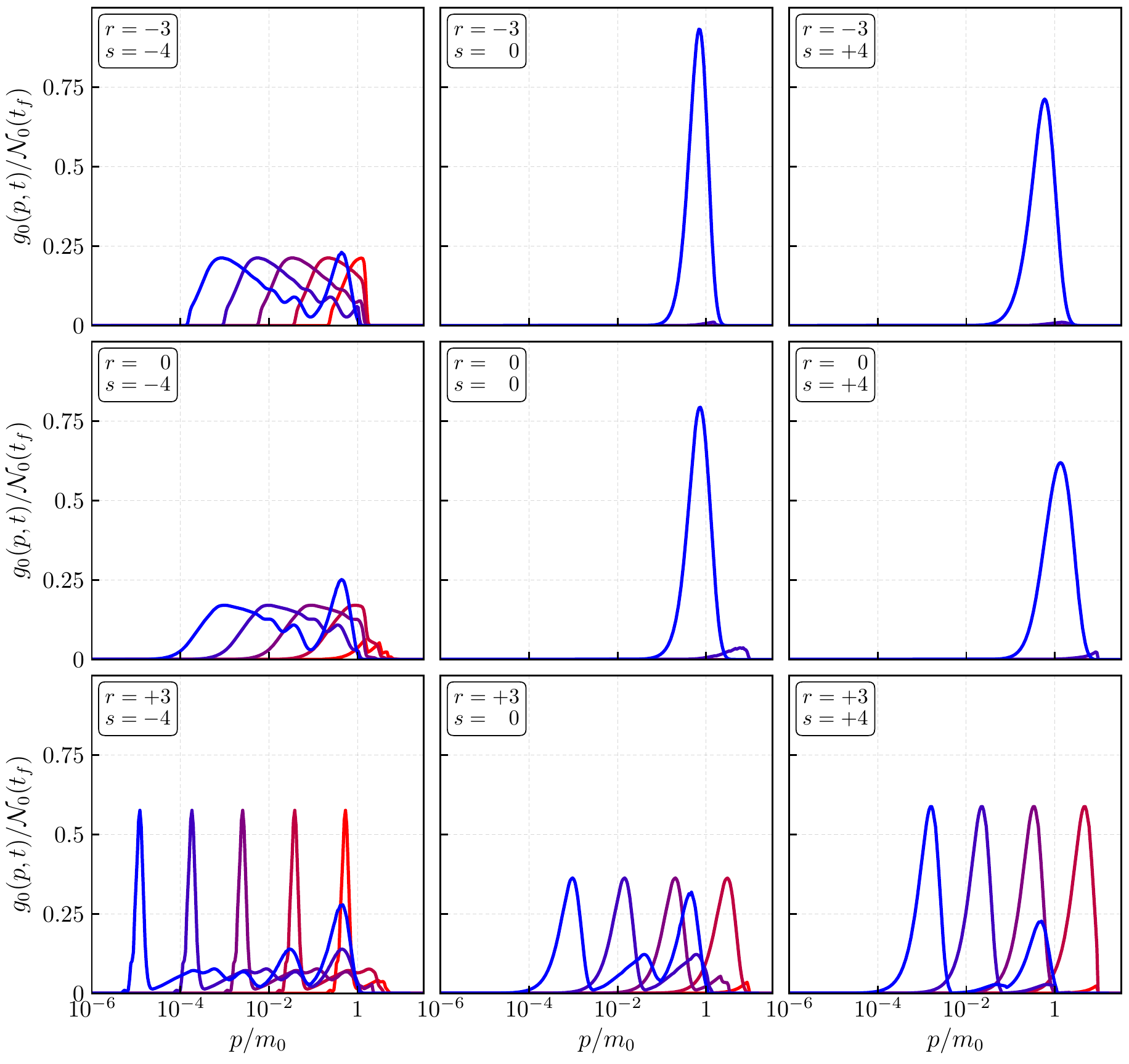}\centering
\caption{The evolution of the ground-state phase-space distribution $g_0(p,t)$ as the decay process unfolds.
For each value of $r$ and $s$ in a given panel,
we show $g_0(p,t)/\calN_0$ at five different ``snapshot'' times 
(ranging from red to blue)
which are
evenly spaced on a logarithmic scale between the initial time $t_I$ and the final
time $t_f$ at which $g_0(p,t)$ reaches $99.9\%$ of its late-time 
asymptotic value.  
Note that for {\il{r\leq 0}} and {\il{s\geq 0}},  
there are no deposits into the ground state until relatively late times.
Consequently fewer phase-space distributions are shown.}
\label{fig:f0_top}
\end{figure*}

Finally, in Fig.~\ref{fig:f0_top}, we illustrate how the phase-space distribution $g_0(p,t)$ 
of the lightest state in the dark sector evolves in time.  The curves shown in each panel
represent ``snapshots'' of $g_0(p,t)$ at five different values of $t$.  For each combination
of $r$ and $s$, 
these snapshot times are determined by dividing the time interval between 
between $t_I$ and the time $t_f$ at which the comoving number density of $\phi_0$ reaches 
$99.9\%$ of its late-time asymptotic value into five time intervals which are evenly spaced 
on a logarithmic scale.  
The red curve in each panel corresponds to the 
earliest snapshot, while the blue curve corresponds 
final snapshot 
when essentially the entire dark-sector abundance has settled into the ground state.
However, we emphasize that since {\il{\tau_9 \gg t_I}} for many of the combinations of $r$ and $s$
shown --- in particular, for those combinations with {\il{r \leq 0}} and {\il{s \geq 0}} --- a
non-negligible population of $\phi_0$ particles is not generated until late times.  Thus, 
in the corresponding panels of Fig.~\ref{fig:f0_top}, the only curve which deviates 
considerably from {\il{g_0(p,t) \approx 0}} is the one corresponding to {\il{t = t_f}}.    

We have seen in Sect.~\ref{sec:toy_model} that 
the present-day phase-space distribution for $\phi_0$ is unimodal
for {\il{r \leq 0}} and {\il{s \geq 0}}. 
The results shown in the corresponding panels of Fig.~\ref{fig:f0_top} provide some additional
insight into how such a phase-space distribution develops.  In particular, since 
all relevant decay chains through which $\phi_9$ ultimately decays to the ground state 
have similar characteristic timescales, the deposits 
from these different decay chains arrive on the $g_0(p,t)$ ``conveyor belt'' 
at roughly the same time.  Prior to this 
time --- which for the particular parameter choices adopted in these panels is quite late
in comparison with $t_I$ --- we have {\il{g_0(p,t) \approx 0}}.  Thus 
the deposits from all of these decay chains all arrive together at this late time, 
without a significant intervening time interval over which 
substantial redshifting can occur.
This then generates a single, narrow peak for $g_0(p,t)$.  

By contrast, when {\il{r > 0}} and/or {\il{s > 0}}, we have seen in 
Sect.~\ref{sec:toy_model} that the present-day phase-space distribution for $\phi_0$
is highly non-trivial and multi-modal.  The results shown in the corresponding panels
of Fig.~\ref{fig:f0_top} illustrate how this multi-modality arises as a 
consequence of sequential deposits from different decay pathways with different 
characteristic timescales.  Sizable deposits to $g_0(p,t)$ arrive at early times
and experience significant redshifts during the time interval between $t_I$ and $t_f$,
while additional deposits arrive at subsequent times.     
This is consistent with our general expectations from Sects.~\ref{sec:phase_space_evolution}
        and \ref{sec:toy_model}.

As an example, let us consider the sequence of snapshots that emerges for the case
with {\il{(r,s)=(3,4)}}, as indicated in the
lower right panel of
Fig.~\ref{fig:f0_top}.
At early times,
we essentially have only a single large peak (red curve) 
which then simply rides along the momentum conveyor
belt towards smaller momenta at later times.
Indeed, by the time of our final snapshot, this large peak
has been transported to values of $p/m_0$ in the 
approximate range $10^{-4}\lsim p/m_0 \lsim 10^{-2}$
(large bright blue peak).
However, by this time,
we see from Fig.~\ref{fig:f0_top} that another much smaller deposit onto the conveyor
belt has been made, corresponding to the small bright blue peak 
shown within the approximate range $10^{-1}\lsim p/m_0\lsim 1$.
Together, these deposits result in the final
bi-modal phase-space  
distribution shown in 
the lower right panel of Fig.~\ref{fig:f0_all}.
Indeed, this sequence of deposits onto the conveyor belt
is also consistent with the results shown in the lower right panel of Fig.~\ref{fig:decay_chain},
in which two dominant reduced decay chains make deposits onto the conveyor belt at different times. 
However, the snapshots in Fig.~\ref{fig:f0_top} 
now allow us to confirm 
that the smaller of the two peaks within 
the final bi-modal distribution
was deposited later than 
than the larger peak.
Thus the dark-sector $\phi_0$ particles 
which populate the larger peak within the $\phi_0$ phase-space distribution
were produced earlier than those which populate the smaller peak.

Interestingly, if we look even more closely at the lower right panel of Fig.~\ref{fig:f0_top}, we observe
that there is actually a very small bump (bright red) within the approximate range
$1\lsim p/m_0\lsim 10$ which was deposited even before
the larger red peak.
(Indeed, the decay chain corresponding to this deposit is too subdominant to appear in Fig.~\ref{fig:decay_chain}.)
After redshifting, this eventually contributes to the very small bump
shown within the approximate range $10^{-7}\lsim p/m \lsim 10^{-6}$ in 
the lower right panel of Fig.~\ref{fig:f0_all}.
We thus learn that the $\phi_0$ particles within this small subdominant bump were actually the first to be produced
from amongst the entire phase-space distribution.

We conclude, then, that the time-evolution of our dark sector can be highly non-trivial.
Although we have focused in this paper
on the rich consequences of the late-time 
phase-space distribution $g_0(p,t)$ for {\il{t\geq t_f}} and
its consequences for the matter power spectrum,
we now see that the internal dynamics within the 
dark sector can also exhibit its own richness.
For example, we have seen in
Fig.~\ref{fig:w_all}
that the equation of state of the dark sector can have
a non-trivial time-evolution at early times --- a time dependence which may also leave imprints in 
the cosmological evolution.   Likewise, 
although we have assumed in this paper that all of the dark-sector decays
have concluded long before the present cosmological era, 
there do exist other interesting cosmological frameworks (such as that associated with 
Dynamical Dark Matter~{\mbox{\cite{Dienes:2011ja,Dienes:2011sa}}})
in which such ``intra-ensemble'' decays 
within the dark sector 
are occurring even 
at the present time.
With only minor changes in the appropriate timescales,
the analysis we have performed here 
should therefore be relevant for such other scenarios as well.


\bibliography{references}

\end{document}